\documentclass[draft]{agujournal2019}
\usepackage{url} %
\usepackage{lineno}
\usepackage[inline]{trackchanges} 
\usepackage{soul}
\usepackage{flafter}
\usepackage{subcaption}
\usepackage{caption}
\usepackage{graphicx}
\usepackage{booktabs} %
\usepackage{multirow}
\usepackage{makecell}
\usepackage{array}
\usepackage{amsfonts}
\usepackage{amsmath}
\usepackage{nameref}
\usepackage{amssymb}
\usepackage{siunitx}
\usepackage{algorithm}
\usepackage{changepage}
\usepackage{algpseudocode}
\usepackage{rotating}
\usepackage{pifont}
\usepackage{xcolor}
\usepackage[colorinlistoftodos,prependcaption,textsize=tiny]{todonotes}
\usepackage{morefloats}
\usepackage{xr}
\usepackage[utf8]{inputenc}

\usepackage{tikz}
\usetikzlibrary{bayesnet}
\usetikzlibrary{arrows.meta, positioning, shapes, fit, backgrounds, calc}
\usetikzlibrary{arrows, positioning, calc, decorations.pathreplacing}

\journalname{JGR: Solid Earth}

\usepackage{amsmath,amsfonts,bm}

\def\eqref#1{equation~\ref{#1}}

\def\1{\bm{1}}

\def\vtheta{{\bm{\theta}}}

\def\vc{{\bm{c}}}

\def\ve{{\bm{e}}}

\def\vx{{\bm{x}}}
\def\vy{{\bm{y}}}

\DeclareMathAlphabet{\mathsfit}{\encodingdefault}{\sfdefault}{m}{sl}
\SetMathAlphabet{\mathsfit}{bold}{\encodingdefault}{\sfdefault}{bx}{n}

\newcommand{\pdata}{p_{\rm{data}}}

\newcommand{\Ls}{\mathcal{L}}

\def\dd{{\mathrm{d}}}

\begin{document}

\title{Broadband Ground-Motion Synthesis by Conditioning Denoising Diffusion Models on Low-Frequency Waveforms}

\authors{Kadek Hendrawan Palgunadi\affil{1, \dagger \ddagger}, Simon Dirmeier\affil{2, \dagger}, Maria Koroni\affil{1}, Laura Ermert\affil{3}, Men-Andrin Meier\affil{1,4}}

\affiliation{1}{Swiss Seismological Service (SED), ETH Zürich, Switzerland}
\affiliation{2}{Swiss Data Science Center (SDSC), ETH Zürich, Switzerland}
\affiliation{3}{Univ. Grenoble Alpes, Univ. Savoie Mont Blanc, CNRS, IRD, Univ. Gustave Eiffel, ISTerre, Grenoble, France}
\affiliation{4}{Earth and Planetary Science Department, ETH Zürich, Switzerland}
\affiliation{\dagger}{Equal Contribution}
\affiliation{\ddagger}{Now: Department of Geophysical Engineering, Institut Teknologi Sepuluh Nopember, Indonesia}
\correspondingauthor{Kadek Hendrawan Palgunadi}{kadek.palgunadi@its.ac.id}

\begin{keypoints}
\item HighFEM-2 is a generative diffusion model that synthesizes 0.1--50 Hz three-component ground motion conditioned on a low-frequency waveform.
\item Conditioning on per-event low-frequency waveforms yields broadband synthetics for entire earthquakes that share source and path terms.
\item Synthetics reproduce observed peak amplitudes, envelopes, spectra and integrated displacement of Italian records with small bias.
\end{keypoints}

\begin{abstract}
Seismic hazard and structural analyses require broadband ground-motion time histories, yet the high-frequency band (above 1~Hz) is the hardest to obtain: physics-based simulations are too costly at these frequencies, and stochastic methods lose the deterministic phase, directivity, and long-period coherence that control structural demand. We present HighFEM-2, a latent denoising-diffusion waveform model trained with rectified flow matching on a multi-modal diffusion transformer. It synthesizes three-component acceleration up to 50~Hz, conditioned on a low-frequency waveform below the source corner frequency and on magnitude, hypocenter and station location, and $V_{S30}$. The low-frequency waveform supplies the deterministic source, path, and directivity, so the network learns only the regional high-frequency scattering. Conditioning each station on its own waveform lets HighFEM-2 generate records for all stations of an earthquake that share its source and low-frequency path terms. Trained on 195{,}138 Italian INSTANCE records ($M_w$ 1.4--6.5, within 200~km), HighFEM-2 produces realistic waveforms: it reproduces observed peak amplitudes with small bias ($0.042 \pm 0.088$ log units for PGA, $0.057 \pm 0.090$ for PGV) and matches observed envelopes, Fourier spectra, and integrated velocity and displacement.
The conditioning waveform can come from a simulation or a filtered record, and ensembles of realizations sample the high-frequency variability given that input. HighFEM-2 thus provides fast, regionally calibrated broadband time histories for hazard and risk assessment, including near-fault scenarios where recordings are scarce.

\end{abstract}

\section*{Plain Language Summary}
Earthquakes shake the ground across a wide range of frequencies, and engineers need realistic predictions of this shaking to design safe structures and assess seismic hazard. Detailed computer simulations can accurately reproduce the slower, long-period part of the shaking, but they are far too expensive to routinely compute the rapid, high-frequency part that damages many structures. Simpler statistical methods are fast but lose important physical detail. We developed a generative machine-learning model, HighFEM-2, that combines the physical realism of simulations with the speed of statistical methods. Given the long-period, low-frequency part of a ground-motion record together with basic information about the earthquake and the recording site, the model generates the full broadband shaking, including realistic high-frequency content up to 50~Hz. Trained on nearly 200,000 Italian earthquake recordings, HighFEM-2 reproduces the observed shaking amplitudes and waveforms well, and can produce many possible versions of the shaking to represent its natural variability. Because the low-frequency input can come from a recording or a physics-based simulation, the method is a fast, flexible way to predict realistic earthquake ground motion, especially where recordings of the strongest shaking are scarce.

\section{Introduction}
Ground-motion prediction underpins seismic hazard assessment, building-code development, and the training of earthquake early-warning algorithms. Two fundamental questions remain unresolved: How can we accurately predict high-frequency ground motions ($> 1$ Hz) in data-scarce regions where limited instrumental recordings preclude empirical calibration? How can we efficiently generate broadband synthetics (0.1--50~Hz) that sample ground-motion variability, at computationally feasible cost, in the high-frequency band where physics-based simulation becomes intractable?

The first question is important because instrumental seismic networks exhibit severe geographic bias, with well-instrumented regions concentrated in a few seismically active, affluent areas (e.g., California, Japan, parts of Europe), while developing countries, high-altitude regions, and tectonically active zones in Central Asia, South America, and Africa remain sparsely instrumented. In addition, the digital strong-motion record is short, typically dating only to the 1980s and often limited to triggered recordings from a handful of stations, further restricting the data available for empirical calibration. This prohibits the accumulation of event recordings required for robust GMMs development and forces reliance on extrapolation from distant tectonic analogues that introduces factors of $\approx 2 - 3$ epistemic uncertainty in predicted amplitudes  \cite{douglas2018capturing, stewart2015selection}. The second question arises because several applications impose hard runtime constraints: training and testing earthquake early-warning algorithms requires large, realistic waveform sets spanning rare magnitude-distance combinations, which recorded data alone cannot supply; probabilistic seismic hazard assessment (PSHA) for building codes requires $10^4 - 10^6$ scenario evaluations to integrate over magnitude-location-rupture distributions with annual exceedance probabilities spanning $10^{-2}$ to $10^{-4}$, making deterministic high-frequency simulations prohibitively expensive; and uncertainty quantification in seismic risk assessment demands ensemble generation to characterize ground-motion variability, requiring hundreds to thousands of realizations that are computationally infeasible through pure physics at engineering-relevant frequencies of 0.1 - 50 Hz where structural response spectra peak and damage accumulates, especially in critical infrastructure (e.g., nuclear power plant).

Existing approaches address these questions only partially. Ground-motion prediction has traditionally relied on three complementary frameworks: physics-based simulations that numerically solve the elastodynamic wave equation (e.g., via finite-difference, finite-element, or spectral-element methods) through heterogeneous 3D velocity models; stochastic methods that synthesize high-frequency motion from omega-squared spectral models parameterized by stress drop, quality factor $Q(f)$, and high-frequency attenuation $\kappa_0$; and empirical Ground Motion Models (GMMs) that regress intensity measures (e.g., PGA, 5\%-damped pseudo-spectral acceleration PSA($T$)) on source, path, and site parameters from strong-motion databases. These frameworks differ not only in cost and accuracy but in the completeness of their output: empirical GMMs predict scalar intensity measures, whereas simulation-based and generative approaches produce full waveforms, a more complete description of the ground motion from which scalar measures, response spectra, and duration metrics can all be derived. In the following, we distinguish scalar GMMs from waveform-generating models, and refer to data-driven models of the latter kind as generative waveform models (GWMs).

Within these frameworks, simulated ground-motion time series come from fully deterministic, fully stochastic, or hybrid methods. Deterministic simulations, for example with spectral-element methods, model wave propagation through 3D velocity structure and surface topography \cite{lee2008, shi2013rupture, taufiqurrahman2022broadband, gallovivc2023broadband, palgunadi2025ground}; regional scenarios reach maximum resolved frequencies of roughly 3--15~Hz, the upper end only in highly optimized or spatially limited cases, and routine regional simulations are resolved to substantially lower frequencies. Stochastic finite-fault methods \cite{atkinson2015implementation, tang2022gmss2, ma2024improvements} filter random Gaussian noise with Brune's omega-squared source spectrum and empirical attenuation relationships, which is efficient but discards deterministic features such as coherent phase, wave arrivals, and site-specific amplification. Hybrid methods combine a deterministic simulation at low frequencies ($\leq 1$~Hz) with a stochastic or semi-stochastic method at high frequencies \cite{akinci2017physics, rezaeian2017simulation, paolucci2021bb, li2025probability}, as in the widely used Graves--Pitarka method, which uses kinematic rupture models with heterogeneous slip, rupture speed, and rise time \cite{graves2010broadband}. Pushing deterministic methods above $\sim$5--10~Hz remains prohibitive, because cost and memory requirements grow sharply with frequency, and because the small-scale velocity heterogeneity and scattering structure that generate realistic high-frequency radiation are generally not constrained well enough to be modeled deterministically.

Three developments now make data-driven broadband prediction feasible: physics-based waveform simulation for the deterministic low frequencies, GWMs for stochastic high-frequency synthesis, and curated strong-motion datasets for model training. \citeA{paolucci_broadband_2018} developed artificial neural network models to compute response spectra and Fourier amplitude spectra at high frequencies based on their low-frequency counterparts from physics-based simulations, extending the frequency range of deterministic simulations. Generative adversarial networks entered seismological practice early, e.g., for augmenting earthquake early-warning classifiers with synthetic signals \cite{li2018seismic}. \citeA{gatti2020towards} used autoencoders and generative adversarial networks to generate hybrid simulations combining physics-based numerical simulations and seismic databases, transforming synthetic earthquake ground motion accurate up to 5 Hz into broader-band predictions that mimic recorded seismograms. \citeA{okazaki2021simulation} introduced an approach combining neural networks with t-distributed stochastic neighbourhood embedding (t-SNE) to establish relationships between high-frequency and low-frequency acceleration time envelopes, though this method requires partial retraining for each specific station. Directly generative approaches followed: \citeA{florez_data-driven_2022} trained a GAN to produce three-component broadband accelerograms conditioned on magnitude, distance, and $V_{S30}$, and \citeA{esfahani_tfcgan_2023} generated Fourier amplitude spectra with a conditional GAN, recovering time histories via phase retrieval. Extending this line of work to full waveform generation, our earlier model HighFEM-1 \cite{palgunadi2025high}, trained on Japanese strong-motion data, synthesizes broadband three-component records with a latent denoising-diffusion model conditioned on scalar source, path, and site parameters. Fully stochastic GWMs, including HighFEM-1, produce realistic single-station records but sample each record independently. They cannot represent the source term shared by all stations of an earthquake, nor the correlation between those stations. Deterministic simulations capture both, but only at low frequencies. In HighFEM-2, a band-limited deterministic waveform supplies the source, path, and directivity, and the generative model adds only the high-frequency content that deterministic methods cannot resolve.

A parallel line of research has employed advanced computational methods to accelerate physics-based predictions: \citeA{rekoske2023instantaneous} developed a reduced-order modelling approach based on interpolated proper orthogonal decomposition to enable instantaneous prediction of peak ground velocities from regional 3D physics-based wave propagation simulations. Fourier Neural Operators (FNOs), which, unlike conventional solvers on a fixed spatial grid with explicit time-stepping, learn resolution-invariant mappings between function spaces, have since been applied to seismic wave propagation: \citeA{lehmann20243d} developed a Factorized Fourier Neural Operator (F-FNO) for 3D elastic wave propagation, achieving memory-efficient predictions from heterogeneous geological models, which was subsequently extended by \citeA{lehmann2025multiple} into the Multiple-Input Fourier Neural Operator (MIFNO), which includes source properties and predicts surface ground motion from any earthquake point source with varying position and orientation; it was the first surrogate model with the flexibility of an earthquake simulator for varying sources and material properties. \citeA{aquib2024broadband} introduced a machine-learning framework using Fourier Neural Operators that generates broadband ground motions (up to 20 Hz) from low-frequency waveforms ($\leq$ 1 Hz) through two separate models: an Amplitude Model that establishes the relationship between low-frequency and high-frequency Fourier amplitude spectra, and a Phase Model that learns phase characteristics from filtered normalized waveforms in the time domain, with predictions conditioned on magnitude, epicentral distance, $V_{S30}$, and terrain ruggedness index, incorporating a Wasserstein GAN framework to generate multiple realizations and quantify ground-motion variability. \citeA{pan2025lfw2bbp} developed LFW2BBP, which takes a complementary parameter-focused approach using residual neural networks to map physics-based simulated low-frequency waveforms directly to broadband ground motion parameters (PGA, PSA) through multi-domain feature extraction from time-domain waveforms, time-frequency spectra, and spectral acceleration, a computationally efficient alternative for engineering applications that require intensity measures rather than full time series. \citeA{perrone2025integrating} integrated these approaches by combining a neural operator that approximates the elastodynamics Green's operator with a denoising diffusion probabilistic model trained to correct mid-frequency spectral falloff in generated ground-motion time series, which improved frequency biases and goodness-of-fit scores at low inference cost.

In this study, we introduce HighFEM-2, a GWM for local-scale broadband synthesis that differs from HighFEM-1 in two respects: (i) it conditions on a deterministic low-frequency waveform, so each synthetic inherits the source, path, and directivity information of a specific earthquake rather than being an independent stochastic realization; and (ii) it is trained and evaluated on Italian strong-motion data, yielding a regionally calibrated model. HighFEM-2 is a latent denoising-diffusion waveform model, trained with rectified flow matching on a multi-modal diffusion transformer (Section~\ref{sec:methods}).  Conditioning on the low-frequency waveform together with magnitude, hypocenter and station location, and site conditions (parameterized by $V_{S30}$), HighFEM-2 generates broadband waveforms with frequency content up to 50~Hz. Our objective is to synthesize ground-motion records that remain physically consistent at low frequencies and statistically consistent with real records, across the evaluated seismological and machine-learning metrics, over the broadband frequency range. The low-frequency waveform supplies the deterministic, source- and path-controlled part of the motion; here we obtain it by low-pass filtering observed records at the magnitude-dependent corner frequency, but in application it can also be drawn from a physics-based simulation (demonstrated in Section~\ref{sec:dynamic_rupture}), a recorded analogue event, or, in principle, a waveform-generating GMM (untested here; Section~\ref{sec:limitation}).

\section{Methods}
\label{sec:methods}
Following \citeA{palgunadi2025high}, we develop a conditional GWM aiming to model the conditional probability distribution of high-frequency waveforms $\vy$ given a deterministic low-frequency waveform $\vx$ and conditioning parameters $\vc$ 
\begin{equation}
\pdata(\vy | \vx, \vc)
\label{eqn:data-distribution}
\end{equation}
In the following, we briefly introduce relevant methodology to describe HighFEM-2.

\subsection{Flow matching}
\label{sec:flow-matching}
We model \eqref{eqn:data-distribution} using a continuous normalizing flow \cite{lipman2023flow} trained using rectified flow matching \cite{liu2023flow}. Continuous flow models are time dependent maps describing the trajectory of a base point $\vy$ via an ordinary differential equation
\begin{equation}
\begin{split}
  \dd \phi_t(\vy) &= u_t(\phi_t(\vy)) \dd  t \\ 
    \phi_0(\vy)  &= \vy
\end{split}
\label{eqn:ode}
\end{equation}
where $u: [0, 1] \times \mathbb{R}^{d_{\vy}} \rightarrow \mathbb{R}^{d_{\vy}}$ is a velocity field that induces the trajectory of $\vy$ and $\vy_t = \phi_t(\vy)$ is the solution of the ODE when integrating it from time $0$ up to time $t$. Intuitively, flow matching learns a velocity field that moves a sample of random noise, step by step, into a realistic high-frequency waveform. Rectified flow chooses straight paths between noise and data, which can be followed accurately with few steps; this is why sampling needs only 25 network evaluations.

Given a simple user-defined base distribution $p_0(\vy)$, e.g., a Gaussian distribution, and a complex target distribution $p_1(\vy)$ which is close to the data distribution $\pdata(\vy | \vx, \vc)$, the idea of rectified flow matching is to regress a velocity field $v_\vtheta(\vy, \vx, \vc, t)$ that is parameterized by a neural network onto the velocity field $u_t$ by minimizing a simple mean-squared error objective
\begin{equation}
    \Ls(\vtheta) = \mathbb{E}_{t\sim \mathcal{U}(0, 1), \vy_1 \sim \pdata }\left[  \left\lVert  u_t(\vy_t) - v_\vtheta(\vy_t,  \vx, \vc, t)\right\rVert^2\right]
\end{equation}

While there are many approaches to choose $u_t$, here we set $\vy_t = t \vy_1 + (1  -t) \vy_0$, yielding
\begin{equation}
     \dot{\vy}_t = \vy_1 - \vy_0
\end{equation}
in order to construct a "straight" ODE which would allow sampling with only a few neural network evaluations.

As a consequence, this yields the simplified mean-squared error loss
\begin{equation}
    \Ls(\vtheta) = \mathbb{E}_{t\sim \mathcal{U}(0, 1), \vy_1 \sim \pdata, \vy_0 \sim p_0 }\left[  \left\lVert (\vy_1 - \vy_0) - v_\vtheta(\vy_t,  \vx, \vc, t)\right\rVert^2\right]
\end{equation}
The objective is minimized by first sampling a time point $t \sim \mathcal{U}$ from a uniform distribution, then sampling a batch from the data $\vy_1 \sim p_{\text{data}}$ and a batch from the base distribution $\vy_0 \sim p_0$, then computing the differential $\dot{\vy}$, and finally estimating the mean-squared error.

With a trained model $v_\vtheta$, we can generate synthetic waveforms by solving the ODE in \eqref{eqn:ode} starting from a base point $\vy \sim p_0$, a deterministic low-frequency waveform $\vx$ and some condition $\vc$, and integrating it until $t=1$. This can be done using any contemporary ODE solver, for instance, as implemented in the Python package \texttt{scipy}. Here, we use a simple custom Euler solver using $25$ equidistantly separated time points $t \in [0, 1]$. The conditioning waveform is an input to the velocity field at every step of the integration.

\subsection{Neural network}
\label{sec:neural-network}
\begin{sidewaysfigure}
\centering
\begin{subfigure}[b]{0.48\textwidth}
    \scalebox{.446}{
\begin{tikzpicture}[            
        block/.style={rectangle, rounded corners, minimum width=70pt, minimum height=15pt, draw, thick, font=\ttfamily\small},
        trainableblock/.style={block, fill=blue!10},
        frozenblock/.style={block,fill=blue!10},
        fnblock/.style={block,fill=green!10},
        operation/.style={draw, fill=red!10, circle, minimum width=2em,thick},
        tensor/.style={draw, fill=red!20, circle, minimum width=20pt, minimum height=15pt,thick},
        input/.style={block,fill=red!40,thick},
    ]

    \pgfdeclarelayer{arrowlayer}
    \pgfdeclarelayer{residuallayer}
    \pgfdeclarelayer{decorations}
    \pgfdeclarelayer{bg}
    \pgfsetlayers{bg,arrowlayer,residuallayer,main,decorations}

    \node[] (context) {%
         \includegraphics[width=2cm]{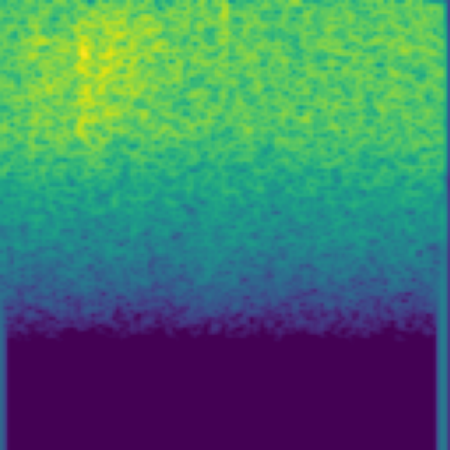}%
    };
    \node [trainableblock, above=0.25cm of context] (c_vae_enc) {VAE Encoder};
    \node [fnblock, above=0.2cm of c_vae_enc] (c_stft) {STFT + log};
    \node [tensor, rectangle, above=0.3cm of c_stft] (c_wave) {LF waveform};
    \node [fnblock, below=0.5cm of context] (c_patch) {Patching};
    \node [trainableblock, below=0.2cm of c_patch] (c_lin) {Linear};
    \node [operation, below=0.5cm of c_lin] (c_image_plus_pos) {$+$};
    \node [trainableblock, right=0.75cm of c_image_plus_pos,align=center] (c_pos_emb) {Sinusoidal\\Embedding};
    \node [tensor,rectangle, below=0.5cm of c_image_plus_pos] (c_in) {$\text{context}$};
    \node[right=2.5cm of context] (target) {%
        \includegraphics[width=2cm]{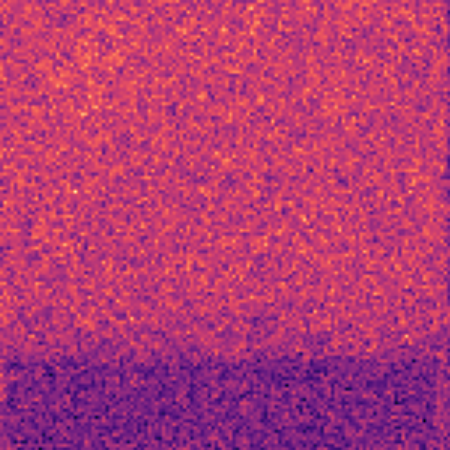}%
    };
    \node [trainableblock, above=0.25cm of target] (t_vae_enc) {VAE Encoder};
    \node [fnblock, above=0.2cm of t_vae_enc] (t_stft) {STFT + log};
    \node [tensor, rectangle, above=0.3cm of t_stft] (t_wave) {HF waveform};
    \node [fnblock, below=0.5cm of target] (patching) {Patching};
    \node [trainableblock, below=0.2cm of patching] (patch_proj) {Linear};
    \node [operation, below=0.5cm of patch_proj] (image_plus_pos) {$+$};
    \node [tensor, rectangle, below=0.5cm of image_plus_pos] (x_in) {$\text{target}$};
    \node [tensor, rectangle, left=1cm of c_in] (emb_final) {$\text{emb}$};
    \node [operation, left=1cm of emb_final] (y_plus) {$+$};
    \node [trainableblock, below=0.5cm of y_plus] (t_emb) {MLP};
    \node [fnblock, below=0.2cm of t_emb] (t_enc) {Fourier Embedding};
    \node [tensor, rectangle, below=0.5cm of t_enc] (timestep) {$t$};
    \node [trainableblock, above=0.5cm of y_plus] (y_mlp) {MLP};
    \node [tensor, rectangle, above=0.5cm of y_mlp] (condition) {${\vc}$};
    
    \begin{pgfonlayer}{arrowlayer}
        \draw [->] (context) -- (c_image_plus_pos);
        \draw [->] (c_pos_emb) -- (c_image_plus_pos);
        \draw [->] (c_image_plus_pos) -- (c_in);
        \draw [->] (c_wave) -- (c_stft);
        \draw [->] (c_vae_enc) -- (context);
        \draw [->] (t_wave) -- (t_stft);
        \draw [->] (t_vae_enc) -- (target);
    \end{pgfonlayer}

    \begin{pgfonlayer}{arrowlayer}
        \draw [->] (timestep) -- (t_emb) -- (y_plus);        
        \draw [->] (condition) -- (y_mlp) -- (y_plus);
        \draw [->] (y_plus) -- (emb_final);
    \end{pgfonlayer}
        
    \begin{pgfonlayer}{arrowlayer}
        \draw [->] (target) -- (image_plus_pos);
        \draw [->] (c_pos_emb) -- (image_plus_pos);
        \draw [->] (image_plus_pos) -- (x_in);
    \end{pgfonlayer}

    \node [trainableblock, below=1.5cm of $(c_in)!0.5!(x_in)$, minimum width=5cm] (attn1) {MM-DiT-Block 1};
    \node [trainableblock, below=0.3cm of attn1, minimum width=5cm] (attn2) {MM-DiT-Block 2};
    \node [below=0.4cm of attn2, minimum width=5cm] {$\ldots$};
    \node [trainableblock, below=1.3cm of attn2, minimum width=5cm] (attnN) {MM-DiT-Block N};

    \coordinate (attn1_mid_left) at ($(attn1.north west)!0.5!(attn1.north)$);
    \coordinate (attn1_mid_right) at ($(attn1.north east)!0.5!(attn1.north)$);

    \coordinate (attnN_mid_left) at ($(attnN.north west)!0.5!(attnN.north)$);
    \coordinate (attnN_mid_right) at ($(attnN.north east)!0.5!(attnN.north)$);

    \node [fnblock, below=1.5cm of attnN_mid_right] (layernormout) {LayerNorm};
    \node [trainableblock, below=0.2cm of layernormout] (out_mod) {Modulation};
    \node [trainableblock, below=0.2cm of out_mod] (out_mlp) {Linear};
    \node [fnblock, below=0.2cm of out_mlp] (depatch) {Unpatching};
    \node [tensor, rectangle, below=0.5cm of depatch] (output) {output};
    \node [trainableblock, below=0.3cm of output] (vae_dec) {VAE Decoder};
    \node [fnblock, below=0.2cm of vae_dec] (phase_rec) {Phase reconstruction};
    \node [tensor, rectangle, below=0.3cm of phase_rec] (ts_out) {time series};

    \coordinate (attn1_left) at ($(attn1.west)+(-2, 0)$);

    \begin{pgfonlayer}{arrowlayer}
        \draw [->] (x_in) -| (attn1_mid_right) -- (output);
        \draw [->] (c_in) -| (attn1_mid_left) -- (attnN_mid_left);

        \draw [->] (emb_final) -| (attn1_left) -- (attn1);
        \draw [->] (attn1_left) |- (attn2);
        \draw [->] (attn1_left) |- (attnN);
        \draw [->] (attn1_left) |- (layernormout);
        \draw [->] (output) -- (vae_dec);
        \draw [->] (phase_rec) -- (ts_out);
    \end{pgfonlayer}

    \begin{pgfonlayer}{bg}
        \draw [fill=blue!5] ($(attn1.north east)+(1, 0.5)$) -| ($(attnN.south west)+(-1, -0.5)$) -| cycle;
    \end{pgfonlayer}

\end{tikzpicture}
}
\caption{Neural network architecture.}
\end{subfigure}
    ~ 
    \hfill
\begin{subfigure}[b]{0.48\textwidth}
    \scalebox{.562}{
\begin{tikzpicture}[
        block/.style={rectangle, rounded corners, minimum width=70pt, minimum height=15pt, draw, thick, font=\ttfamily\small},
        trainableblock/.style={block, fill=blue!10},
        frozenblock/.style={block,fill=blue!10},
        fnblock/.style={block,fill=green!10},
        operation/.style={draw, fill=red!10, circle, minimum width=2em,thick},
        tensor/.style={draw, fill=red!20, circle, minimum width=20pt, minimum height=15pt,thick},
        input/.style={block,fill=red!40,thick},
    ]

    \pgfdeclarelayer{arrowlayer}
    \pgfdeclarelayer{residuallayer}
    \pgfsetlayers{arrowlayer,residuallayer,main}
    
    \newcommand{\inblockdistance}{0.2cm}
    \newcommand{\outblockdistance}{0.5cm}
    \newcommand{\attnskip}{4.75cm}
    \newcommand{\leftmodpos}{-2.6}
    \newcommand{\rightmodpos}{7.2}

    \node [tensor, rectangle] (c) {context};
    \node [tensor, rectangle, right=4cm of c] (x) {target};

    \node [below=\outblockdistance of c, fnblock] (c_ln1) {LayerNorm};
    \node [below=\inblockdistance of c_ln1, trainableblock] (c_mod) {Modulation};
    \node [below=\inblockdistance of c_mod, trainableblock] (c_lin) {Linear};

    \path let \p1=($(x)!0.5!(c)$) in node [fnblock, minimum width=7cm, minimum height=1.5cm, align=center] at (\x1, -6) (attn) {\vphantom{Attention}\\ \Large Attention};
    \coordinate [below=0.3cm of attn] (attn_split);

    \node [below=\attnskip of c_lin, trainableblock] (c_proj) {Linear};
    \node [below=\inblockdistance of c_proj, fnblock] (c_proj_gated) {Scale};
    \node [below=\outblockdistance of c_proj_gated, operation] (c_attn_out) {$+$};
    \node [below=\outblockdistance of c_attn_out, fnblock] (c_ln2) {LayerNorm};
    \node [below=\inblockdistance of c_ln2, trainableblock] (c_mod2) {Modulation};
    \node [below=\inblockdistance of c_mod2, trainableblock] (c_mlp) {MLP};
    \node [below=\inblockdistance of c_mlp, fnblock] (c_mlp_gated) {Scale};
    \node [below=\outblockdistance of c_mlp_gated, operation] (c_out) {$+$};

    \coordinate [above left=2cm and 1.75cm of c] (cx) {};
    \node [fnblock, below=0.6cm of cx] (y_silu) {SiLU};
    \node [tensor, rectangle, above=0.5cm of y_silu] (y) {emb};
    \node [trainableblock, below=\inblockdistance of y_silu] (y_linear) {Linear};
    \coordinate (y_split) at ($(y_linear.south)+(-0, -0.3)$);

    \coordinate (c_attn_out_left) at ($(c_attn_out) + (-2, 0)$);

    \begin{pgfonlayer}{arrowlayer}
          \draw [->] (y) -- (y_silu); 
       \draw [-] (c) -- (c_lin);
        \draw [->] (c) -- (c_ln1);
        \draw [->] (attn) -- (attn_split) -| (c_proj);
        \draw [->] (c_proj) -- (c_attn_out);
        \draw [->] (c_attn_out) -- (c_out);

        \draw [<-] (y.north) --++ (0, 0.4);
        \draw (y) -- (y_linear) |- (y_split);

        \draw [->] (y_split) |- ($(c_mod.north west)!1!(c_mod.west)$);
        \draw [->] (y_split) |- ($(c_proj_gated.south west)!1!(c_proj_gated.west)$);        
        \draw [->] (y_split) |- ($(c_mlp_gated.north west)!1!(c_mlp_gated.west)$);
        \draw [->] (y_split) |- ($(c_mod2.south west)!1!(c_mod2.west)$);

    \end{pgfonlayer}
    \begin{pgfonlayer}{residuallayer}
        \draw [<-, line width=2mm, draw=white] (c) --++ (0, 3);
        \draw [->, line width=2mm, draw=white] (c) -| (c_attn_out_left) -- (c_attn_out);
        \draw [->, line width=2mm, draw=white] (c_attn_out_left) |- (c_out);
        \draw [->, line width=2mm, draw=white] (c_out) --++ (0, -1);
        \draw [<-] (c) --++ (0, 3.1);
        \draw [->] (c) -| (c_attn_out_left) -- (c_attn_out);
        \draw [->] (c_attn_out_left) |- (c_out);
        \draw [->] (c_out) --++ (0, -1);
    \end{pgfonlayer}

    \node [below=\outblockdistance of x, fnblock] (x_ln1) {LayerNorm};
    \node [below=\inblockdistance of x_ln1, trainableblock] (x_mod) {Modulation};
    \node [below=\inblockdistance of x_mod, trainableblock] (x_lin) {Linear};

    \node [below=\attnskip of x_lin, trainableblock] (x_proj) {Linear};
    \node [below=\inblockdistance of x_proj, fnblock] (x_proj_gated) {Scale};
    \node [below=\outblockdistance of x_proj_gated, operation] (x_attn_out) {$+$};
    \node [below=\outblockdistance of x_attn_out, fnblock] (x_ln2) {LayerNorm};
    \node [below=\inblockdistance of x_ln2, trainableblock] (x_mod2) {Modulation};
    \node [below=\inblockdistance of x_mod2, trainableblock] (x_mlp) {MLP};
    \node [below=\inblockdistance of x_mlp, fnblock] (x_mlp_gated) {Scale};
    \node [below=\outblockdistance of x_mlp_gated, operation] (x_out) {$+$};

    \coordinate [above right=2cm and 1.75cm of x] (yx) {};
    \node [fnblock, below=0.6cm of yx] (y_silu) {SiLU};
    \node [trainableblock, below=\inblockdistance of y_silu] (y_linear) {Linear};
    \coordinate (y_split) at ($(y_linear.south)+(0, -0.3)$);

    \coordinate (x_attn_out_left) at ($(x_attn_out) + (2, 0)$);

    \begin{pgfonlayer}{arrowlayer}        
        \draw [-] (x) -- (x_lin);
        \draw [->] (x) -- (x_ln1);
        \draw [->] (attn) -- (attn_split) -| (x_proj);
        \draw [->] (x_proj) -- (x_attn_out);
        \draw [->] (x_attn_out) -- (x_out);

        \draw (y) -| (yx) -- (y_linear) |- (y_split);        
        \draw [->] (yx) -- (y_silu);        
        \draw [->] (y_split) |- ($(x_mod.south east)!1!(x_mod.east)$);
        \draw [->] (y_split) |- ($(x_proj_gated.south east)!1!(x_proj_gated.east)$);
        \draw [->] (y_split) |- ($(x_mod2.south east)!1!(x_mod2.east)$);
        \draw [->] (y_split) |- ($(x_mlp_gated.south east)!1!(x_mlp_gated.east)$);
    \end{pgfonlayer}
    
    \begin{pgfonlayer}{residuallayer}
        \draw [<-, line width=2mm, draw=white] (x) --++ (0, 3);
        \draw [->, line width=2mm, draw=white] (x) -| (x_attn_out_left) -- (x_attn_out);
        \draw [->, line width=2mm, draw=white] (x_attn_out_left) |- (x_out);
        \draw [->, line width=2mm, draw=white] (x_out) --++ (0, -1);
        \draw [<-] (x) --++ (0, 3.1);
        \draw [->] (x) -| (x_attn_out_left) -- (x_attn_out);
        \draw [->] (x_attn_out_left) |- (x_out);
        \draw [->] (x_out) --++ (0, -1);
    \end{pgfonlayer}

    \node [below=0.1cm of attn.north] (k) {\Large \emph{K}};
    \node [left=2.5cm of k.center] (q) {\Large \emph{Q}};
    \node [right=2.5cm of k.center] (v) {\Large \emph{V}};
    \node [above=0.25cm of q, operation] (q_join) {$\odot$};
    \node [above=0.75cm of k, operation] (k_join) {$\odot$};
    \node [above=1.25cm of v, operation] (v_join) {$\odot$};
    
    \begin{pgfonlayer}{arrowlayer}
        \draw (k_join) -- (k);

        \draw [->] ($(c_lin)-(0.25, 0)$) -| (q_join);  
        \draw [line width=2mm, white] ($(x_lin)-(0.9, 0)$) |- (q_join);
        \draw [->] ($(x_lin)-(0.9, 0)$) |- (q_join);
        \draw (q_join) -- (q);

        \draw [->] ($(c_lin)+(0.4, 0)$) |- (k_join);
        \draw [line width=2mm, white] ($(x_lin)-(0.4, 0)$) |- (k_join);
        \draw [->] ($(x_lin)-(0.4, 0)$) |- (k_join);

        \draw [line width=2mm, white] ($(c_lin)+(0.9, 0)$) |- (v_join);
        \draw [->] ($(c_lin)+(0.9, 0)$) |- (v_join);
        \draw [->] ($(x_lin)+(0.25, 0)$) -| (v_join);
        \draw (v_join) -- (v);
    \end{pgfonlayer}
\end{tikzpicture}
}
\caption{MM-DiT block.}
\end{subfigure}
\caption{Neural network architecture of HighFEM-2, based on the multi-modal diffusion transformer (MM-DiT) of \protect\citeA{esser2024scaling}. Waveforms are represented as log-spectrograms and compressed by variational autoencoders into $32{\times}32{\times}8$ latents; the network operates in this latent space, and generated latents are decoded and reconstructed to time series (Section~\ref{sec:neural-network}).
(a)~The low-frequency context and noisy high-frequency target are patched and embedded as separate token streams; conditioning parameters $\vc$ and flow-matching time $t$ form a joint embedding; the network stacks $N{=}12$ MM-DiT blocks.
(b)~Within each MM-DiT block, the two streams are processed with separate weights, interact through joint attention, and are modulated by the conditioning embedding. LF waveform: conditioning waveform.} 
\label{fig:mmdit-architecture}
\end{sidewaysfigure}

To model the bi-modality of the task, we employ a multi-modal diffusion transformer architecture (MM-DiT, \citeA{esser2024scaling}). The network is based on a vision transformer architecture taking as an input the high-frequency waveform $\vy$, but in addition conditions on a low-frequency waveform $\vx$ and time $t$ \cite{peebles2023scalable,dosovitskiy2021an} (see Figure~\ref{fig:mmdit-architecture} for an overview and \ref{app:model-architecture} for details).

In short, the high-frequency and low-frequency waveforms are first converted to spectrogram representations. Both are then transformed using two different variational autoencoders (VAEs), yielding two latent representations, $\vy \in \mathbb{R}^{32 \times 32 \times 8}$ for the high-frequency waveform and $\vx \in \mathbb{R}^{32 \times 32 \times 8}$ for the conditioning waveform, respectively (where $8$ is the number of latent channels of the autoencoder). The network $v_\vtheta$ first converts both representations $\vy$ and $\vx$ into lists of ``patches'' yielding two tensors $\vy' \in \mathbb{R}^{256 \times 32}$ (the patches of the high-frequency waveform) and $\vx' \in \mathbb{R}^{256 \times 32}$ (the patches of the conditioning waveform). Both time $t$ and physical parameters $\vc$ are preprocessed using multi-layer perceptrons (MLPs) and summed together yielding an embedding $\ve$. The variables $\vy'$ and $\vx'$ are further transformed and, together with $\ve$, then fed through a set of customized transformer blocks. In each of these, $\vy'$ and $\vx'$ are first processed in parallel, fed through a scaled-dot product attention mechanism and finally post-processed. The conditioning information $\ve$ is modulated at different stages of the transformer block using adaLN-zero and scaling operations (see \citeA{peebles2023scalable} for definitions and details). After 12 transformer blocks, the high-frequency component of the output of the last block (corresponding to $\vy'$) is then reshaped to its original image dimensionality ($32 \times 32 \times 8$) and returned as a result.

The spectrogram representation is obtained by applying a short-time Fourier transform to each acceleration component (256 frequency channels, hop size 32; \ref{app:model-architecture}) and taking the logarithm of the amplitudes. After ODE integration, the generated latent is decoded by the VAE into a log-spectrogram, from which the three-component time series is recovered using the phase-reconstruction procedure of \citeA{palgunadi2025high}.

\section{Data and Model Training}
\label{sec:data_description}
\subsection{Data Selection}
\label{sec:data-selection}
We use open access Italian three-component, INSTANCE (\url{https://www.pi.ingv.it/banche-dati/instance/}, last access October 2025), strong-motion seismic waveforms recorded from 1 January 2005 to 31 January 2020 \cite{michelini2021instance}. The dataset comprises 315,255 events and more than three million waveforms, with event magnitudes ranging from 0.0 to 6.5 (only a single event reaches $M_w$ 6.5, the 2016 Norcia earthquake). It also offers broad station and network coverage: 19 networks and 620 seismic stations across Italy. However, the dataset is not uniformly clean or ready for training a generative waveform model. It includes low–signal-to-noise traces and single-window traces that contain multiple events (e.g., an event followed by aftershocks or overlapping background activity).

The magnitude types in the INSTANCE dataset vary: local magnitude ($M_L$), moment magnitude ($M_w$), and duration magnitude ($M_d$). We homogenize all magnitudes to $M_w$ using published Italian conversions (\ref{app:magnitude}) and exclude events outside their calibrated ranges.
All subsequent steps use these homogenized $M_w$ values.

We keep only high-quality data within the area of interest and apply three layers of preprocessing to filter the original dataset. The first filter imposes the following criteria: minimum magnitude $M_w$ 1.4; maximum magnitude $M_w$ 6.5; maximum hypocentral distance 200 km; maximum event depth 100 km; minimum $V_{S30}$ 116 m/s; P-phase arrival time between 500 and 3500 samples; and minimum signal-to-noise ratio 3. The second filter removes waveforms that contain more than one event per window: we fit an exponential model to the coda (peak-to-peak), and discard traces whose amplitudes exceed the theoretical exponential decay, indicating a new event. Details of the second filter are provided in Supplementary Text S1.

\begin{figure}[!htp]
    \centering
    \includegraphics[width=0.9\linewidth]{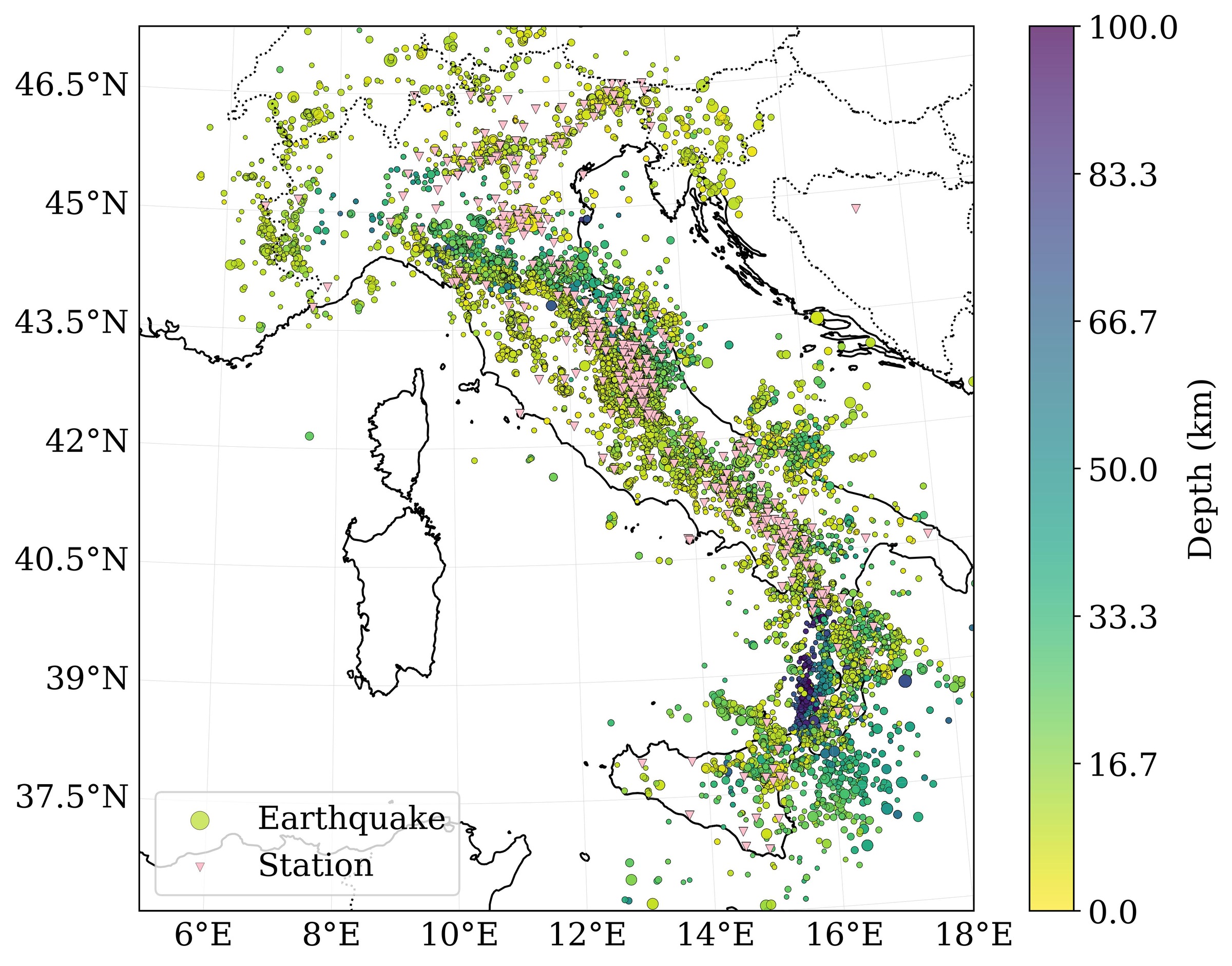}
    \caption{Data used in this study. Inverted pink triangles show station locations. Colored dots mark earthquake hypocentral depth; dot size scales with magnitude.}
    \label{fig:data_used}
\end{figure}

However, some overlapping events evade the second filter (Figure~\ref{fig:filter_step3}). To address this, we apply a third filter based on the number of phase picks produced by the deep-learning picker PhaseNet \cite{zhu2019phasenet}, trained on the INSTANCE dataset \cite{michelini2021instance}. We remove traces with more than one pick occurring within a 5~s window. After the three-stage filtering, 195{,}138 waveforms remain (Figure~\ref{fig:statistics_data}). The earthquakes and stations used in this study are shown in Figure~\ref{fig:data_used}. Model training was conducted on four NVIDIA H100 GPUs, requiring approximately 48 hours for each of the variational autoencoders and 48 hours for the diffusion model.

All retained INSTANCE waveforms are provided at a uniform 100~Hz sampling rate; from each record we extract a fixed 9{,}000-sample (90~s) window with the P-wave arrival placed at 5~s (sample 500), so that all traces share a common length and alignment before further processing (Supplementary Text~S1).

\subsection{Conditioning Inputs}
\label{sec:conditioning-inputs}
We build the low-frequency conditioning waveform directly from the data by low-pass filtering each windowed record at its magnitude-dependent corner frequency $f_c$, the frequency at which the source spectrum transitions to an $\omega^{-2}$ decay. We apply a fourth-order zero-phase Butterworth low-pass filter at $f_c$. Below $f_c$ the displacement spectrum is essentially flat and controlled by the seismic moment, so a record filtered at $f_c$ retains the deterministic long-period motion, and HighFEM-2 synthesizes the content above $f_c$. We compute $f_c$ from stress drop $\Delta\sigma$ using Brune-model relations, drawing $\Delta\sigma$ from the Central Italy compilation of \citeA{bindi2020reliability}
\begin{align}
    \Delta \sigma &= \frac{7}{16}\,\frac{M_0}{r^3}, \label{eq:stress_drop}\\
    f_c &= \frac{2.34\,\beta}{2\pi r}, \label{eq:fc}
\end{align}
where $M_0$ is the seismic moment derived from $M_0 = 10^{(1.5 M_w + 9.1)}$, $r$ is the source radius, and $\beta$ is the shear-wave speed, here assumed to be 3500~m/s. Following \citeA{bindi2020reliability}, the median stress drop is taken as
$0.1$, $0.2$, $0.8$, $1.5$, and $7.9$~MPa for the magnitude ranges
$M_w<2.0$, $2.0\le M_w<3.0$, $3.0\le M_w<4.0$, $4.0\le M_w<5.0$, and $M_w\ge 5.0$, respectively. Therefore, $f_c$ in our selected dataset ranges from approximately $0.18$ to $14.7$~Hz. Stress drop is generally found to be independent of magnitude over a wide range \cite<e.g.,>{abercrombie1995earthquake, ide2001does, allmann2009global}, and the magnitude dependence in the Central Italy compilation may partly reflect the limited bandwidth available for small events. We nevertheless adopt these regional values, for two reasons. First, $f_c$ here only defines the upper edge of the conditioning band, not a source parameter used elsewhere in the model. Second, a constant stress drop of 3~MPa would change $f_c$ by a factor of 0.7--3.1 across our magnitude range, whereas HighFEM-2 synthetics remain consistent for cutoffs varied by a factor of about 15 (Section~\ref{sec:freq-band}). For the smallest events, a constant stress drop would also place $f_c$ close to the Nyquist frequency, leaving little of the band to be synthesized.

We use ``conditioning waveform'' for the low-frequency input throughout, whether it is a low-pass-filtered record or a simulation. Table~\ref{tab:conditioning-inputs} lists all conditioning inputs, which we extract together with the selected traces.

\begin{table}[!htp]
\centering
\caption{Conditioning inputs of HighFEM-2.}
\label{tab:conditioning-inputs}
\small
\begin{tabular}{@{}lll@{}}
\toprule
\textbf{Input} & \textbf{Range} & \textbf{Role} \\
\midrule
Low-frequency waveform (3 components) & low-pass at $f_c$ & Phase, timing and long-period amplitude \\
$M_w$ & 1.4--6.5 & Spectral level and duration \\
Hypocenter latitude, longitude, depth & \makecell[l]{35.14°--47.80°~N, 6.10°--18.81°~E,\\ 0.1--100~km} & \makecell[l]{With station location: distance, azimuth,\\ depth, regional path} \\
Station latitude, longitude & 35.84°--46.56°~N, 6.68°--19.26°~E & As above \\
Station elevation & 0--2605~m & Topographic site effect \\
$V_{S30}$ & 116--1131~m/s & Site amplification \\
\bottomrule
\end{tabular}
\end{table}

\subsection{Model training}
Of the 195{,}138 retained waveforms, a random $90\%$ is used for training (with $5\%$ of that portion reserved for validation, i.e., $\sim$167k/9k records) and the remaining $10\%$ ($\sim$19.5k records) is held out as a test set, unseen during training and used exclusively for model evaluation.

We train HighFEM-2 using the AdamW optimizer with a learning rate of $10^{-4}$, momentum parameters of $\beta_1 = 0.9$ and $\beta_2 = 0.999$, and a weight decay of $10^{-5}$ \cite{loshchilov2019decoupled}. The model is trained on 4 H100 GPUs with a batch size of $64$ per GPU for $300,000$ steps. We employ a linear warmup of $1,000$ steps followed by a cosine decay schedule, which reduces the learning rate to $10^{-6}$. These values worked well as defaults; the model is not sensitive to their precise settings (\ref{app:model-evaluations}).

During training, we maintain an exponential moving average (EMA) of the model weights using a decay rate of $\alpha = 0.999$, which is then used for synthetic data sampling.

\section{Results}
\label{sec:results}
Unlike the fully stochastic model, HighFEM-1 \cite{palgunadi2025high}, HighFEM-2 conditions on a deterministic low-frequency waveform together with a stochastic component for the high-frequency content. HighFEM-2 receives the conditioning inputs of Section~\ref{sec:conditioning-inputs}. In the following section, we examine how closely HighFEM-2 can reproduce the characteristics of observed data and compare its performance with alternative generative waveform model architectures, including modified diffusion models and generative adversarial networks.

\begin{figure}
    \centering
    \includegraphics[width=0.99\linewidth]{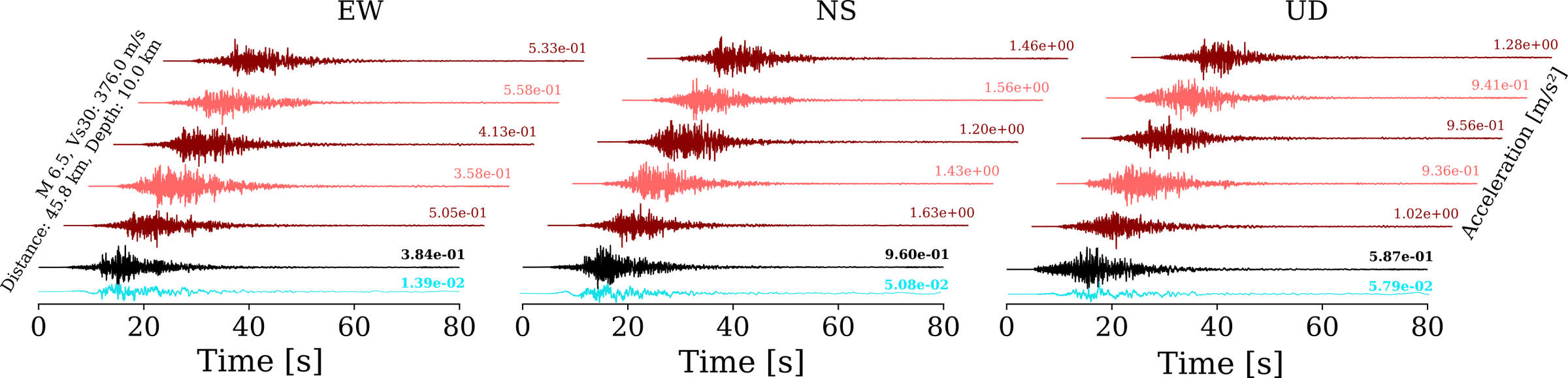}
    \caption{Observed three-component accelerograms (black), low-frequency conditioning waveforms (cyan; low-pass filtered at $f_c$), and five randomly selected HighFEM-2 synthetics (alternating red shades for visibility, offset in time) for a single station ($V_{S30}=376$~m/s, hypocentral distance 45.8~km) during the Norcia, 30 October 2016, $M_w$ 6.5 event. Peak absolute amplitudes are shown above each trace.}
    \label{fig:wavegen}
\end{figure}

HighFEM-2 can generate arbitrary numbers of realizations for a given set of conditioning inputs (Figure~\ref{fig:wavegen}), which we evaluate in the following subsections against observed records using both seismological and machine-learning metrics. Section~\ref{sec:results} proceeds in four steps. We first show that the synthetics have realistic waveform characteristics (Section~\ref{sec:waveform-realism}), then compare their amplitude statistics with competing generative models and empirical GMMs (Section~\ref{sec:amplitude-statistics}), then test how they respond to different low-frequency inputs (Section~\ref{sec:different-inputs}), and finally test whether a model trained in one region transfers to another (Section~\ref{sec:regional-transfer}). Throughout Sections~\ref{sec:waveform-realism}--\ref{sec:different-inputs}, illustrative records are drawn at random from the test set, complemented by records from the largest events to cover the sparse high-magnitude range; all aggregate statistics use the full test set.

\subsection{Waveform Realism}
\label{sec:waveform-realism}

\subsubsection{Long-period features of the waveforms}
\label{sec:long-period}
\begin{figure}
    \centering
    \includegraphics[width=1\linewidth]{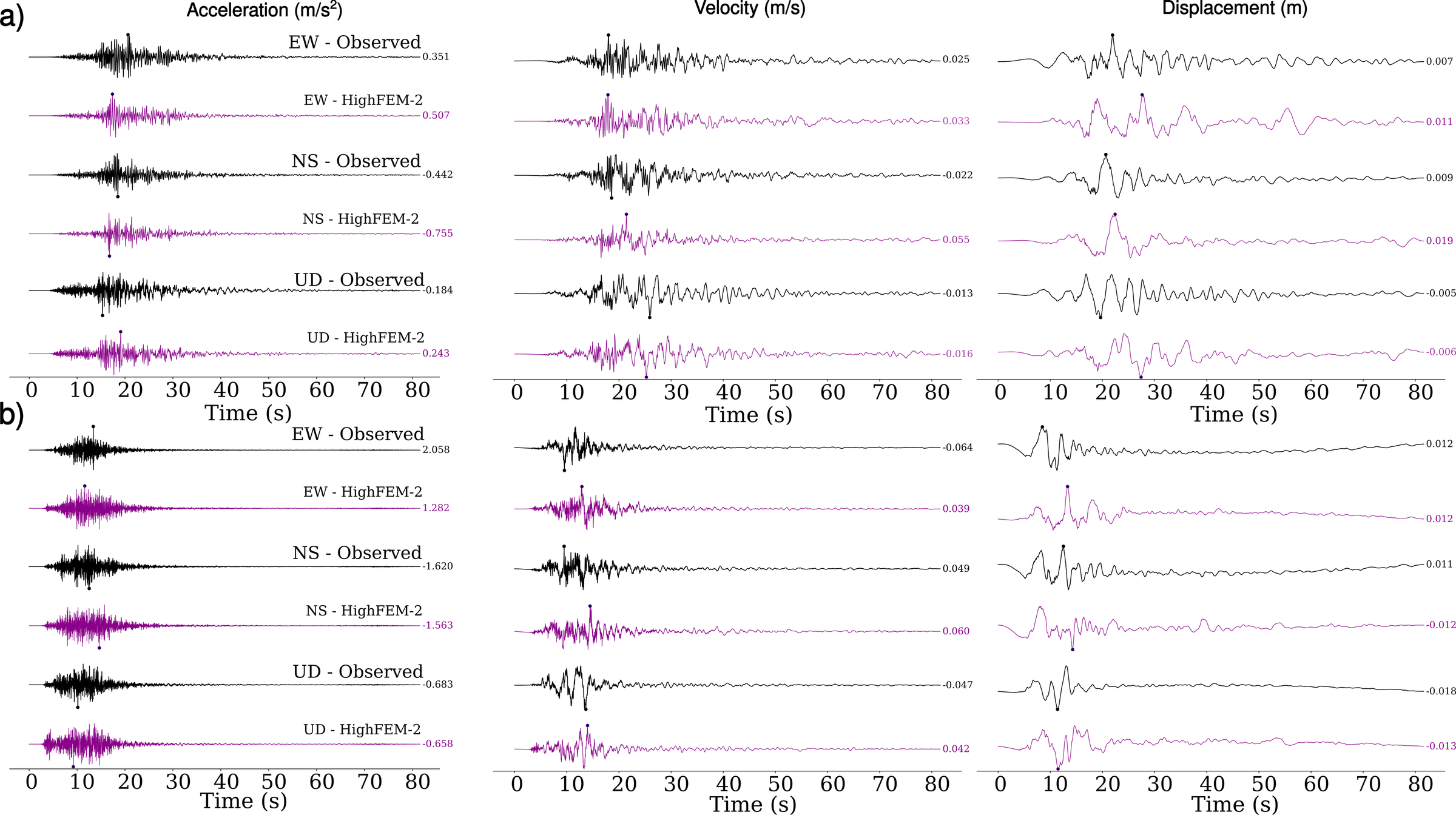}
    \caption{Waveform comparison between observed data (black) and HighFEM-2 (purple) for one $M_w$ 5.9 event at two recording distances, (a) 62~km ($V_{S30}=552$~m/s) and (b) 25~km ($V_{S30}=464$~m/s): acceleration (left), velocity (middle), and displacement (right). All traces are aligned with the P onset at 5~s.}
    \label{fig:integration_waveforms}
\end{figure}

To assess whether HighFEM-2 captures long-period ground motion behavior, we compare velocity and displacement waveforms obtained by time-integrating the observed accelerograms and the corresponding GWM synthetics (Figure \ref{fig:integration_waveforms}). The synthetics reproduce the overall long-period motion, including the gradual rise in amplitude following P-wave onset, the energy growth associated with S- and surface-wave arrivals, and the subsequent coda decay. The velocity traces show similar amplitude scaling and duration across all three components, while the displacement waveforms preserve the broad waveform envelope and reach their peak motion at comparable times in the record. Although wiggle-by-wiggle matching of the high-frequency phase is not expected from a stochastic model, the synthetic records retain the low-frequency characteristics. Other examples of different conditioning parameters show consistent observations (Figure \ref{fig:integration_waveforms_supp}).

Beyond a single realization, HighFEM-2 can generate ensembles of velocity and displacement time series for the same set of conditioning parameters, using only the low-frequency acceleration input as the conditioning waveform. The ensembles illustrate the variability learned by the model while preserving the essential long period structure of the waveform (Figures \ref{fig:wavegen_vel} and \ref{fig:wavegen_displ}). Across all realizations, the first-order characteristics remain consistent with the observed records. Although the individual waveforms differ in their high-frequency details, the low-frequency content remains stable across realizations, because it is set by the conditioning inputs. The model therefore produces ensembles that carry both the deterministic long-period structure and the stochastic high-frequency variability of ground motion.

\subsubsection{Preservation of the Low-Frequency Input}
\label{sec:lf-preservation}
\begin{figure}
    \centering
    \includegraphics[width=\linewidth]{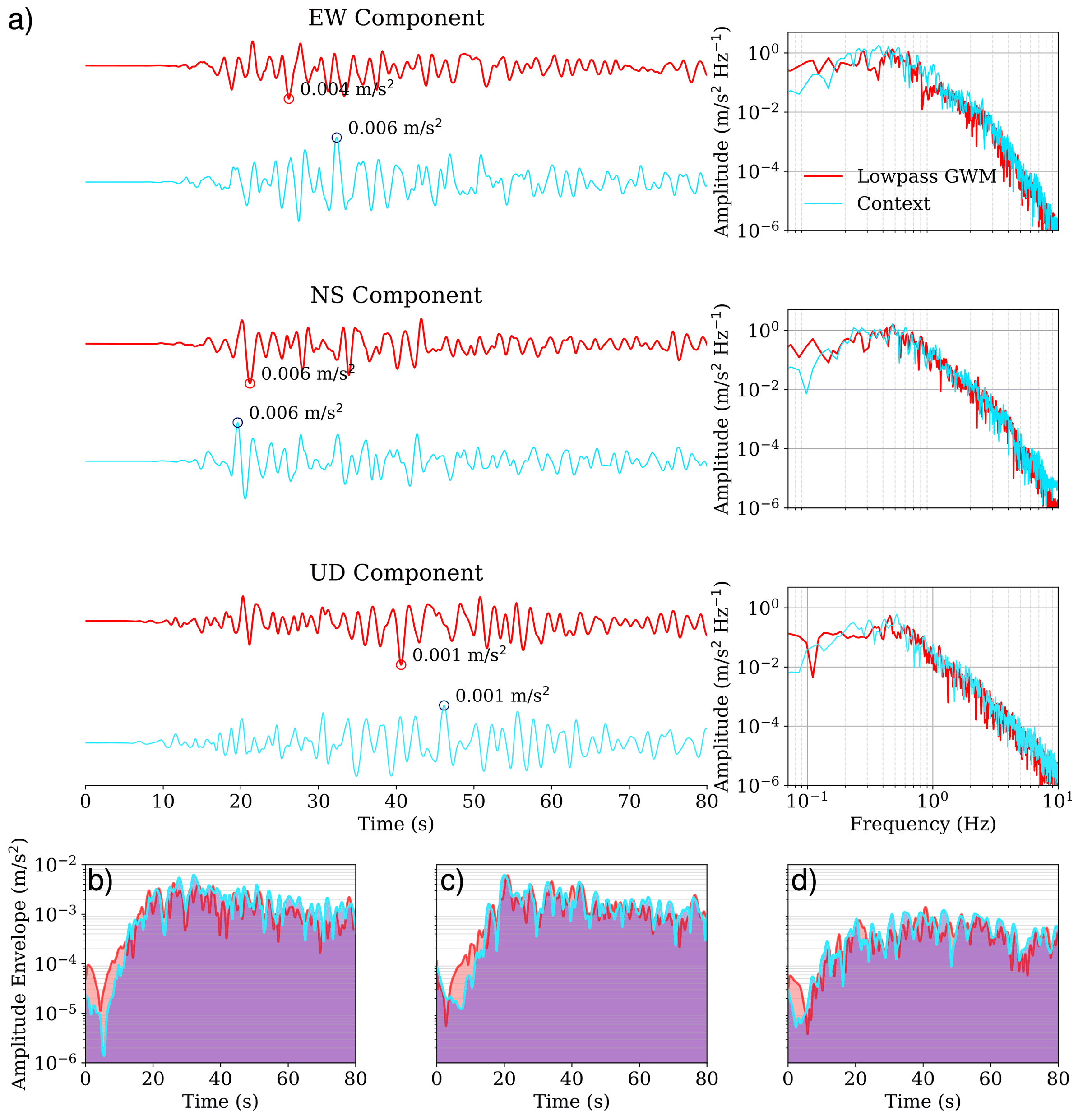}
    \caption{Comparison of low-pass-filtered HighFEM-2 synthetics at their corner frequency $f_c$ with the corresponding conditioning waveform for all three components, for conditioning parameters: $M_w=5.3$, $V_{S30}=293$ m/s; station at $11.50^\circ$E, $44.72^\circ$N, elevation 1.0 m; hypocenter depth 4.35 km at $10.98^\circ$E, $44.87^\circ$N; hypocentral distance 51.5 km. (a) Time-domain waveforms and corresponding amplitude spectra: HighFEM-2 synthetics (red) vs. conditioning waveform (cyan). (b–d) Amplitude envelopes for the three components, with synthetics in red and conditioning waveform in cyan. The close agreement in waveforms, spectra, and envelopes shows that HighFEM-2 preserves the conditioning low-frequency content without substantial distortion. Legend terms, Context: conditioning waveform; Lowpass GWM: low-pass-filtered HighFEM-2 synthetic.}
    \label{fig:compare_low_freq}
\end{figure}

A generative model could alter the long-period motion it is given. That would bias the response of long-period structures and remove the physical constraint the conditioning is meant to impose. We therefore low-pass filter the synthetics at $f_c$ and compare them with the conditioning waveform, which isolates the long-period band used as input to the model (Figure \ref{fig:compare_low_freq}). The time domain comparisons in Figure \ref{fig:compare_low_freq}a show that the synthetics closely follow the broad waveform structure of the conditioning waveform across all three components, including the gradual amplitude growth around major phase arrivals (at $t < 20$ s), the sustained long period motion, and the overall signal duration. The spectral amplitudes also agree within the low-frequency band, so the model keeps the spectral amplitudes of the conditioning waveform in this band.

The amplitude envelope comparisons in Figures \ref{fig:compare_low_freq}b–d show that the synthetics replicate the first-order temporal evolution of low-frequency motion. The envelope shapes, peak levels, and duration of long-period energy remain closely aligned with the conditioning waveform, suggesting that HighFEM-2 retains the low-frequency information and carries it into the broadband synthetic (see also Section \ref{sec:signal_envelope}). The largest mismatch is before the first arrival, where amplitudes are low (Figure~\ref{fig:compare_low_freq}b--d); see Section~\ref{sec:limitation} on onset timing. These results show that the model uses the low-frequency input as a constraint and produces long-period ground motion that matches the conditioning waveform.

\subsubsection{Time domain signal envelopes and Fourier amplitude spectra} \label{sec:signal_envelope}

\begin{figure}
    \centering
    \includegraphics[width=\linewidth]{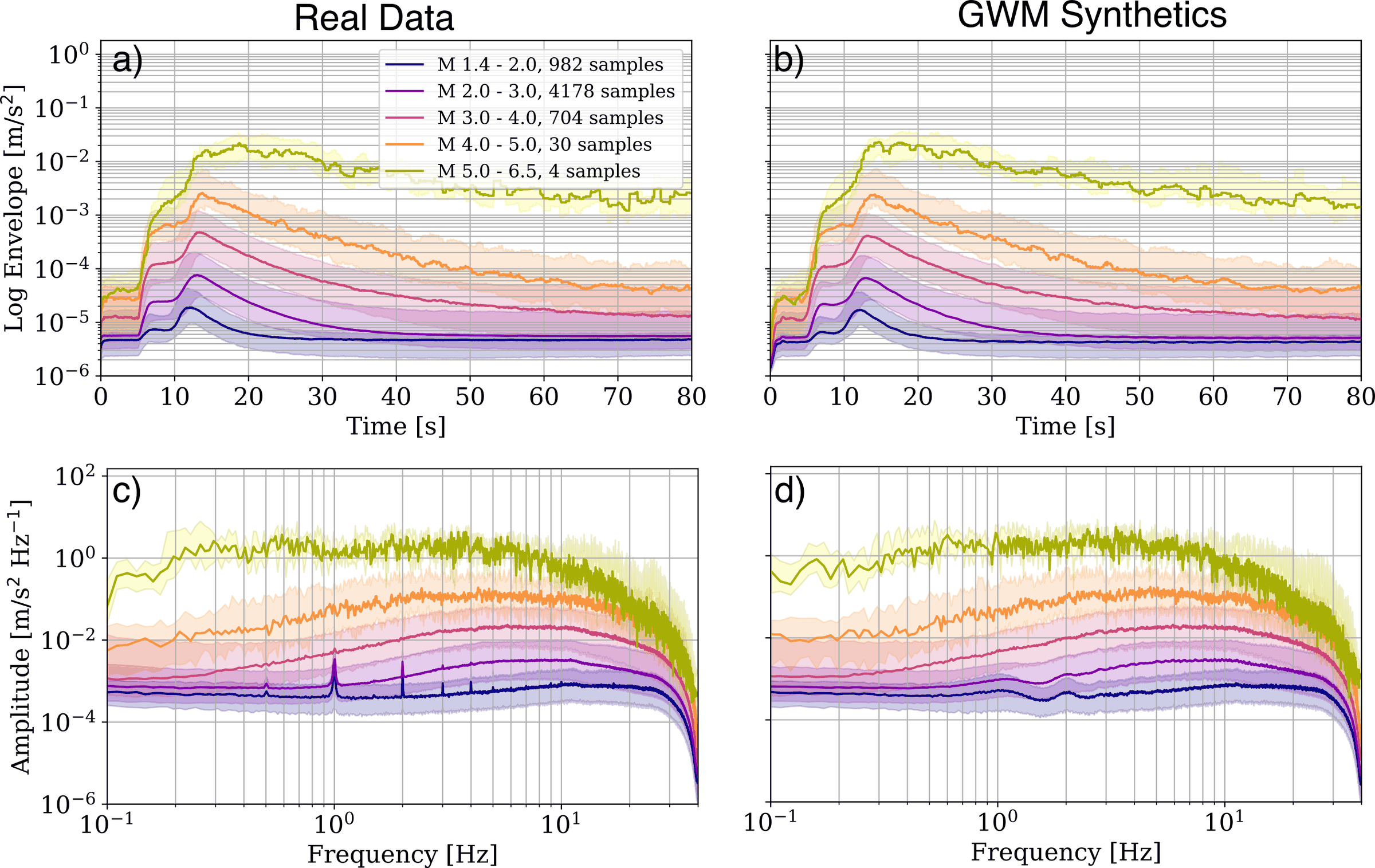}
    \caption{First-order seismogram characteristics of real data (left) and HighFEM-2 synthetics (right) for hypocentral distances of 30–50 km across five magnitude bins. (a–b) Distributions of time-domain envelopes for the north–south acceleration component, showing the mean (solid line) and standard deviation (shaded area). (c–d) Distributions of Fourier spectral log-amplitudes for the north–south acceleration component. The number of samples in each bin (from smallest to largest magnitude) is 982, 4178, 704, 30, and 4. Narrowband peaks at 1~Hz and its multiples in the observed spectra of the low-magnitude bins reflect anthropogenic and instrumental noise present in the original INSTANCE recordings; the synthetics do not reproduce these narrowband features.}
    \label{fig:envelope_NS}
\end{figure}

To evaluate how well the HighFEM-2 synthetics reproduce first-order waveform behavior, we compare time domain signal envelopes and Fourier amplitude spectra of real accelerograms with those generated by the HighFEM-2. Following \citeA{palgunadi2025high}, the envelopes are computed by taking a moving average of absolute waveform amplitudes using a 128 sample kernel, followed by a logarithmic transformation applied uniformly to both real and synthetic datasets. For hypocentral distances of 30--50~km (Figure~\ref{fig:envelope_NS}), the synthetic envelopes capture the main temporal evolution of ground motion across all magnitude groups (Figures \ref{fig:envelope_NS}a and \ref{fig:envelope_NS}b) in North-South component. Low pre-arrival noise is followed by a clear rise in motion near P wave onset, with peak amplitudes that scale with magnitude. Smaller events reach peak amplitude shortly after first arrival, whereas larger events continue developing energy later in time, and this behavior is preserved in the generated records. The spread of envelope amplitudes within each magnitude bin is comparable between real and synthetic datasets and remains approximately symmetric in log space. Despite the reduced number of high magnitude samples, the synthetic distributions continue to reflect the observed envelope shape and variability. These observations are still consistent for the other components and distance bins (Figures \ref{fig:envelope_supp_EW} - \ref{fig:envelope_supp_UD}).

We then compare Fourier amplitude spectra computed through a Fourier transform of the time domain acceleration series, followed by log scaling of spectral magnitudes. The real and synthetic spectra show similar distributional structure across frequency, including consistent mean levels and variance envelopes across most magnitude bins (Figures \ref{fig:envelope_NS}c and \ref{fig:envelope_NS}d) for the corresponding parameter bins. Both datasets exhibit increasing amplitude toward mid-frequency ranges and decaying spectral content near the high-frequency limit. Slightly lower spectral levels appear in the largest magnitude bin due to limited sample count, yet the synthetics remain within the variability bounds of the real data. Additional magnitude and distance bins are provided in Supplementary Figures \ref{fig:spectra_supp_EW} - \ref{fig:spectra_supp_UD}.

\subsection{Amplitude Statistics}
\label{sec:amplitude-statistics}

\subsubsection{Comparison of generative models}
\label{sec:model-comparisons}
We evaluate the following four models (we refer to \ref{app:model-evaluations} for evaluations using different hyperparameters):
\begin{enumerate}
    \item \textbf{HighFEM-2}: the proposed model in this study, which parameterizes the flow-matching velocity field $v_{\vtheta}$ (cf.\ Section~\ref{sec:methods}) with a multi-modal diffusion transformer; FM (DiT) in Table~\ref{tab:model-performance}.

    \item \textbf{EDM (UNet)}: a diffusion model \cite{karras2022edm} with a UNet architecture, the HighFEM-1 formulation \cite{palgunadi2025high}, retrained on the Italian data with the same conditioning as HighFEM-2.

    \item \textbf{CM (DiT)}: a distillation approach \cite{geng2025consistency} that learns to generate samples with only 2 function evaluations (instead of 25, cf. Section~\ref{sec:flow-matching}).

    \item \textbf{WGAN-GP}: a Wasserstein-GAN with gradient penalty \cite{gulrajani2017improved} that learns to translate the spectrograms of low-frequency waveforms to spectrograms of high-frequency waveforms.
\end{enumerate}

\begin{figure}
    \centering
    \includegraphics[width=\linewidth]{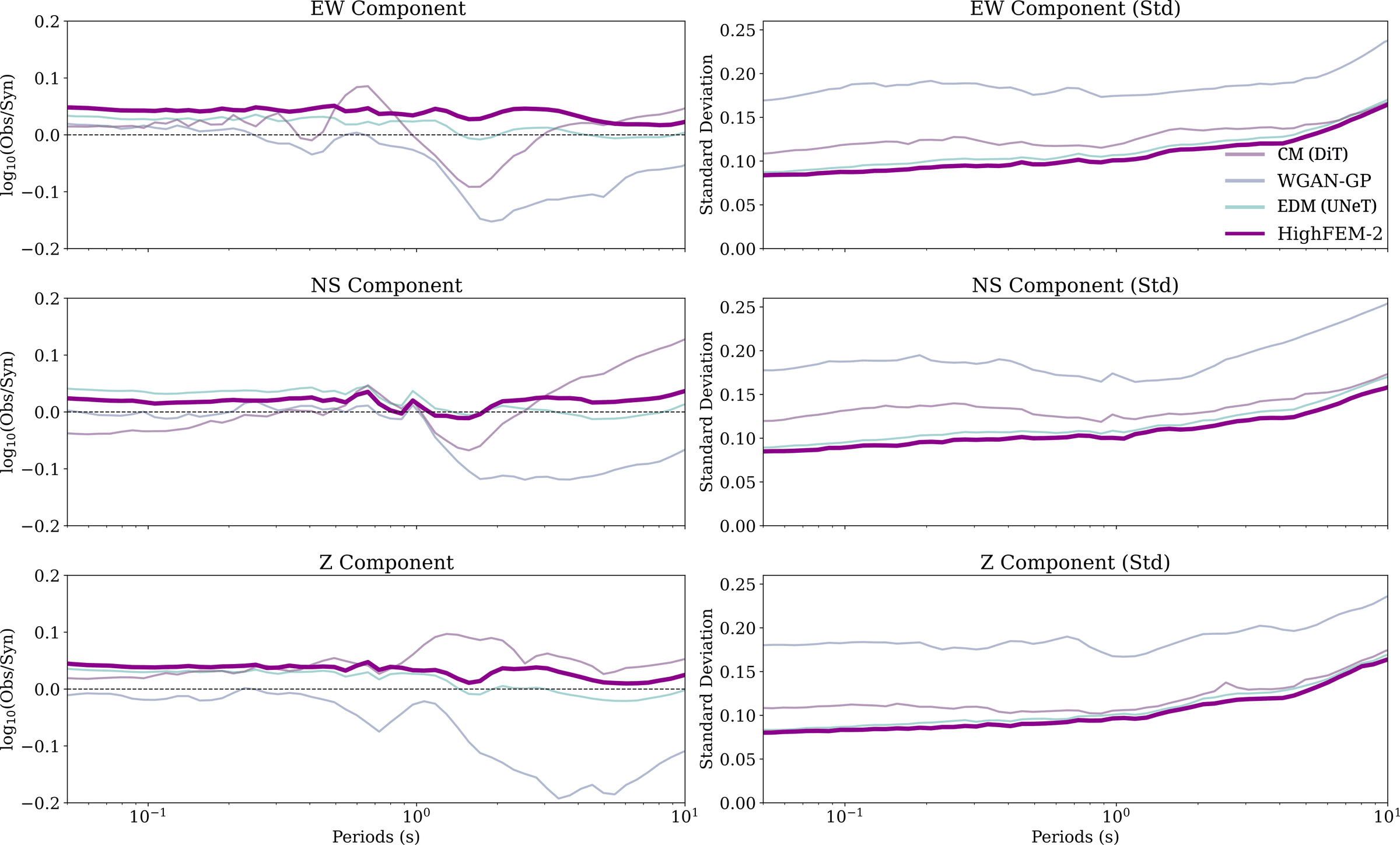}
    \caption{Bias (left) and standard deviation (right) of $\log_{10}(\mathrm{PSA}_{obs}/\mathrm{PSA}_{syn})$ for $5\%$-damped PSA across periods, for the four generative models on the test set, per component (rows: East--West, North--South, vertical). The thickest line marks HighFEM-2.}
    \label{fig:SA_comparison}
\end{figure}

For engineering use, synthetic waveforms need realistic amplitudes across the period range of interest. We therefore compare the bias of $5\%$-damped PSA of observations relative to the generative models as a function of period (Figure~\ref{fig:SA_comparison}) for the three components. HighFEM-2 and EDM (UNet) perform nearly identically: EDM (UNet) yields marginally smaller residuals (e.g., 0.015 versus 0.024 log units on the East--West component, corresponding to roughly a 2\% difference in pseudospectral acceleration), and both clearly outperform the remaining models, with the smallest PSA bias and scatter. HighFEM-2 additionally gives the most consistent envelopes in the sparse high-magnitude bins, including events up to $M_w$~6.5 and distances down to 6~km (Section~\ref{sec:signal_envelope}; Supplementary Figures~\ref{fig:compare_model_19}--\ref{fig:compare_model_10572}), and, within the same transformer architecture, needs half the network evaluations of diffusion sampling (25 versus 50; 1.29~s versus 2.57~s per three-component synthetic for EDM (DiT); Table~\ref{tab:model-performance}).

\ref{app:model-evaluations} compares further architectures and hyperparameters. WGAN-GP shows the largest deviations from observations where training data are limited (i.e., $M_w \geq 6$; Figures~\ref{fig:compare_model_8399} and~\ref{fig:compare_model_10572}).

The standard deviation comparison shows a clearer separation between the models than the bias. HighFEM-2 has the lowest standard deviation of the four models at all periods, closely followed by EDM (UNet). The bias of HighFEM-2 and EDM (UNet) stays within about $\pm 0.05$ log units at all periods, whereas WGAN-GP and CM (DiT) drift at periods longer than about 1~s, most visibly on the vertical component: WGAN-GP overpredicts long-period amplitudes by up to 0.1--0.2 log units, and CM (DiT) departs from zero in both directions. The standard deviation of all models grows toward long periods; for HighFEM-2 it roughly doubles, from about 0.09 log units at 0.1~s to about 0.16 at 10~s.

The remaining evaluations use HighFEM-2 only. All results are evaluated relative to observed ground motion records to ensure consistent references.

\subsubsection{Peak ground motion and comparison with scalar GMMs}
\label{sec:intensity-measures}
\begin{figure}
    \centering
    \includegraphics[width=\linewidth]{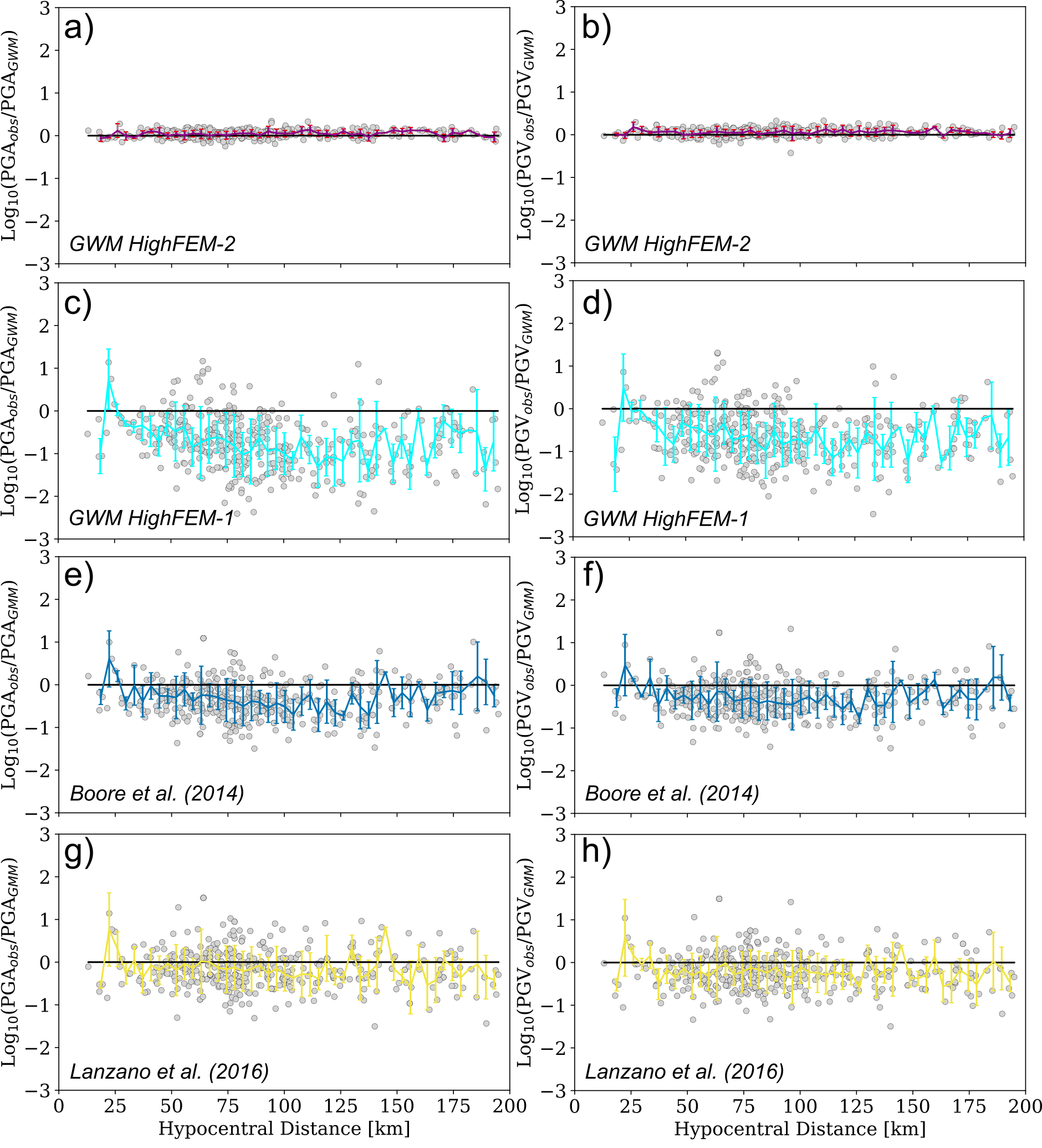}
    \caption{Residuals $\log_{10}(\mathrm{PGA}_{obs}/\mathrm{PGA}_{pred})$ (a, c, e, g) and $\log_{10}(\mathrm{PGV}_{obs}/\mathrm{PGV}_{pred})$ (b, d, f, h) against hypocentral distance for the Italian test set: HighFEM-2 (purple), HighFEM-1 \protect\cite{palgunadi2025high} (cyan), \protect\citeA{boore2014nga} (blue), and \protect\citeA{lanzano2016ground} (yellow). Lines: mean residual per distance bin; bars: one standard deviation.}
    \label{fig:bias}
\end{figure}

For each record in the test set, we compute a single HighFEM-2 synthetic using the conditioning inputs of the real record (Section~\ref{sec:conditioning-inputs}). Using one realization per record leaves the reported standard deviations comprising both model dispersion and single-draw sampling variability; the mean biases are unaffected. As in \citeA{palgunadi2025high}, we compare the PGA and PGV values between the real data, the GWM synthetics, and scalar GMMs. We compute the model bias as the logarithm of the ratio between observed and predicted peak amplitude. Figure \ref{fig:bias} presents these logarithmic residuals across hypocentral distances of 0-200 km for the Italian test dataset, comparing HighFEM-2 (panels a-b), HighFEM-1 (panels c-d; Section~\ref{sec:regional-transfer}), \citeA{boore2014nga} GMM (panels e-f), and \citeA{lanzano2016ground} GMM (panels g-h).

Because the conditioning waveform is specific to each earthquake--station pair, the ensembles generated by HighFEM-2 sample the variability of the high-frequency motion conditional on the low-frequency input, rather than the total ground-motion variability that empirical GMMs describe. This distinction is important for hazard applications and explains the small residual spread relative to the scalar GMMs in Figure~\ref{fig:bias}: the observed low-frequency waveform already fixes the event term and much of the path and site contribution, components that a GMM must absorb into its aleatory standard deviation. The residual spread reported here must therefore not be interpreted as, or substituted for, a GMM sigma.

\subsection{Response to Different Low-Frequency Inputs}
\label{sec:different-inputs}

\subsubsection{Conditioning on physics-based simulations}\label{sec:dynamic_rupture}
\begin{figure}
    \centering
    \includegraphics[width=0.99\linewidth]{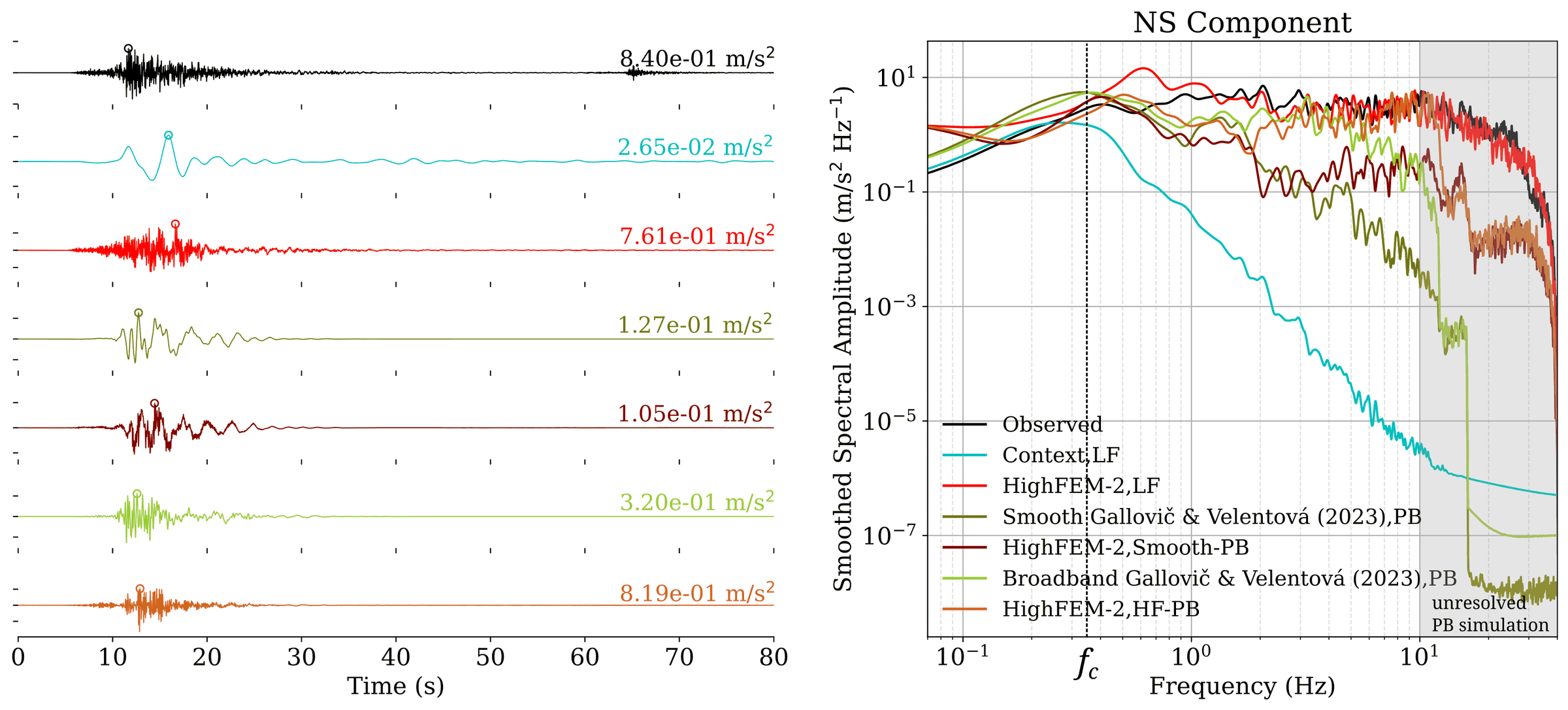}
    \caption{HighFEM-2 synthetics conditioned on three inputs for the 2016 Amatrice earthquake ($M_w$ 6.0 used for conditioning; Section~\ref{sec:dynamic_rupture}) at station IV.TERO, North--South component (other components in Figures \ref{fig:compare_acc_ew} and \ref{fig:compare_acc_ud}). Inputs: observed record low-pass filtered at $f_c$, and the smooth and rough dynamic-rupture simulations of \protect\citeA{gallovivc2023broadband}. The simulations are resolved up to 10~Hz (gray area: unresolved band). The dotted line marks $f_c$. Each input is shown directly above its synthetic. Left: acceleration time series, aligned 5~s after the start of each trace. Right: smoothed Fourier amplitude spectra. Legend terms, Context,LF: conditioning waveform from the filtered record; Smooth and Broadband Gallovi\v{c} \& Velentov\'a (2023),PB: smooth and rough physics-based simulations; HighFEM-2,LF, HighFEM-2,Smooth-PB, and HighFEM-2,HF-PB: synthetics conditioned on the filtered record, the smooth simulation, and the rough simulation.}
    \label{fig:compare_acc_ns}
\end{figure}

In application, the conditioning waveform will usually come from a physics-based simulation rather than a filtered record. We therefore ask how HighFEM-2 behaves when given a simulation, and how errors in the simulation carry into the synthetics. We use two dynamic-rupture simulations of the 2016 Amatrice earthquake ($M_w$~6.2 in \citeA{gallovivc2023broadband}; $M_w$~6.0 in INSTANCE, the value used for conditioning), resolved up to about 10~Hz. The smooth simulation uses a planar fault with smooth distributions of strength and initial stress; the rough simulation adds fractal variations of both, which generate richer high-frequency radiation. We condition HighFEM-2 on each simulation at station IV.TERO and compare the results with the observed record and with a synthetic conditioned on the observed record filtered at $f_c$ (Figure~\ref{fig:compare_acc_ns}).
The simulations are used as provided, so their bandwidth, up to 10~Hz, exceeds any cutoff tested in Section~\ref{sec:freq-band}. HighFEM-2 was trained only on filtered records, so this also tests transfer to a new type of input.

The observed NS peak acceleration at IV.TERO is 0.84~m/s$^2$ (Figure~\ref{fig:compare_acc_ns}, left). The filtered record peaks at only 0.027~m/s$^2$, yet its synthetic reaches 0.76~m/s$^2$, close to the observation. The rough simulation peaks at 0.32~m/s$^2$ and its synthetic at 0.82~m/s$^2$, also close to the observation. The smooth simulation peaks at 0.13~m/s$^2$, and its synthetic at 0.11~m/s$^2$ reproduces the pulse shape but underestimates the observed peak by a factor of about eight.

The spectra show the same pattern (Figure~\ref{fig:compare_acc_ns}, right). Below 0.2--0.5~Hz, all three synthetics lie within the range of the observed spectrum. Near the dominant peak at 0.7~Hz they diverge: the synthetic from the filtered record overshoots the observed amplitude by a factor of about 3, the synthetic from the smooth simulation undershoots it by a factor of 2--3, and the synthetic from the rough simulation matches it most closely. Between 0.7 and 5~Hz, the rough-simulation synthetic inherits the $\sim$30\% amplitude underprediction of the simulation itself.

HighFEM-2 thus does not correct deficiencies of its input. With the rough simulation, which radiates realistic energy up to its resolved frequency, the synthetic reaches realistic amplitudes; with the smooth simulation, it stays low. For coupling to simulations, this behavior is useful, because errors in the simulation remain visible in the synthetics and can be traced back to, and improved in, the simulation. It also means that HighFEM-2 cannot compensate for a simulation that lacks high-frequency radiation.

\subsubsection{Response of HighFEM-2 to different frequency-band inputs}
\label{sec:freq-band}
Building on the previous result that different conditioning waveforms (a filtered record, a smooth simulation, and a rough simulation) lead to different spectral amplitudes in the resulting HighFEM-2 synthetics, we now investigate whether this sensitivity persists when the conditioning waveform comes from the same record filtered to different maximum frequencies. To do so, we provide HighFEM-2 with multiple low-pass-filtered versions of the same observed record, where the cutoff is progressively varied from the natural $f_c$ to fixed cutoffs of 0.5, 1, 2, 3, 4, and 5 Hz.

\begin{figure}
    \centering
    \includegraphics[width=\linewidth]{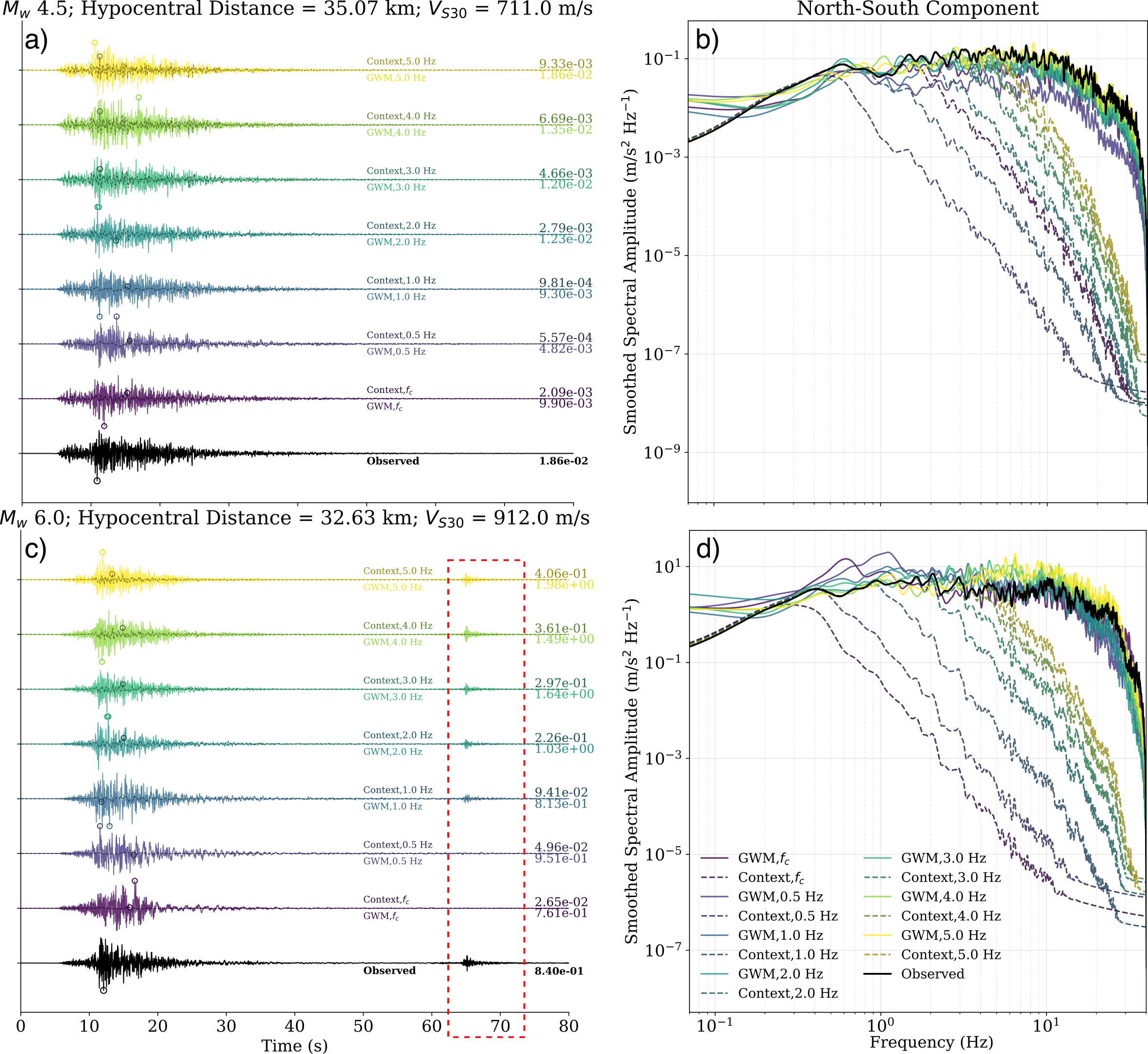}
    \caption{Comparison of HighFEM-2 synthetics as a function of the maximum frequency of the conditioning waveform for two example earthquakes. Conditioning waveforms are low-pass filtered at a range of upper cut-off frequencies (corner frequency $f_c$ (purple), as well as  at 0.5, 1, 2, 3, 4, and 5~Hz). (a,b) $M_w=4.5$ event, recorded at 35.07~km, $V_{S30}=711$~m/s, time-domain waveforms and smoothed spectra of the conditioning waveforms and the corresponding HighFEM-2 synthetics. (c,d) Same as (a,b) but for $M_w=6.0$, hypocentral distance 32.63~km, and $V_{S30}=912$~m/s. The dashed red rectangle ($t = 62$--73~s) marks an aftershock present in the conditioning waveform for cutoffs $\geq 1$~Hz, which the synthetics reproduce. Synthetic spectra remain similar across all input bandwidths. Legend terms, Context: conditioning waveform; GWM: HighFEM-2 synthetic.}
    \label{fig:context_diff_freq}
\end{figure}

For the $M_w~4.5$ case, the key result is in the spectra (Figure~\ref{fig:context_diff_freq}b): between 0.8 and 6~Hz, all synthetic spectra lie within a factor of two of each other and follow the shape of the observed spectrum, regardless of input bandwidth. The peak amplitudes show the same convergence (Figure~\ref{fig:context_diff_freq}a). The conditioning waveforms span more than an order of magnitude: the 0.5~Hz conditioning waveform peaks at $\sim 5.6 \times 10^{-4}$ m/s$^2$, while the 5~Hz input reaches $\sim 9.3 \times 10^{-3}$ m/s$^2$, a factor of about 17. The HighFEM-2 synthetics span a much narrower range, with peaks from $\sim 4.8 \times 10^{-3}$ to $1.9 \times 10^{-2}$ m/s$^2$ (a factor of about 4; observed $1.9 \times 10^{-2}$ m/s$^2$), so the model increases peak amplitudes by factors of about 2 (5~Hz input) to 9 (0.5 and 1~Hz inputs) relative to the conditioning waveform. The input mainly sets phase and timing; the spectral level appears to follow the scalar conditioning.

The same behavior is observed for the larger $M_w~6.0$ earthquake (Figure \ref{fig:context_diff_freq}c,d), which is the IV.TERO record of the 2016 Amatrice earthquake used in Section~\ref{sec:dynamic_rupture}. Here, conditioning amplitudes vary from $2.6 \times 10^{-2}$ to $4.1 \times 10^{-1}$ m/s$^2$ across the filtering levels, yet HighFEM-2 again generates synthetics within a much narrower range, between $\sim 7.6 \times 10^{-1}$ and $2.0$ m/s$^2$, compared with the observed peak of $8.4 \times 10^{-1}$ m/s$^2$. In the spectral domain, all synthetics lie close to one another up to roughly 8 Hz and remain within a factor of two of the observed spectrum between 1 and 4 Hz. A small aftershock at 62--73~s appears in the conditioning waveform only for cutoffs of 1~Hz and above (Figure~\ref{fig:context_diff_freq}c). The synthetics reproduce it at the same time, although the training windows were filtered to exclude multiple events (Section~\ref{sec:data-selection}). Timing and phase in the synthetics therefore come from the conditioning waveform rather than from patterns learned in training. The same property means that unwanted signals in the input, such as noise bursts or overlapping events, also appear in the output.

Varying the cutoff frequency thus strongly alters the conditioning waveform, but the synthetics keep consistent broadband spectra. The conditioning waveform controls timing and deterministic features and has little effect on the overall spectral shape, so the model is insensitive to its bandwidth. Within the tested range, no lower limit is reached: even the narrowest input, filtered at the natural corner frequency ($f_c \approx 0.32$~Hz for the $M_w$~6.0 case), yields synthetics consistent with those from broader-band inputs. We expect the scalar inputs, in particular magnitude and distance, to control this spectral level; an ablation without them would test this. Identifying the minimum information content required of the conditioning waveform is left for future work. This insensitivity applies to the bandwidth of the conditioning waveform, not to its amplitude: amplitude biases in the conditioning waveform propagate into the synthetics, as observed for the rough-simulation input in Section~\ref{sec:dynamic_rupture}, whose $\sim$30\% underprediction between 0.7 and 5~Hz carries over to the corresponding synthetic.

\subsection{Regional Specificity: Transfer of a Japan-Trained Model to Italy}
\label{sec:regional-transfer}
Regions differ in attenuation and stress drop, so a generative model may not transfer between them. We test whether the earlier, fully stochastic HighFEM-1 model \cite{palgunadi2025high}, trained in Japan, can predict Italian ground motion (Figure~\ref{fig:bias}c--d). Unlike EDM (UNet) in Section~\ref{sec:model-comparisons}, this model is not retrained. The comparison shows that regional retraining is necessary: applied to Italian data without retraining, HighFEM-1 overpredicts substantially, with mean residuals of $-0.724 \pm 0.636$ log units for PGA (predictions about five times the observations) and $-0.613 \pm 0.589$ for PGV (about four times), consistent with known differences in high-frequency attenuation and stress-drop characteristics between the two regions \cite{bindi2020reliability}. This cross-regional bias, which is comparable in sign but larger in magnitude than that of the global scalar GMM of \citeA{boore2014nga}, shows that high-frequency generation is region-specific and motivates the regional training of HighFEM-2. A direct comparison of Japanese and Italian strong-motion data and regional GMMs (\ref{app:japan-italy}) shows systematically higher PGA in Japan at the same magnitude and distance, consistent with this overprediction.

\section{Discussion} 
\label{sec:discussion}
HighFEM-2 does not push deterministic simulation to higher frequencies, where cost grows rapidly. Given a band-limited waveform that carries the deterministic source, path, and directivity information, it adds the missing high-frequency content stochastically, learned from regional data. It therefore sits between fully deterministic wavefield simulations and the fully stochastic generation of HighFEM-1, with the physical constraint of a simulation at low frequencies and the speed and variability sampling of a stochastic model at high frequencies.

\subsection{Whole-Earthquake Generation and Amplitude Accuracy}
\label{sec:variability-correlation}
Because the conditioning waveform is specific to each station, HighFEM-2 generates consistent records for an entire network, which fully stochastic generative models such as HighFEM-1 cannot. The resulting synthetics are realistic in all metrics evaluated here, including peak amplitudes, response spectra, envelopes, and integrated displacement.

The small residual scatter of HighFEM-2 relative to scalar GMMs (Figure~\ref{fig:bias}) reflects the variability of the high-frequency motion conditional on the low-frequency input, not the total ground-motion variability that GMMs describe (Section~\ref{sec:intensity-measures}). At low frequencies, synthetics for different stations of one earthquake share its source term by construction; whether their high-frequency parts reproduce observed inter-station correlation remains to be tested. A decomposition into inter- and intra-event variability is left for future work.

\subsection{Application of HighFEM-2}\label{sec:application_highfem2}
HighFEM-2 is designed to generate broadband ground-motion ensembles whenever a low-frequency conditioning waveform is available, irrespective of how that waveform is obtained, for example from a physics-based simulation resolved up to some maximum frequency, a low-frequency ground-motion model, or, as in this study, low-pass-filtered records. Because the model returns full three-component time series rather than only intensity measures, and can draw many realizations per conditioning set (Section~\ref{sec:flexible_lf}), it is well suited to applications that require physically consistent waveforms together with a quantified ground-motion variability. These include providing input motions for nonlinear structural and soil response analyses, augmenting sparse strong-motion databases for engineering and machine-learning studies, generating scenario shaking for a prescribed event geometry, and supplying the high-frequency complement to deterministic low-frequency simulations in hybrid broadband workflows. As discussed for HighFEM-1 \cite{palgunadi2025high}, fast conditional generation could also help construct larger and more balanced waveform datasets for applications that rely on waveform libraries, such as waveform-based PSHA and engineering studies, within the magnitude, distance, and site ranges covered by the training data. Generating the synthetics is computationally cheap. Each three-component broadband record requires 25 network evaluations, approximately 1.3~s on a single GPU (Table~\ref{tab:model-performance}), so ensembles of hundreds of realizations per scenario cost little compared with producing the low-frequency input itself.

The breadth of these applications is, however, bounded by the data used to train the model. HighFEM-2 has learned the ground-motion characteristics of the Italian (Apennine) crustal seismicity sampled by INSTANCE, spanning $M_w$ 1.4--6.5, hypocentral distances up to 200~km, focal depths up to 100~km, and site conditions of $V_{S30}\approx116$--1131~m/s, with broadband content up to 50~Hz. Within this envelope, the model reproduces observed amplitudes, envelopes, and spectra with small bias (Section~\ref{sec:results}); outside it, in particular for $M_w>6.5$, for other tectonic regimes (e.g., subduction or stable continental regions), or for site and distance ranges not represented in training, predictions should be treated as extrapolation and validated before use (Section~\ref{sec:limitation}). HighFEM-2 is therefore a region- and dataset-specific broadband waveform generator, most useful for moderate-magnitude events, where the corner frequency is low and a large, engineering-relevant band above it must be synthesized from comparatively limited low-frequency information.

\subsection{Flexible Low-Frequency Conditioning Waveform}\label{sec:flexible_lf}
HighFEM-2 accepts conditioning waveforms of substantially different bandwidth and amplitude (Section~\ref{sec:freq-band}, Figure~\ref{fig:context_diff_freq}). The conditioning waveform is an input to the velocity field at every step of the sampling trajectory (Section~\ref{sec:flow-matching}), so the model learns to generate high-frequency content compatible with whatever low-frequency structure is provided, without a fixed input specification of the kind required by deterministic upsampling or by stochastic approaches such as HighFEM-1. The conditioning waveform thus constrains the overall waveform structure, including temporal features, phase arrivals, and long-period characteristics. We expect the scalar inputs (magnitude, hypocenter location, $V_{S30}$, and station location) to let the model set the spectral level appropriate to magnitude, distance and site, independent of how much bandwidth the conditioning waveform carries. This is relevant for coupling to physics-based simulations, whose maximum resolved frequency varies with computational constraints and velocity model resolution. A possible extension is a model with a tunable level of high-frequency energy, so that the high-frequency content of the synthetics could be adjusted for a given conditioning waveform.

\subsection{Limitations}\label{sec:limitation}
HighFEM-2 has four main limitations. First, the model's training dataset includes earthquakes with magnitudes up to $M_w$~6.5, which constrains its reliable application to events within this magnitude range. The training set inherits the strong magnitude skew of its parent catalogue. Of the 315,255 events in the full INSTANCE catalogue, roughly 309,000 fall in $1\le M_w\le 2$, only a few hundred reach $M_w\ge 4$, and just three exceed $M_w\,6$. Our magnitude filter ($M_w \geq 1.4$) and quality criteria reduce this to 195,138 waveforms used for training, but this retained set remains dominated by small-to-moderate events, with the high-magnitude range represented by only a handful of events (each contributing several station waveforms). This skew also shapes the learning task. For the small events that dominate the training set, $f_c$ is high and the conditioning waveform already contains much of the target band; for the rare larger events, nearly the entire engineering-relevant band must be synthesized. HighFEM-2 is thus most useful for the larger events, which are the least represented in training.

Second, physics-based inputs enter only through the two Amatrice scenarios of Section~\ref{sec:dynamic_rupture}. Validation against many simulated events is needed before routine coupling to simulation workflows.

Third, all waveforms are represented in a fixed 90~s window with the P arrival at 5~s (Section~\ref{sec:data-selection}). As already noted for HighFEM-1 \cite{palgunadi2025high}, this constrains the usable duration of the synthetics: for large events at long distances, significant shaking duration and the moveout of later phases approach the window length, and events requiring longer records cannot be represented without retraining on longer windows.

Finally, onset timing in the synthetics is reconstructed rather than inherited exactly from the conditioning waveform; future work could refine the phase-reconstruction step to sharpen first arrivals, which matters for applications sensitive to onset timing.

\section{Conclusion}
\label{sec:conclusion}
HighFEM-2 makes it possible to generate, within seconds, ensembles of realistic three-component broadband (0.1--50~Hz) ground-motion records for any earthquake scenario for which a low-frequency waveform and basic source and site parameters are available. Because the conditioning waveform carries the deterministic source, path, and directivity information across a network of seismometers, the model only learns the regional high-frequency scattering.

On the Italian test set, HighFEM-2 reproduces observed PGA and PGV with mean biases of 10\% and 14\%, and matches observed envelopes, Fourier spectra, and integrated velocity and displacement waveforms. The Japan-trained HighFEM-1 overpredicts the same data substantially, so generative waveform models must be trained regionally. The model is well constrained for small-to-moderate crustal events in Italy and unverified beyond that range or in other tectonic settings.

The practical value of this approach is greatest where recordings are scarcest, such as the near-source range, where a physics-based rupture simulation can supply the low-frequency waveform. HighFEM-2 thus opens new possibilities for hazard and risk assessment based on physically consistent broadband waveform ensembles.

\appendix
\section{Model details}
\label{app:model-architecture}
\subsection{Autoencoder} 

We employ a Variational Autoencoder (VAE) to compress high-dimensional log-spectrograms into a compact latent representation. The network follows a fully convolutional U-Net-style encoder-decoder architecture. The encoder maps input log-spectrograms (with $256$ STFT channels, a maximum sequence length of $8\ 161$, and a hop size of $32$) into a continuous latent space of $8$ channels. The base channel dimension of the network is set to 64, which is progressively expanded through a series of channel multipliers $(1, 2, 2, 4)$. Spatial downscaling is performed at three distinct stages using convolutional downsampling with strides of $(2, 2)$, $(2, 2)$, and $(1, 2)$, respectively. Each resolution level comprises three residual blocks. These blocks consist of two consecutive convolutional layers with Group Normalization (using $3$2 groups) and SiLU (Swish) activation functions. A dropout rate of $0.1$ is applied within the residual blocks to prevent over-fitting. The encoder outputs a $16$-channel tensor, which is split equally into shift (mean) and log-scale (variance) parameters to parameterize the Gaussian latent distribution. The decoder mirrors the encoder, using nearest-neighbor upsampling followed by convolutions to reconstruct the original $3$-channel input space. The autoencoder is optimized using the AdamW \cite{loshchilov2019decoupled} optimizer with a learning rate of $10^{-4}$, decaying to $10^{-6}$ over $30\ 000$ steps, a weight decay of $10^{-5}$, and a KL divergence regularization weight of $10^{-6}$. 

\subsection{Diffusion model}

The generative model operates on the $8$-channel latent space of the VAE using a MM-DiT \cite{esser2024scaling}. Both the noisy latent $\vy$ and the context $\vx$ are independently tokenized into $2 \times 2$ patches, projected to a dimension of $D = 384$, and augmented with 2D sinusoidal positional embeddings. The conditioning vector $\text{emb}$ is computed as the element-wise sum of the continuous timestep $t$ (encoded via 1D sinusoidal features) and the meta-data vector $\vc$, after both are projected into a shared $256$-dimensional embedding space via parallel MLPs (see Figure~\ref{fig:mmdit-architecture}
).

The core network comprises $12$ transformer blocks that process the latent and context streams simultaneously. Each block employs Adaptive Layer Normalization (adaLN) to regress shift, scale, and gating parameters from the global conditioning vector. The modulated tokens undergo joint multi-head attention ($6$ heads) across the concatenated sequences, with RMSNorm applied to the query and key tensors to stabilize training. The feed-forward network expands the hidden dimension by a factor of $4$ (to $1536$) using a GELU activation, without dropout. Following the final block, a decoding head applies a terminal adaLN modulation and linear projection to map the tokens back to $32$ channels ($2 \times 2 \times 8$), which are then unpatchified to restore the original spatial tensor structure.

\section{Baseline models}
\subsection{Wasserstein GAN}

The Wasserstein GAN baseline closely follows the Pix2Pix and WGAN-GP frameworks from \cite{isola2017image} and \cite{gulrajani2017improved}.

\paragraph*{Architecture} The generator is structured as a Pix2Pix U-Net mapping context variables directly to the $8$-channel latent space, with standard downsampling and upsampling branches with a dropout rate of $0.5$. The critic follows a PatchGAN configuration with a base channel dimension of $128$ and a dropout rate of $0.1$, terminating in a convolutional layer that outputs local patch validity scores rather than a single scalar. Conditioning vectors (dimension $8$) are embedded into a $256$-dimensional space using an MLP and concatenated into the intermediate layer activations of both networks.

\paragraph*{Training} The framework is optimized using the Wasserstein GAN objective with a gradient penalty weight of $\lambda = 5$ to enforce the 1-Lipschitz constraint on the critic. The critic and generator are trained via the AdamW optimizer for $300\ 000$ steps with an exponential moving average (EMA) tracked for the generator weights. The critic parameters are updated iteratively to maximize the distance between the distribution of real autoencoder latents and synthesized generator latents, while the generator is updated concurrently to minimize this distance.

\subsection{Consistency Distillation}

\paragraph*{Architecture} The architecture is identical to the HighFEM-2 architecture.

\paragraph*{Training} This configuration uses consistency distillation to map any point along the EDM noise trajectory to the clean data distribution in $2$ steps, instead of a multi-step reverse process \cite{geng2025consistency}. The model is trained for $100\ 000$ steps by minimizing a distillation loss that penalizes discrepancies between predictions at adjacent noise scales $t$ and $r$ along the decay trajectory. A target network, updated as an exponential moving average, provides stable reference trajectories, and the student network is trained to satisfy the consistency condition across the noise schedule.

\section{Model evaluations}
\label{app:model-evaluations}

\begin{figure}
\centering
\includegraphics[width=0.99\linewidth]{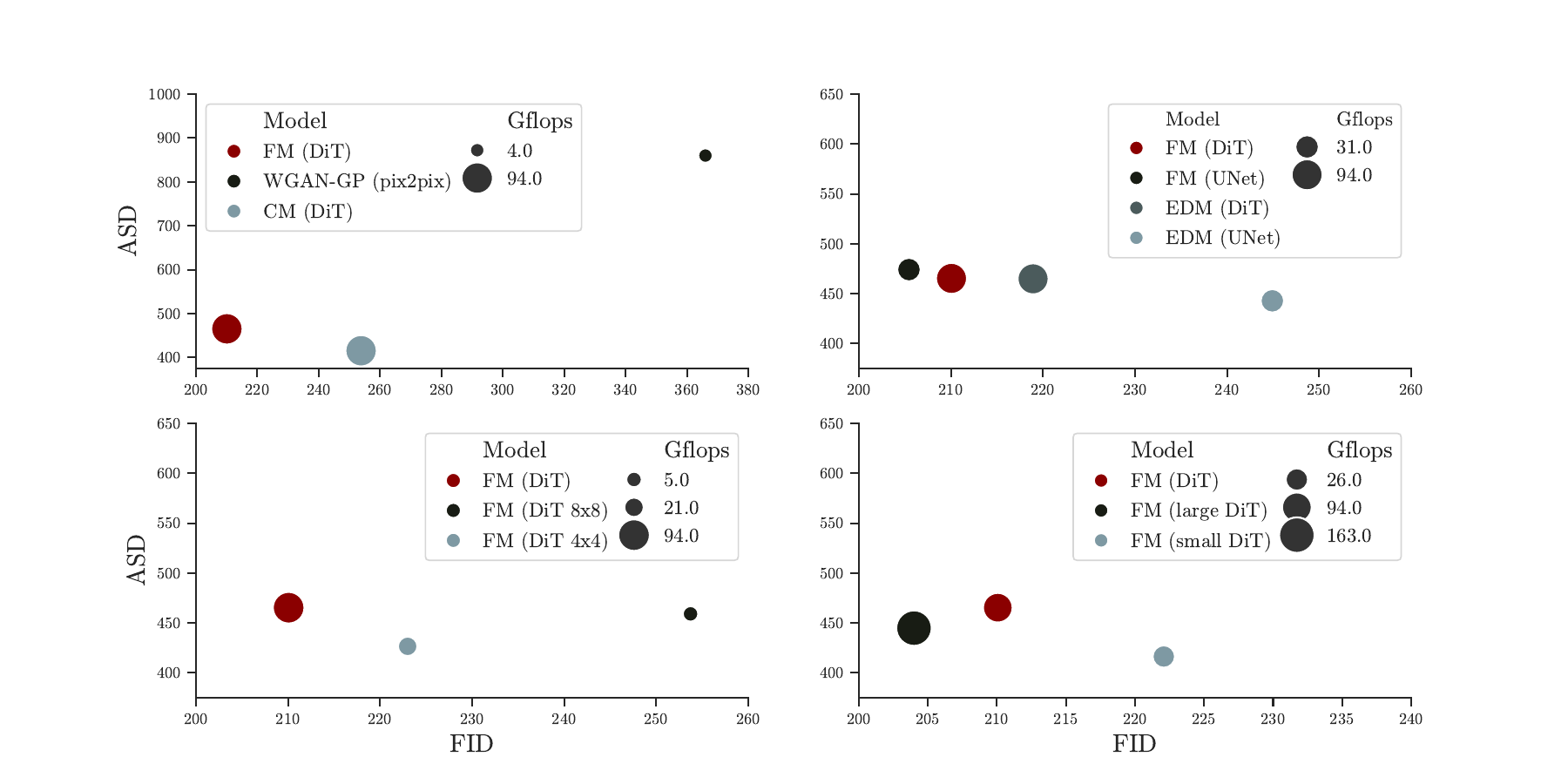}
\caption{Sample quality versus computational cost. Each panel plots ASD FD against FID (Table~\ref{tab:model-performance}; lower left is better); marker size scales with Gflops. Panels compare: baselines (top left), FM versus EDM (top right), patch sizes (bottom left), network sizes (bottom right). HighFEM-2 (FM (DiT)) lies near the quality optimum at intermediate cost.}
\label{fig:asd_vs_fid_vs_gflops}
\end{figure}

\begin{figure}
\centering
\begin{subfigure}[b]{0.48\textwidth}
\caption{{\bf HighFEM-2} (FM (DiT)).}
\vspace{-0.5cm}
\includegraphics[width=1\textwidth]{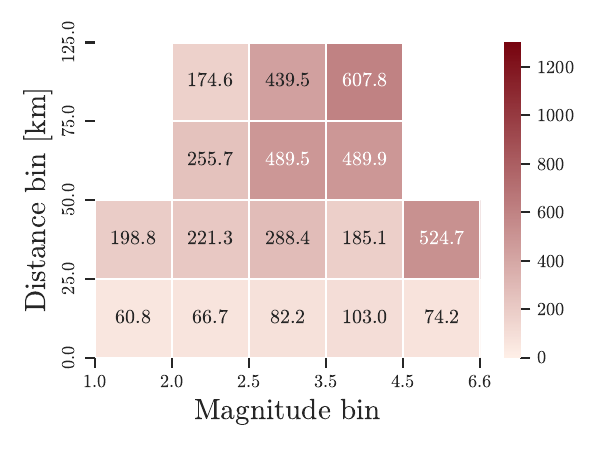}
\end{subfigure}
\hfill
\begin{subfigure}[b]{0.48\textwidth}
\caption{EDM (UNet).}
\vspace{-0.5cm}
\includegraphics[width=1\textwidth]{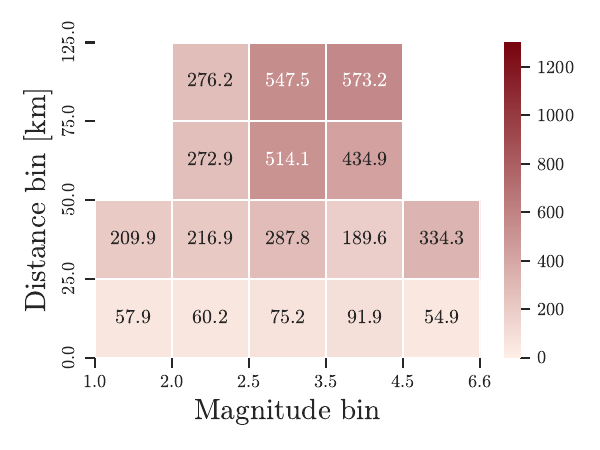}
\end{subfigure}

\vspace{0.3cm} %

\begin{subfigure}[b]{0.48\textwidth}
\caption{Consistency model.}
\vspace{-0.5cm}
\includegraphics[width=1\textwidth]{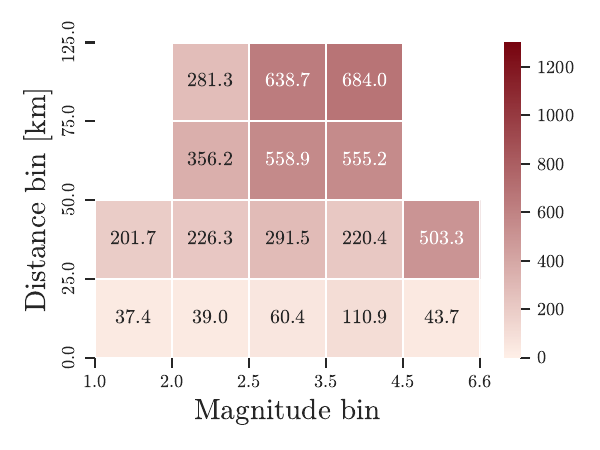}
\end{subfigure}
\hfill
\begin{subfigure}[b]{0.48\textwidth}
\caption{WGAN-GP.}
\vspace{-0.5cm}
\includegraphics[width=1\textwidth]{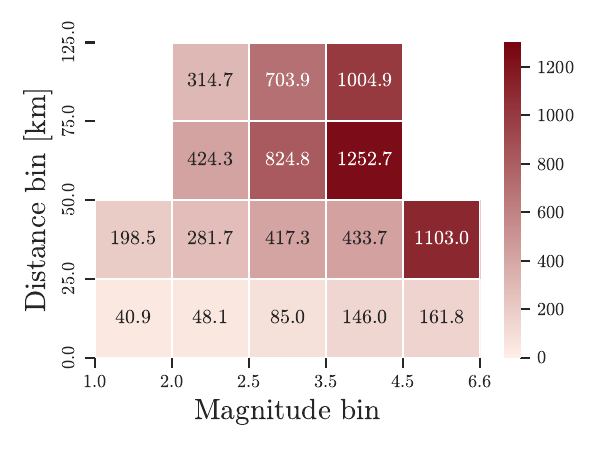}
\end{subfigure}

\caption{FID per magnitude vs. distance bin for HighFEM-2 (a) and three baseline models (b--d). For FID lower is better, empty cells contain no test records. All models degrade toward the sparse high-magnitude bins with WGAN-GP (d) degrading the most.}
\label{fig:model_comparison_fid}
\end{figure}

\begin{figure}
\centering
\begin{subfigure}[b]{1\textwidth}
\caption{{\bf HighFEM-2} (FM (DiT)).}
\vspace{-0.5cm}
\includegraphics[width=1\textwidth]{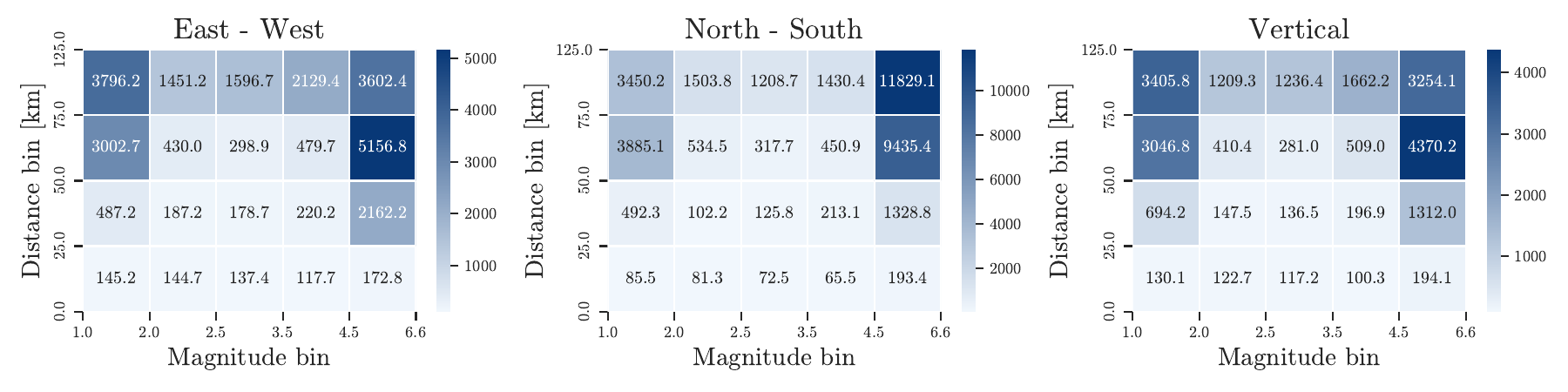}
\end{subfigure}

\begin{subfigure}[b]{1\textwidth}
\caption{EDM (UNet).}
\vspace{-0.5cm}
\includegraphics[width=1\textwidth]{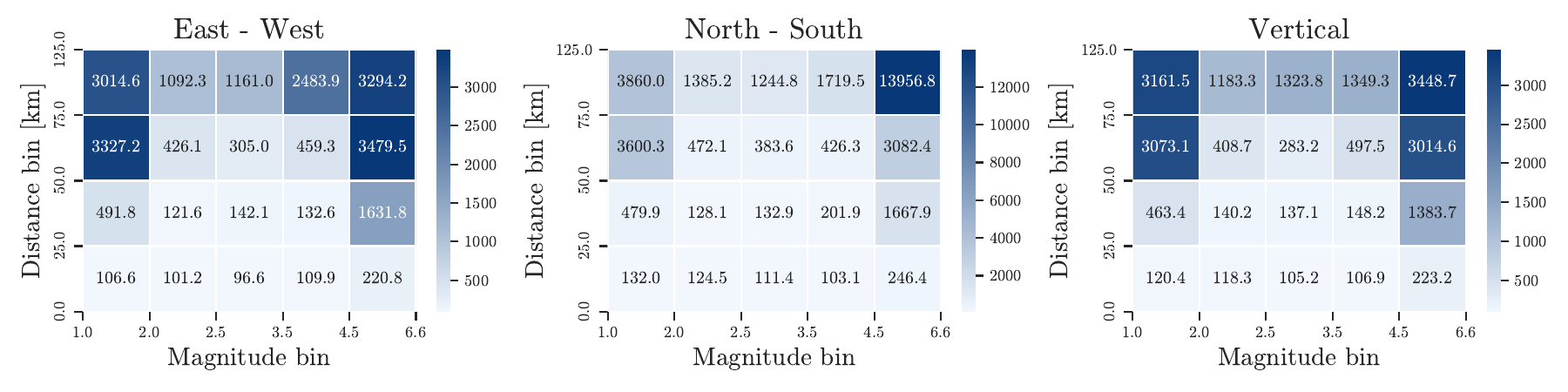}
\end{subfigure}

\begin{subfigure}[b]{1\textwidth}
\caption{Consistency model.}
\vspace{-0.5cm}
\includegraphics[width=1\textwidth]{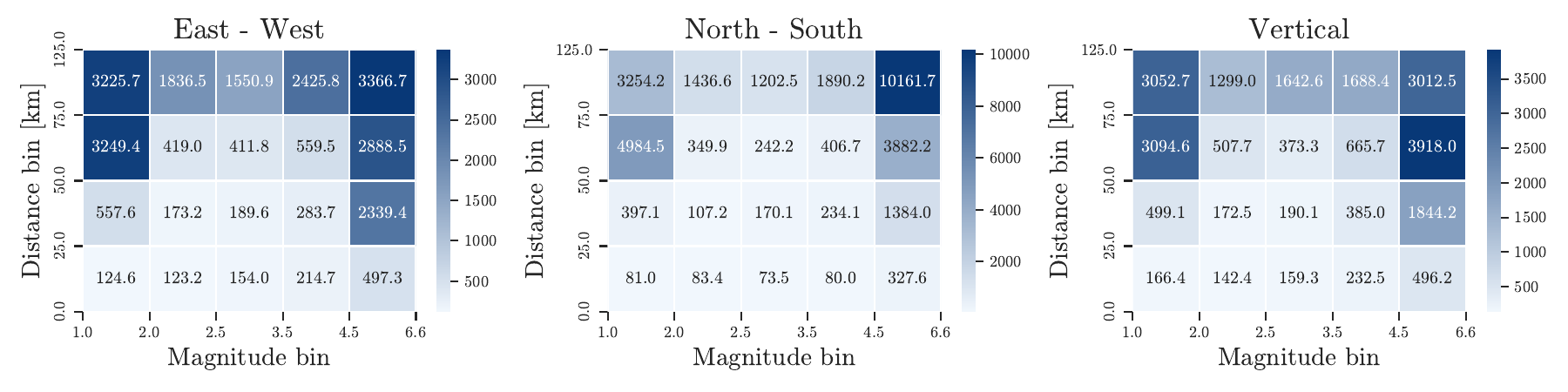}
\end{subfigure}
    
\begin{subfigure}[b]{1\textwidth}
\caption{WGAN-GP.}
\vspace{-0.5cm}
\includegraphics[width=1\textwidth]{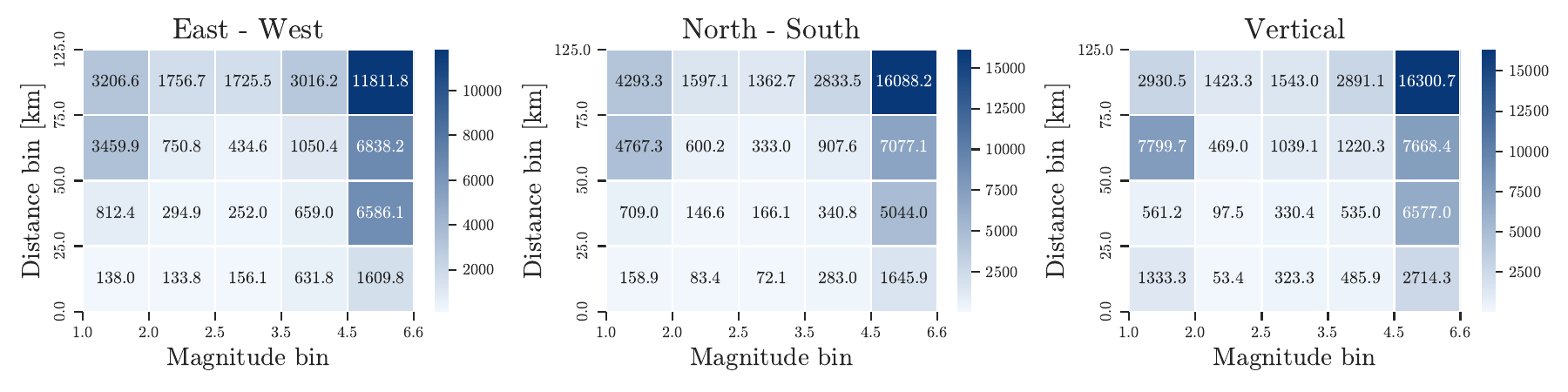}
\end{subfigure}

\caption{ASD FD per magnitude vs. distance bin shown separately per direction. Similarly to Figure~\ref{fig:model_comparison_fid}, all models degrade toward the sparse high-magnitude bins. The differences between the different models is small, with HighFEM-2 being overall the best-performing model.}
\label{fig:model_comparison_asd}
\end{figure}

This appendix compares HighFEM-2 against alternative objectives, architectures, and hyperparameters using machine-learning metrics (Table~\ref{tab:model-performance}, Figures~\ref{fig:asd_vs_fid_vs_gflops}--\ref{fig:model_comparison_asd}). Rankings under these metrics do not always coincide with waveform realism (for example, FM (UNet) below), so the main-text comparison (Section~\ref{sec:model-comparisons}) also considers envelope consistency and spectral stability.

\begin{table}[t]
\centering
\caption{Model comparison. NFE: number of network evaluations per sample. Gflops: cost per evaluation. Runtime: wall-clock time per three-component synthetic on a single NVIDIA H100. FID: Fr\'echet distance between real and synthetic samples in the feature space of a classifier trained to predict magnitude--distance bins. ASD FD: simplified (diagonal-covariance) Fr\'echet distance of the log-amplitude spectra, averaged over the three components. Lower is better for all columns. FM: rectified flow matching; EDM: diffusion \protect\cite{karras2022edm}; CM: consistency distillation; DiT/UNet: network architecture; $8{\times}8$/$4{\times}4$: patch size; small/large: network width. HighFEM-2 is FM (DiT).}
\label{tab:model-performance}
\small
\renewcommand{\arraystretch}{0.9}
\begin{tabular}{@{}l@{\hspace{2cm}}lrrrrr@{}}
\toprule
& \textbf{NFE} $(\downarrow)$&  \textbf{Gflops} $(\downarrow)$&  \textbf{Runtime [s]} $(\downarrow)$&  \textbf{FID} $(\downarrow)$ & \textbf{ASD FD} $(\downarrow)$  \\
\textbf{Diffusion} \\
\midrule
{\bf HighFEM-2} (FM (DiT)) & 25 & 94.43 & 1.294 & 210.06 & 465.27 \vspace{0.15cm}\\
FM (UNet) & 25 & 31.22 & 0.485 & 205.44 & 474.09 \\
EDM (DiT) & 50 & 94.43 & 2.568 & 218.93 & 464.86 \\
EDM (UNet) & 50 & 31.22 & 0.926 & 244.92 & 442.8 \vspace{0.15cm}\\
FM (DiT 8x8) & 25 & 5.42 & 0.107 & 253.72 & 459.06 \\
FM (DiT 4x4) & 25 & 21.84 & 0.291 & 222.99 & 426.54 \vspace{0.15cm}\\
FM (small DiT) & 25 & 26.15 & 0.549 & 222.08 & 416.33 \\
FM (large DiT) & 25 & 163.35 & 1.925 & \textbf{203.99} & 444.79 \vspace{0.15cm}\\
\midrule
\textbf{Baselines} \\
WGAN-GP (pix2pix) & 1 & 4.06 & \textbf{0.003} & 366.01 & 859.97 \\
CM (DiT) & 2 & 94.43 & 0.102 & 253.79 & \textbf{415.41} \\
\bottomrule
\end{tabular}
\end{table}

FM (UNet) reaches similar machine-learning metrics at lower cost (Table~\ref{tab:model-performance}) but produces less consistent envelopes in the sparse high-magnitude bins, so we retain the transformer variant.

\section{Magnitude Homogenization}
\label{app:magnitude}
We homogenize all event magnitudes to $M_w$ using the following conversions
\begin{itemize}
    \item \textbf{From $M_L$ to $M_w$:}
    \begin{align}
        M_w &= 0.79\,M_L + 1.20, && \text{for } 3.5 \leq M_L \leq 5.8~\text{\cite{castello2007local}},\\
        M_w &\approx M_L + 0.20 \pm 0.10, && \text{for } 2.5 \leq M_L < 3.5~\text{\cite{gasperini2013empirical, lolli2020homogenized}},\\
        M_w &= 0.64\,M_L + 0.84, && \text{for } 0 \leq M_L < 2.5~\text{\cite{bindi2005local}}.
    \end{align}

    \item \textbf{From $M_d$ to $M_w$:}
    \begin{equation}
        M_w \approx 1.10\,M_d + 0.10, \qquad \text{for } M_d \geq 3~\text{\cite{gasperini2013empirical}}.
    \end{equation}

    \item Events whose magnitudes fall outside the calibrated ranges of these conversions are excluded from further analysis.
\end{itemize}

\section{Comparison of Japanese and Italian Peak Ground Acceleration}
\label{app:japan-italy}
\begin{figure}
    \centering
    \includegraphics[width=\linewidth]{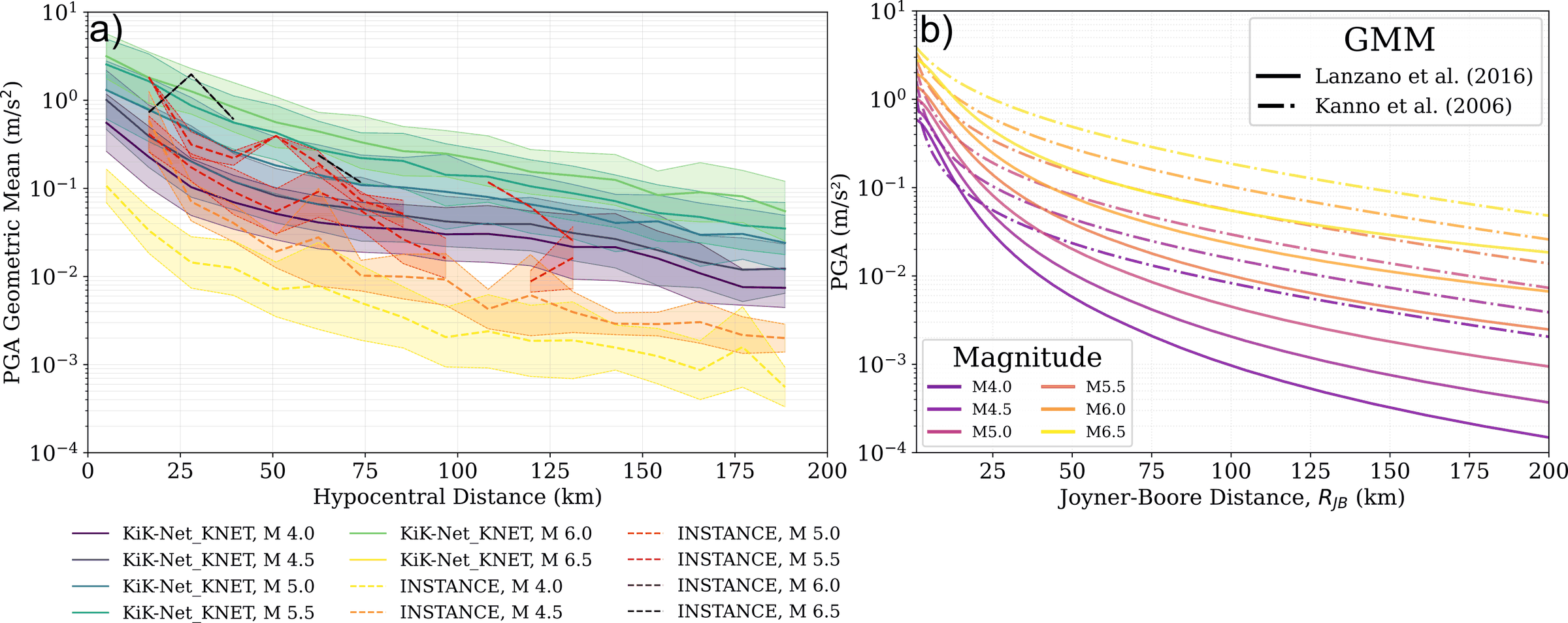}
    \caption{Comparison of peak ground acceleration between two regions, Japan (KiK-net and K-NET \protect\cite{knet2019}) and Italy \protect\cite{michelini2021instance}. (a) Geometric mean of the two horizontal components for the Japan region shown with the viridis colormap and the Italy region shown with the hot colormap. Dashed lines represent the median values within magnitude and hypocentral distance bins for the Italy region, with shading bounded by one standard deviation. Solid lines represent the corresponding means for the Japan region. (b) Ground motion models (GMMs) from the two regions. GMMs for the Italy region from \protect\citeA{lanzano2016ground} are shown as solid lines for different magnitude bins, while GMMs from \protect\citeA{kanno2006new} for the Japan region are shown as dot-dashed lines for different magnitude bins.}
    \label{fig:comparing_japan_italy}
\end{figure}

To understand the overprediction of HighFEM-1 on the Italian test data (Section~\ref{sec:regional-transfer}), we compare the geometric mean of horizontal PGA of onshore crustal earthquakes in the Japanese (KiK-net and K-NET; \citeA{knet2019}) and Italian (INSTANCE; \citeA{michelini2021instance}) datasets (Figure~\ref{fig:comparing_japan_italy}a). For the same magnitude and hypocentral distance, Japanese PGA is systematically higher; for $M$~4.0--4.5, where both datasets contain many records, the difference grows with distance and exceeds an order of magnitude beyond about 100~km. The regional GMMs show the same pattern: \citeA{kanno2006new} for Japan predicts higher PGA than \citeA{lanzano2016ground} for Italy, increasingly so with distance (Figure~\ref{fig:comparing_japan_italy}b). Because ground motion differs strongly between the two regions, a generative model reproduces the amplitudes of the region it was trained on; like GMMs, generative waveform models are regional. Because the two datasets also sample different magnitudes (many $M_w$~4--6 events in Japan, mostly $M_w<2$ in Italy), this comparison concerns the transferability of a trained model, not the relative quality of the two model formulations (Section~\ref{sec:model-comparisons}).

\clearpage
\section*{Open Research Section}\label{sec:data_resources}
The INSTANCE strong-motion dataset used to train, validate, and test HighFEM-2 is openly available from the Istituto Nazionale di Geofisica e Vulcanologia (INGV) at \url{https://www.pi.ingv.it/banche-dati/instance/} (last access October 2025) and is described by \citeA{michelini2021instance}. The Japanese strong-motion records (K-NET and KiK-net) used for the regional amplitude comparison are provided by the National Research Institute for Earth Science and Disaster Resilience (NIED) at \url{https://www.kyoshin.bosai.go.jp/} \cite{knet2019}. Phase picks for the data-cleaning step were computed with PhaseNet \cite{zhu2019phasenet}. The physics-based dynamic-rupture simulations used as alternative low-frequency inputs were generated by \citeA{gallovivc2023broadband}. The empirical ground-motion models used for comparison are those of \citeA{boore2014nga}, \citeA{lanzano2016ground}, and \citeA{kanno2006new}.

The source code for HighFEM-2, the trained model weights, the data preprocessing and conditioning pipeline, and the scripts required to reproduce the figures of this study are developed openly at \url{https://github.com/highfem/tqe}. The models were implemented in Python; the signal-processing and ordinary-differential-equation routines used the \texttt{scipy} library \cite{virtanen2020scipy}.

\section*{Declaration of Competing Interest}
The authors declare that there are no conflicts of interest.

\acknowledgments
This work was supported by grant number C22-10 (HighFEM) of the Swiss Data Science Center (SDSC), Ecole Polytechnique Fédérale de Lausanne and ETH Zürich awarded to M-A. Meier, L. Ermert and M. Koroni. L. Ermert was supported by Swiss National Science Foundation grant 209941. M. Koroni is supported by the Swiss Federal Nuclear Safety Inspectorate (ENSI) under contract number CTR00830. We thank Donat Fäh and Paolo Bergamo for useful discussions on ground motion models and earthquake engineering. We thank CSCS Swiss National Computing Center (Piz Daint and Alps under projects sd28 and s1165) and Swiss Seismological Service "Bigstar" Cluster for providing computational resources for this research.

\clearpage
\bibliography{references}

\end{document}


\title{Supporting Information for "Broadband Ground-Motion Synthesis by Conditioning Denoising Diffusion Models on Low-Frequency Waveforms"}

\authors{Kadek Hendrawan Palgunadi\affil{1, \dagger \ddagger}, Simon Dirmeier\affil{2, \dagger}, Maria Koroni\affil{1}, Laura Ermert\affil{3}, Men-Andrin Meier\affil{1,4}}

\affiliation{1}{Swiss Seismological Service (SED), ETH Zürich, Switzerland}
\affiliation{2}{Swiss Data Science Center (SDSC), ETH Zürich, Switzerland}
\affiliation{3}{Univ. Grenoble Alpes, Univ. Savoie Mont Blanc, CNRS, IRD, Univ. Gustave Eiffel, ISTerre, Grenoble, France}
\affiliation{4}{Earth and Planetary Science Department, ETH Zürich, Switzerland}
\affiliation{\dagger}{Equal Contribution}
\affiliation{\ddagger}{Now: Geophysical Engineering Department, Institut Teknologi Sepuluh Nopember, Indonesia}

\section*{Introduction}
This document provides supporting figures and a detailed description of the data-filtering pipeline referenced in the main text.

\subsection*{Additional Evaluation Figures}

\begin{figure}[!htp]
    \centering
    \includegraphics[width=0.9\textwidth]{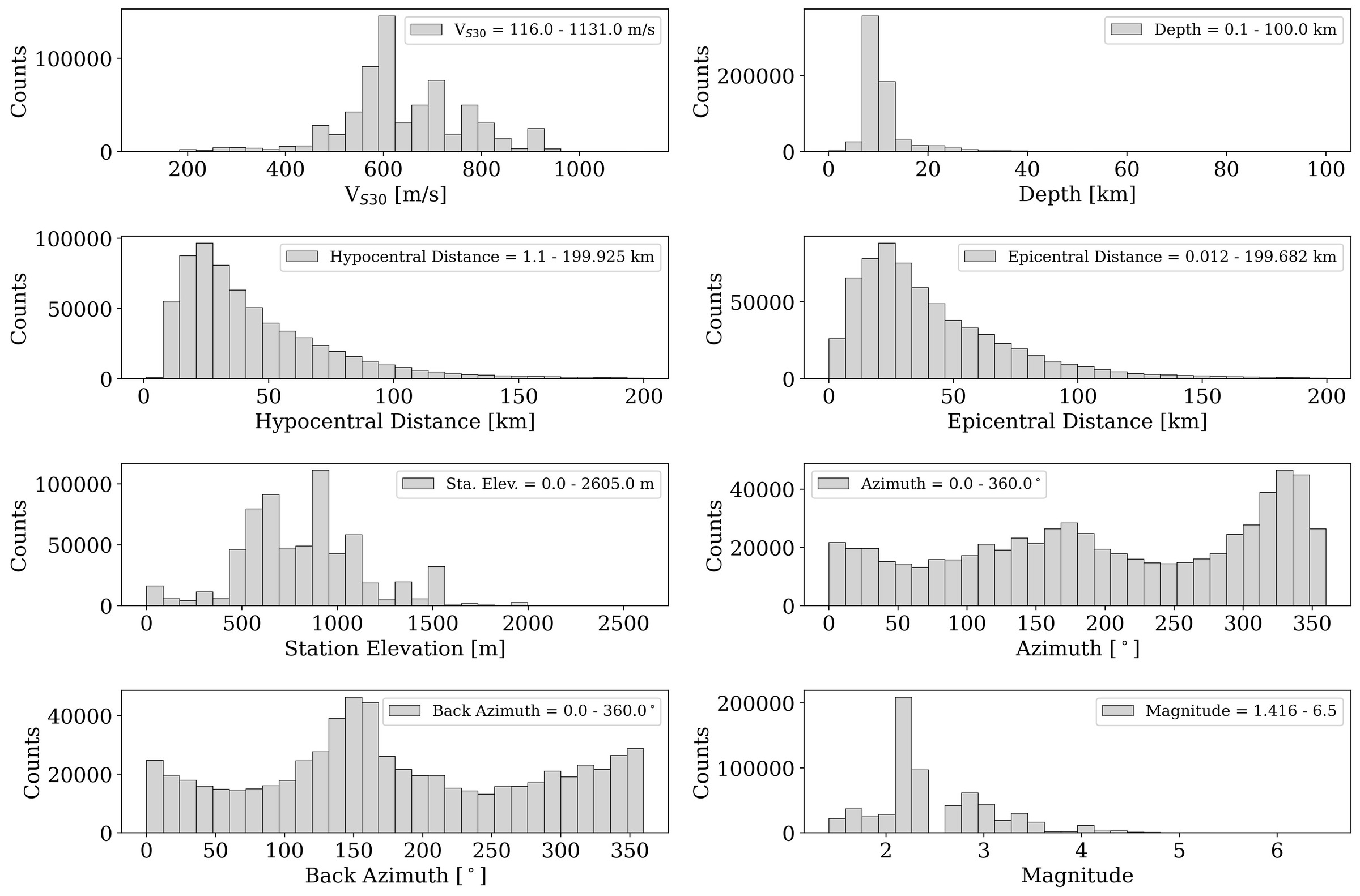}
    \caption{Statistics of the selected metadata used for model training, validation, and test.}
    \label{fig:statistics_data}
\end{figure}

\begin{figure}[!htp]
    \centering
    \includegraphics[width=\linewidth]{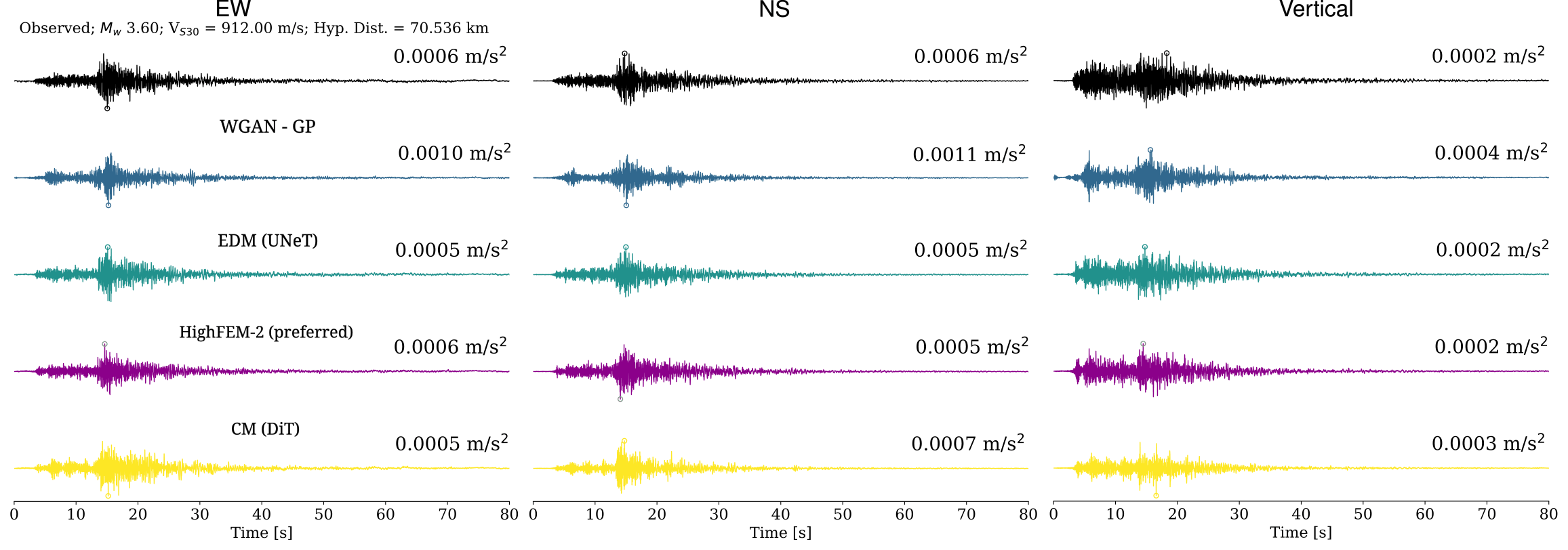}
    \caption{Acceleration waveform comparison between observed data (black) and synthetics from four models, including HighFEM-2 (purple). Waveforms are displayed across three components (left to right): East-West, North-South, and Vertical. The maximum amplitude (m/s$^2$) of each trace is indicated by the number on the right side of the corresponding waveform. The conditioning parameters correspond to a test-set record with $M_w~3.6$, $V_{S30}=912$ m/s, and a hypocentral distance of $70.54$~km.}
    \label{fig:compare_model_19}
\end{figure}

\begin{figure}[!htp]
    \centering
    \includegraphics[width=\linewidth]{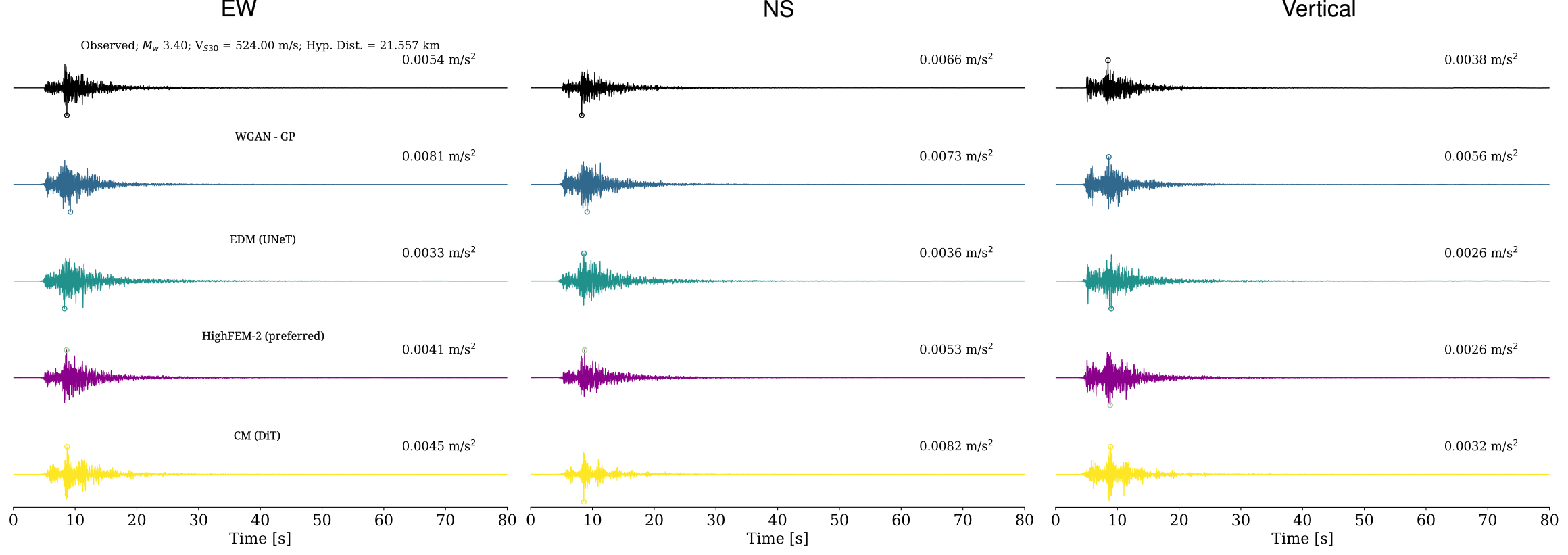}
    \caption{Acceleration waveform comparison between observed data (black) and synthetics from four models, including HighFEM-2 (purple). Waveforms are displayed across three components (left to right): East-West, North-South, and Vertical. The maximum amplitude (m/s$^2$) of each trace is indicated by the number on the right side of the corresponding waveform. The conditioning parameters correspond to a test-set record with $M_w=3.4$, $V_{S30}=524$ m/s, and a hypocentral distance of $21.56$ km.}
    \label{fig:compare_model_0}
\end{figure}

\begin{figure}[!htp]
    \centering
    \includegraphics[width=\linewidth]{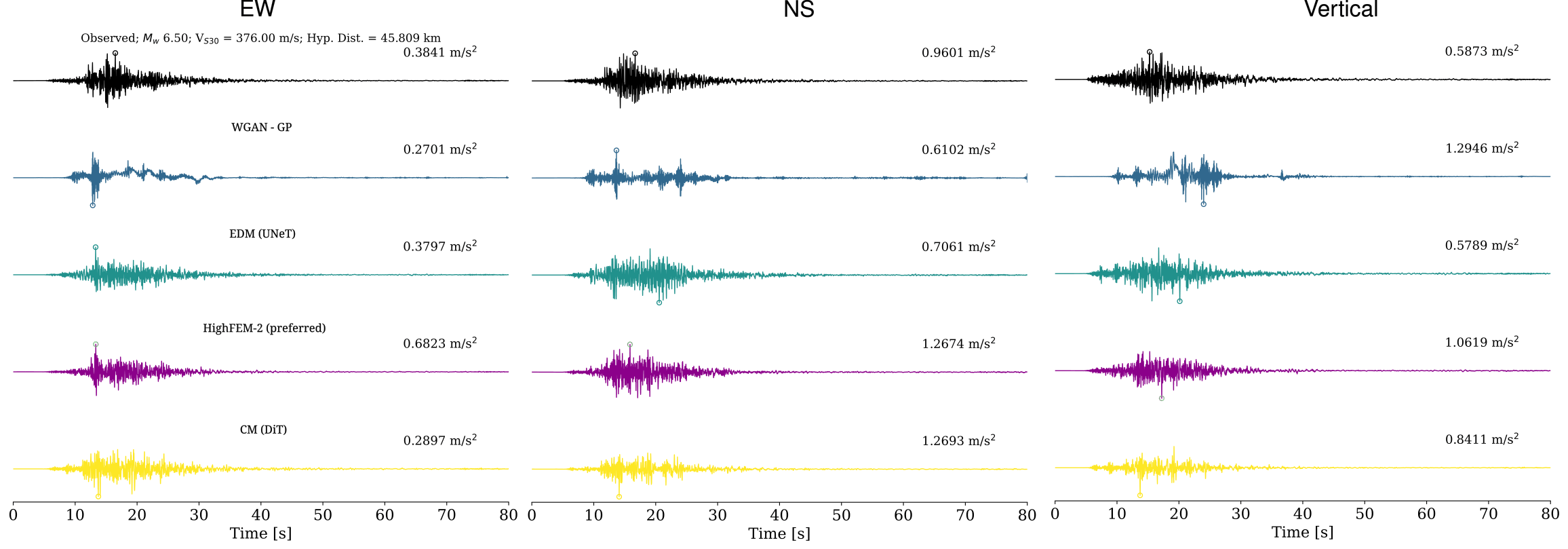}
    \caption{Similar to Figure \ref{fig:compare_model_0} but for conditioning parameters of $M_w~6.5$, $V_{S30}=376$ m/s, and a hypocentral distance of $45.81$ km.}
    \label{fig:compare_model_8399}
\end{figure}

\begin{figure}[!htp]
    \centering
    \includegraphics[width=\linewidth]{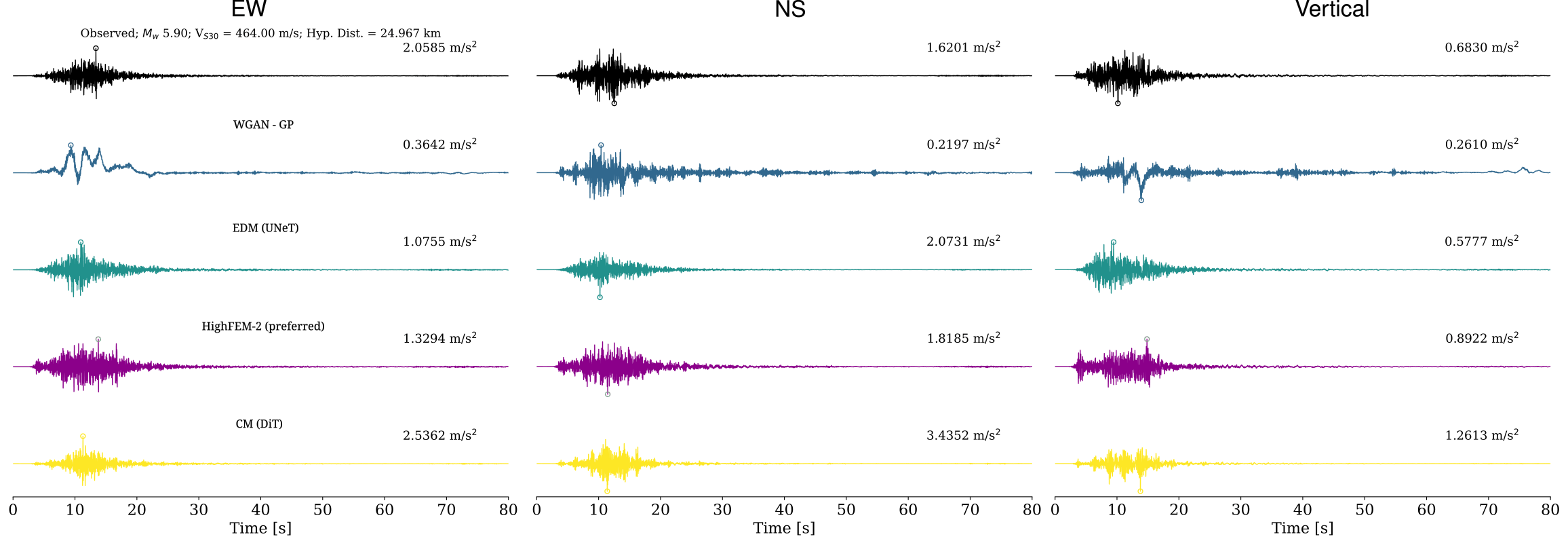}
    \caption{Similar to Figure \ref{fig:compare_model_0} but for conditioning parameters of $M_w~5.9$, $V_{S30}=464$ m/s, and a hypocentral distance of $24.97$ km.}
    \label{fig:compare_model_10572}
\end{figure}

\begin{figure}[!htp]
    \centering
    \includegraphics[width=0.8\linewidth]{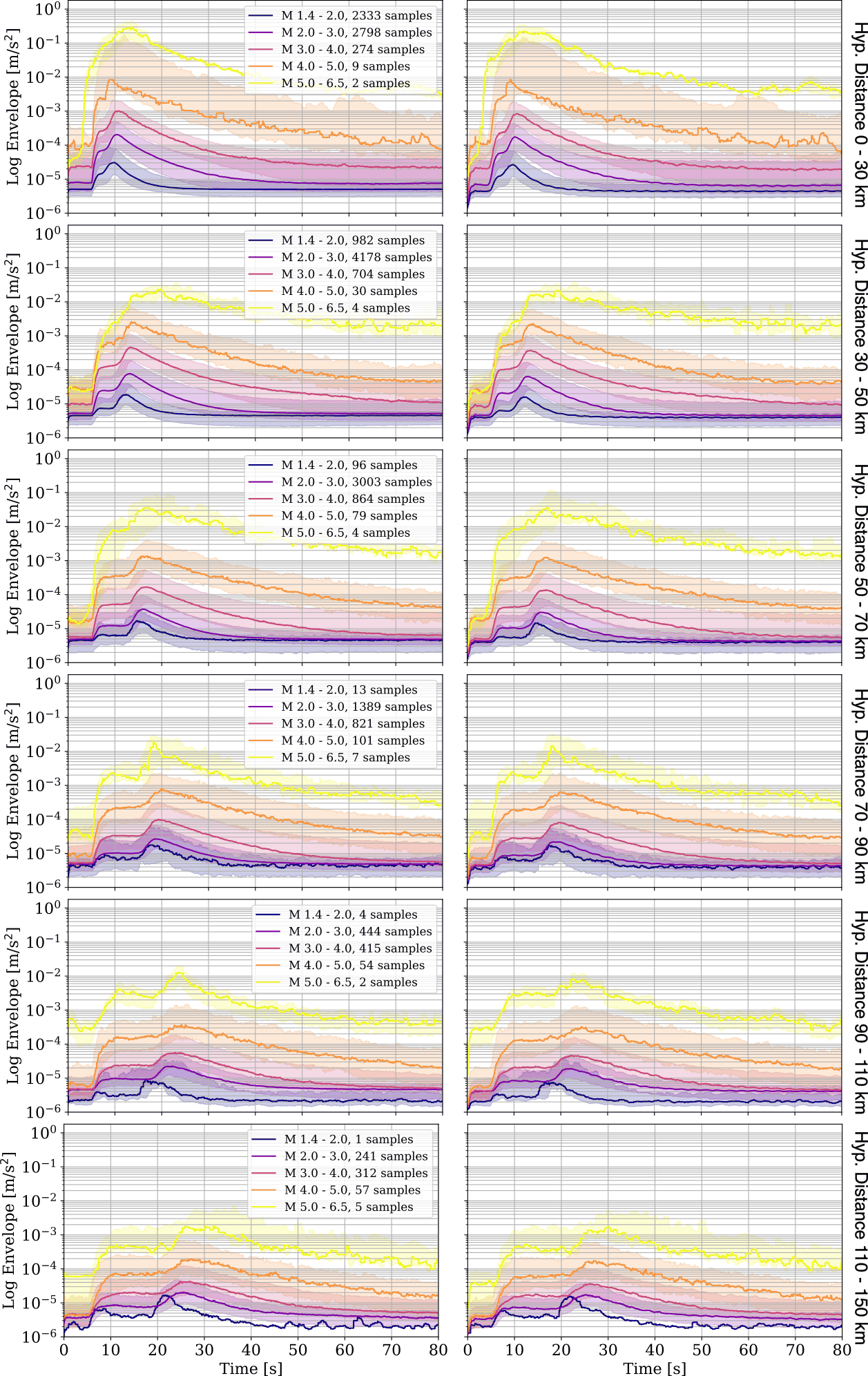}
    \caption{Distribution of time-domain envelopes for East-West component seismograms in different magnitude and distance bin. Target and predicted denote observed (left panel) and generative waveform models HighFEM-2 (right panel).}
    \label{fig:envelope_supp_EW}
\end{figure}

\begin{figure}[!htp]
    \centering
    \includegraphics[width=0.8\linewidth]{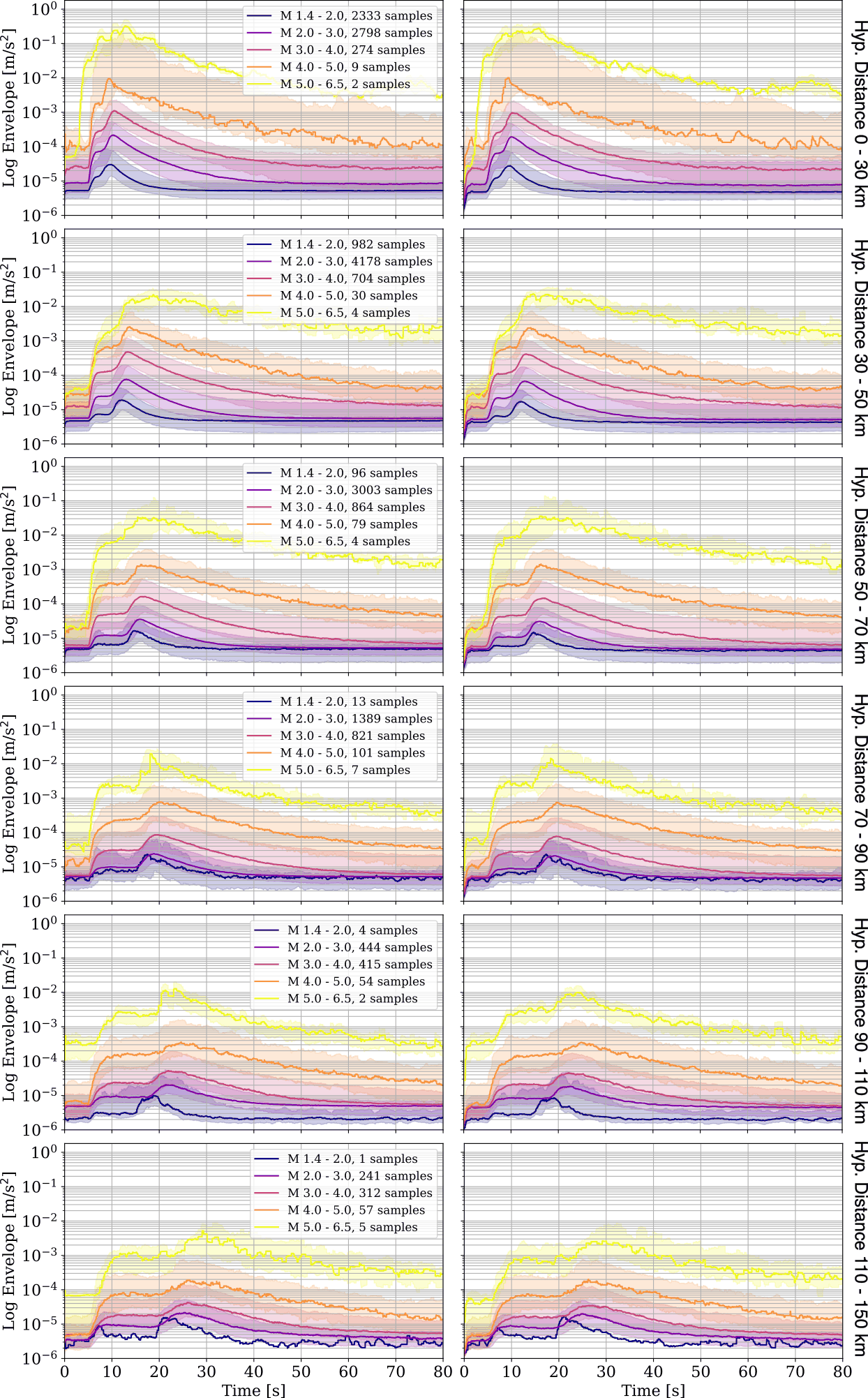}
    \caption{Similar to Figure \ref{fig:envelope_supp_EW} but for North-South component.}
    \label{fig:envelope_supp_NS}
\end{figure}

\begin{figure}[!htp]
    \centering
    \includegraphics[width=0.8\linewidth]{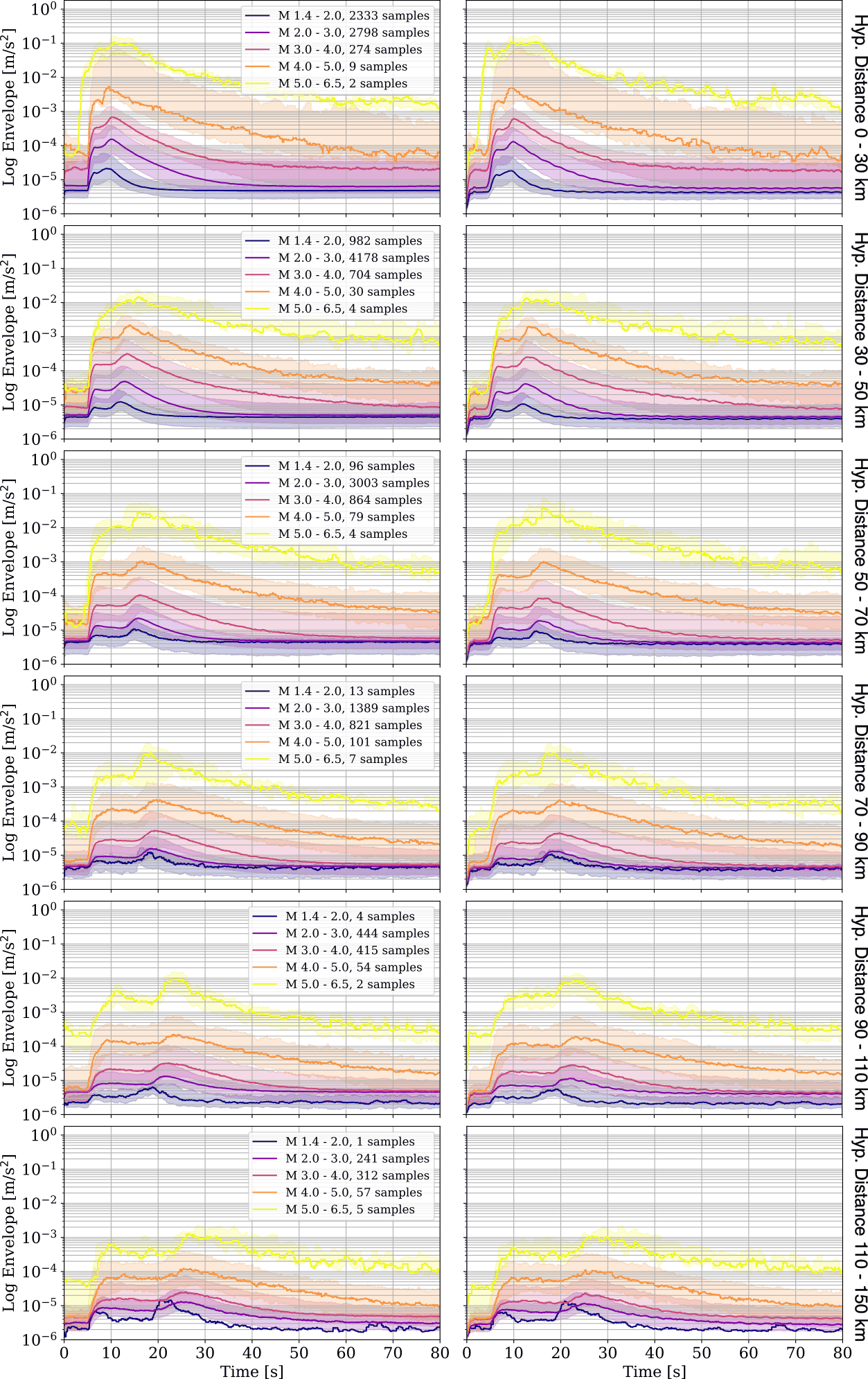}
    \caption{Similar to Figure \ref{fig:envelope_supp_EW} but for vertical component.}
    \label{fig:envelope_supp_UD}
\end{figure}

\begin{figure}[!htp]
    \centering
    \includegraphics[width=0.8\linewidth]{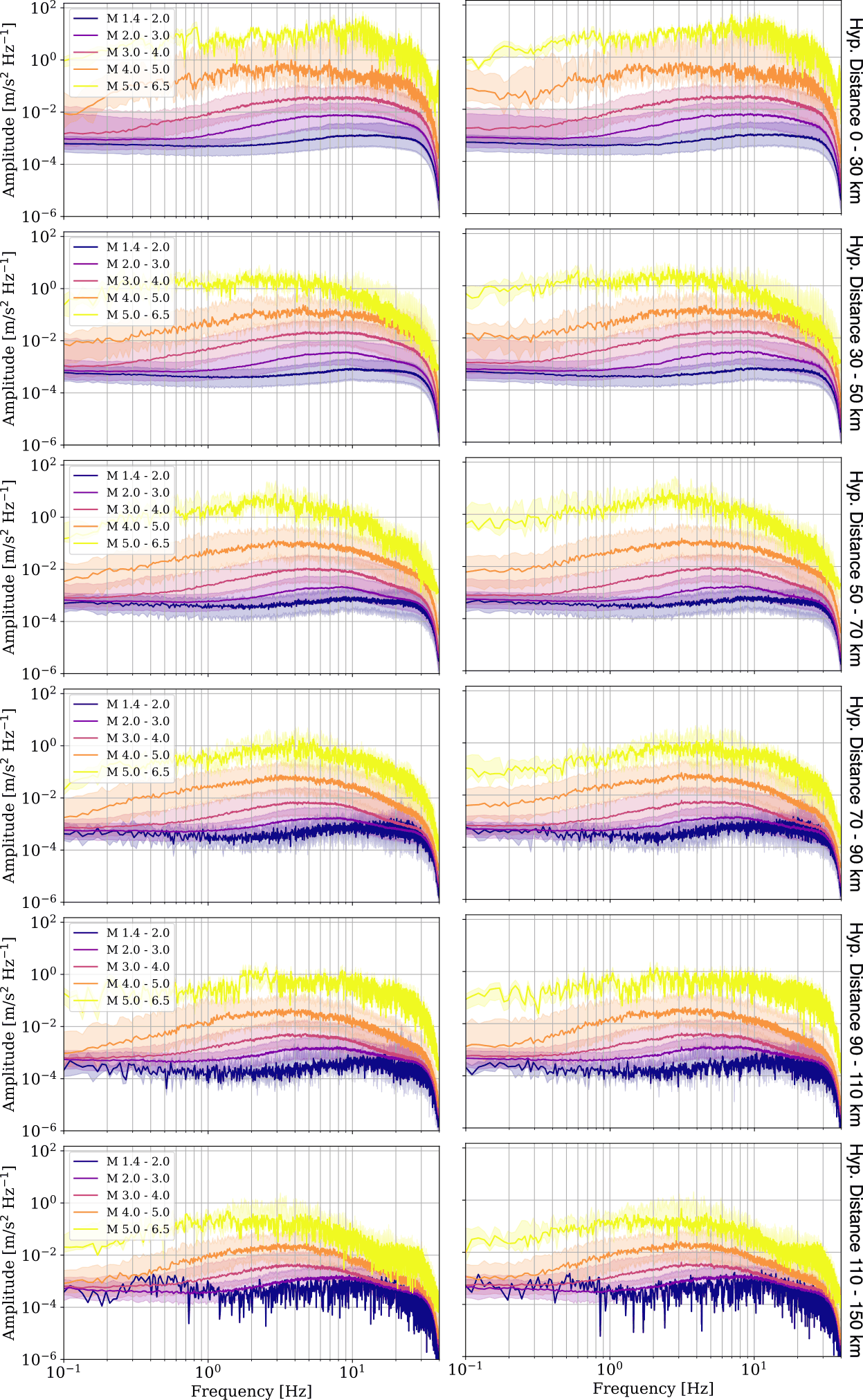}
    \caption{Distribution of Fourier spectra log-amplitude for East-West component seismograms in different magnitude and distance bin. Target and predicted denote observed (left panel) and generative waveform models HighFEM-2 (right panel).}
    \label{fig:spectra_supp_EW}
\end{figure}

\begin{figure}[!htp]
    \centering
    \includegraphics[width=0.8\linewidth]{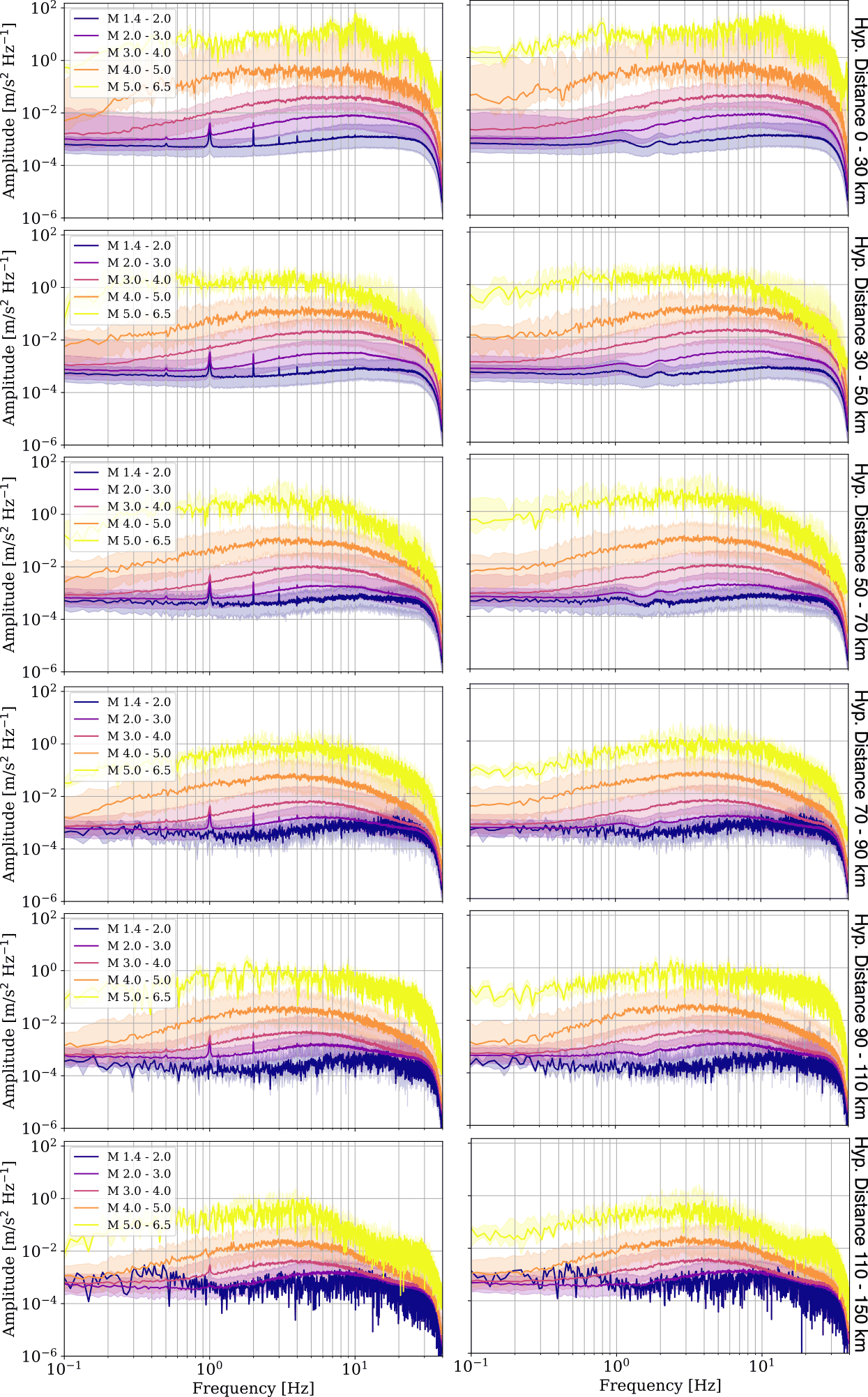}
    \caption{Similar to Figure \ref{fig:spectra_supp_EW} but for North-South component.}
    \label{fig:spectra_supp_NS}
\end{figure}

\begin{figure}[!htp]
    \centering
    \includegraphics[width=0.8\linewidth]{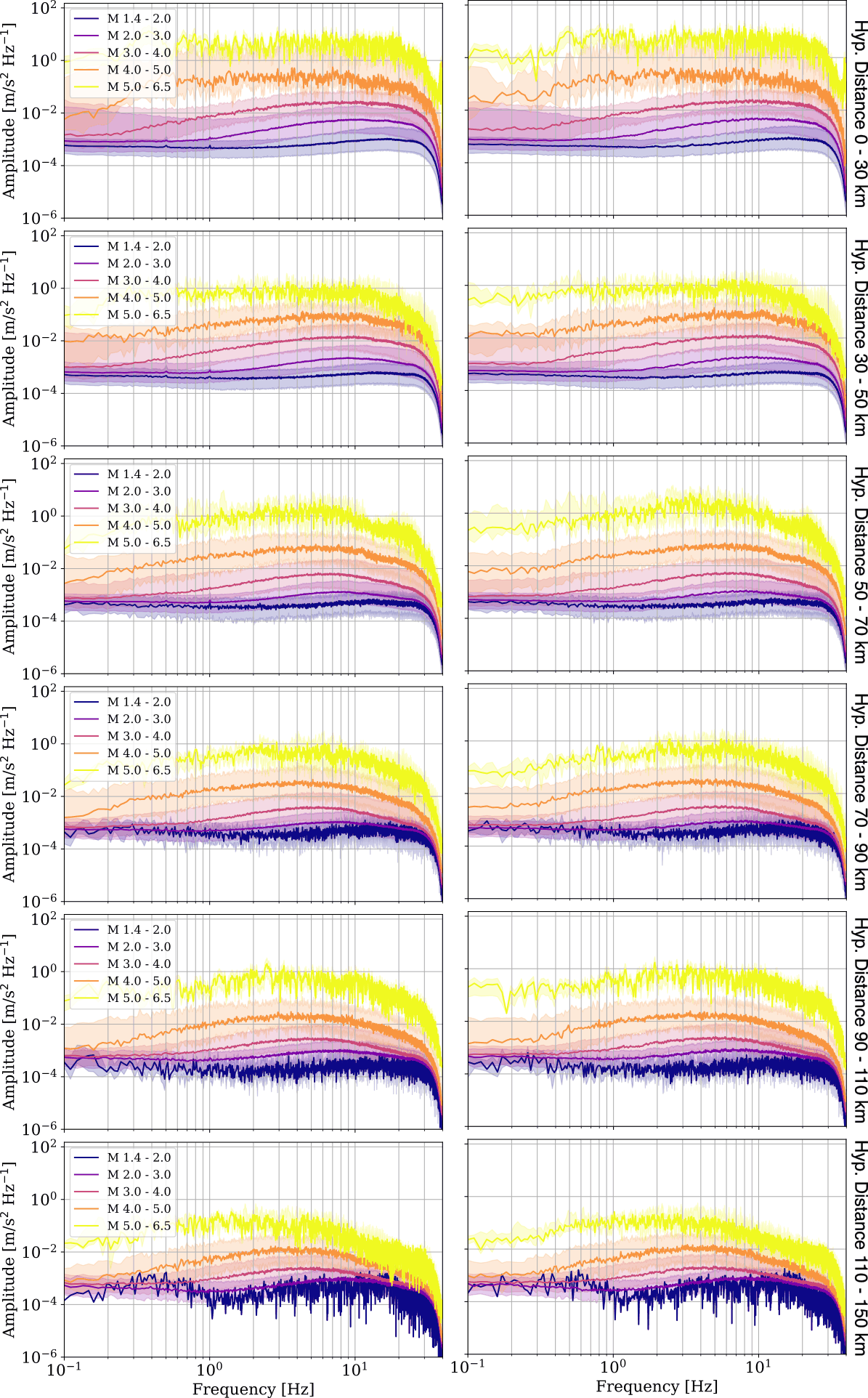}
    \caption{Similar to Figure \ref{fig:spectra_supp_EW} but for vertical component.}
    \label{fig:spectra_supp_UD}
\end{figure}

\begin{figure}[!htp]
    \centering
    \includegraphics[width=1\linewidth]{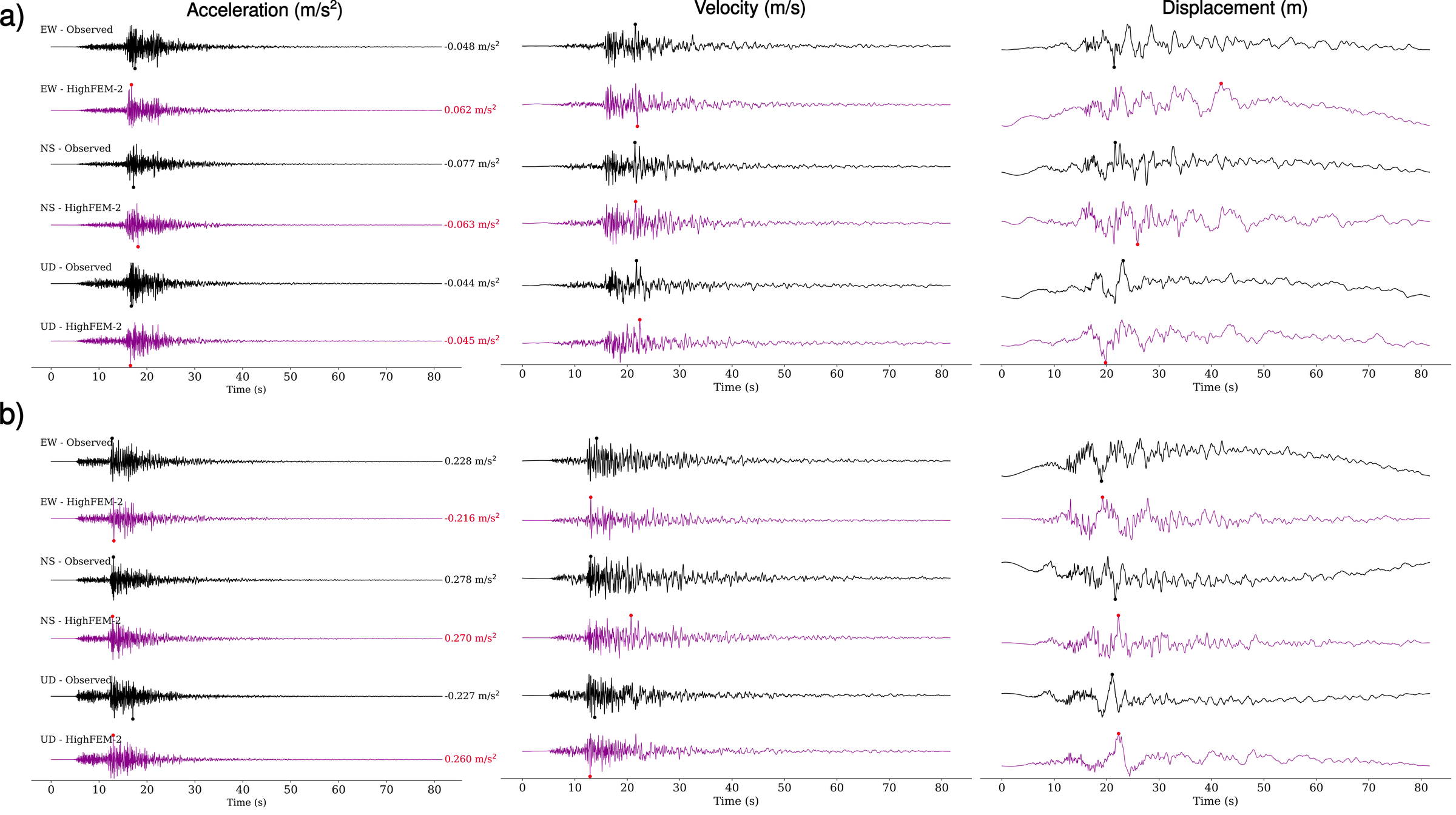}
    \caption{Waveform comparison between observed data (black) and HighFEM-2 (red) under different conditioning parameters. All generated HighFEM-2 waveforms P-phase-onset are aligned 5 s after the start of each trace and high-pass filtered at frequency of 0.1 Hz. (a) $M_w=5.1$, $V_{S30}=534$ m/s; station: $12.93^\circ$E, $43.188^\circ$N, elevation 479 m; hypocenter: depth 10 km, $13.27^\circ$E, $42.545^\circ$N; hypocentral distance 77.61 km. (b) $M_w=5.4$, $V_{S30}=516$ m/s; station: $13.07^\circ$E, $43.24^\circ$N, elevation 484 m; hypocenter: depth 8 km, $13.15^\circ$E, $42.79^\circ$N; hypocentral distance 50.75 km.}
    \label{fig:integration_waveforms_supp}
\end{figure}

\begin{figure}[!htp]
    \centering
    \includegraphics[width=0.99\linewidth]{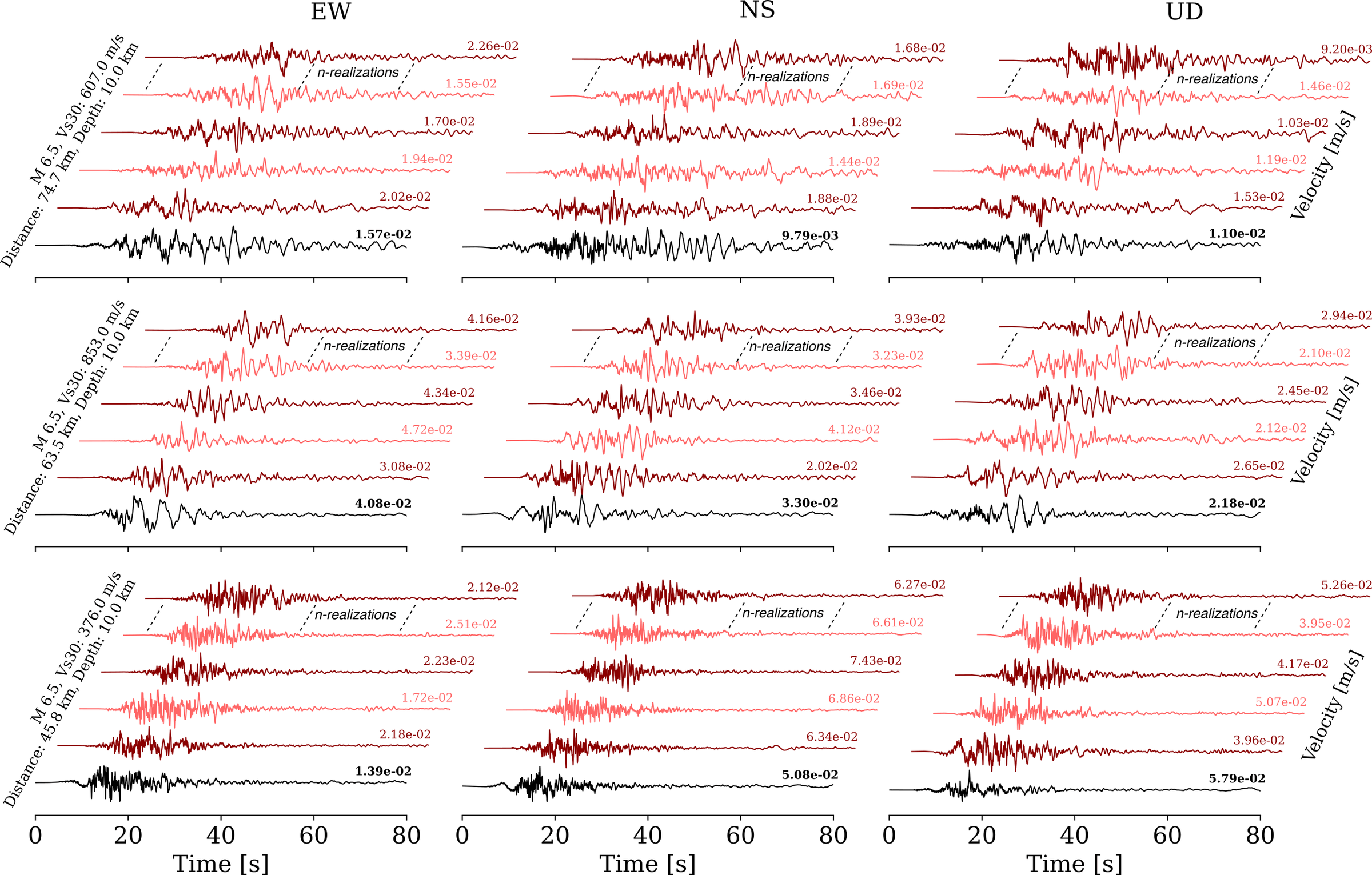}
    \caption{Observed three-component velocity waveforms integrated from the acceleration (black) and five randomly selected HighFEM-2 synthetics (alternating red shades for visibility) for a single station ($V_{S30}=370$~m/s, hypocentral distance 61.3~km) during the Norcia, 30 October 2016, $M_w$ 6.5 event. Peak absolute amplitudes are shown above each trace.}
    \label{fig:wavegen_vel}
\end{figure}

\begin{figure}[!htp]
    \centering
    \includegraphics[width=0.99\linewidth]{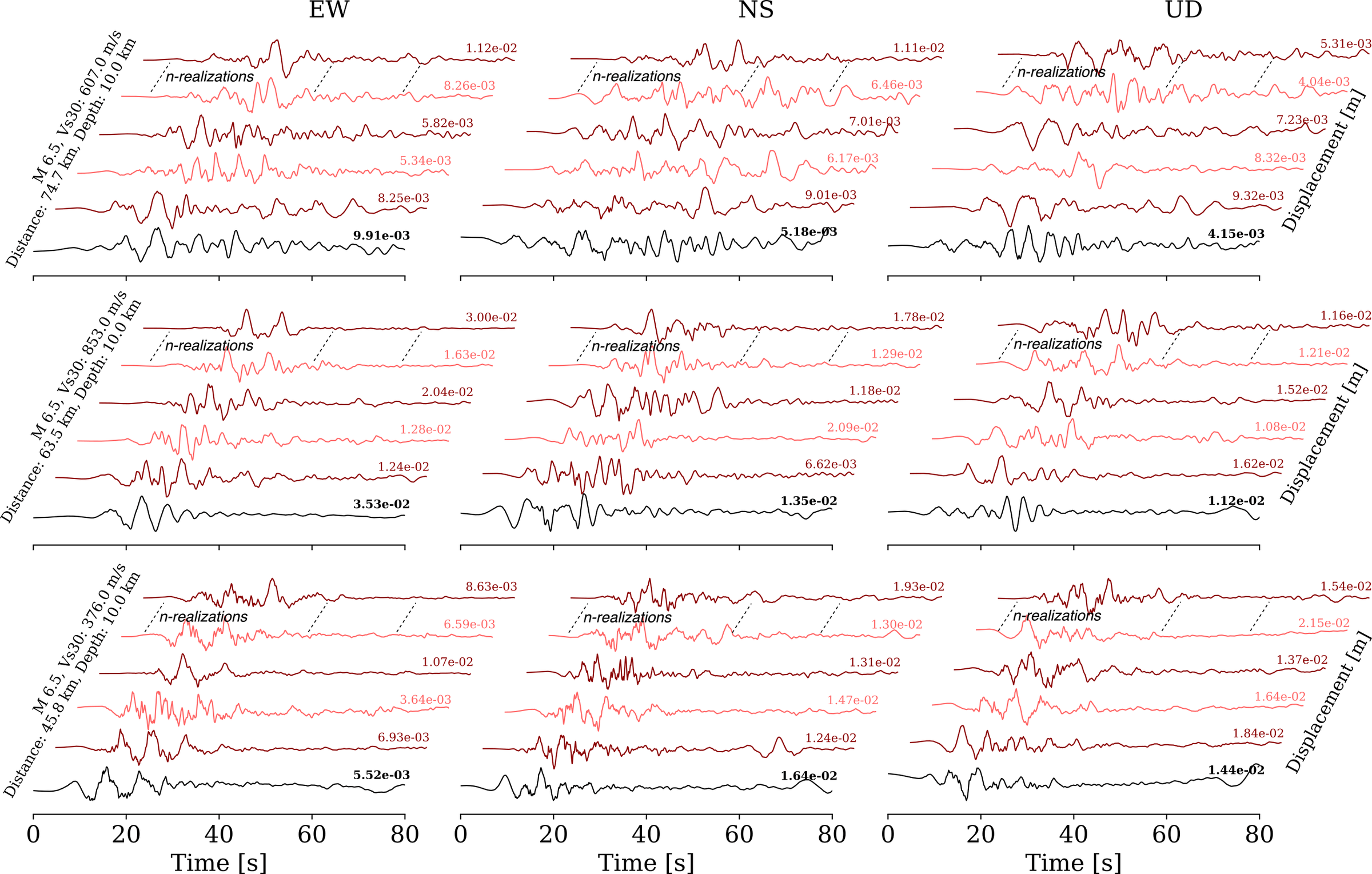}
    \caption{Observed three-component displacement waveforms (black, double-integrated from acceleration) and five randomly selected HighFEM-2 synthetics (alternating red shades for visibility) for a single station ($V_{S30}=370$~m/s, hypocentral distance 61.3~km) during the Norcia, 30 October 2016, $M_w$ 6.5 event. Peak absolute amplitudes are shown above each trace.}
    \label{fig:wavegen_displ}
\end{figure}

\begin{figure}[!htp]
    \centering
    \includegraphics[width=\linewidth]{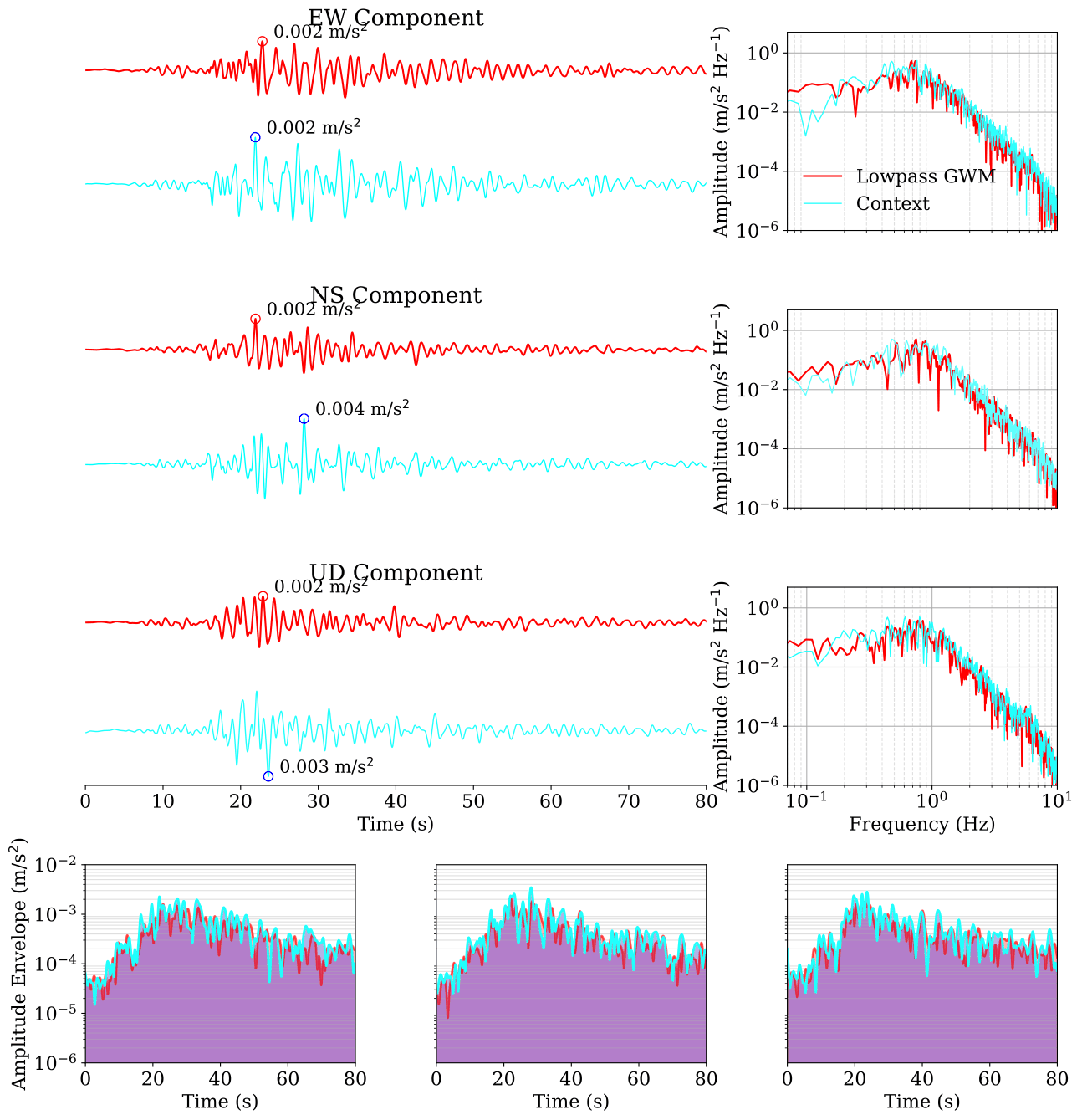}
    \caption{Comparison of low-pass-filtered traces at their corner frequency ($f_c = 0.78$ Hz) with the corresponding conditioning waveform for all three components, for conditioning parameters: $M_w=5.1$, $V_{S30}=534$ m/s; station at $12.93^\circ$E, $43.18^\circ$N, elevation 479.0 m; hypocenter depth 10 km at $13.27^\circ$E, $42.54^\circ$N; hypocentral distance 77.6 km. (upper panel) Time-domain waveforms and corresponding amplitude spectra: HighFEM-2 synthetics (red) vs. conditioning waveform (cyan). (lower panel) Amplitude envelopes for the three components (from left to right: EW, NS, and vertical), with synthetics in red and conditioning waveform in cyan. Legend terms, Context: conditioning waveform; Lowpass GWM: low-pass-filtered HighFEM-2 synthetic.}
    \label{fig:compare_low_freq_12220}
\end{figure}

\begin{figure}[!htp]
    \centering
    \includegraphics[width=\linewidth]{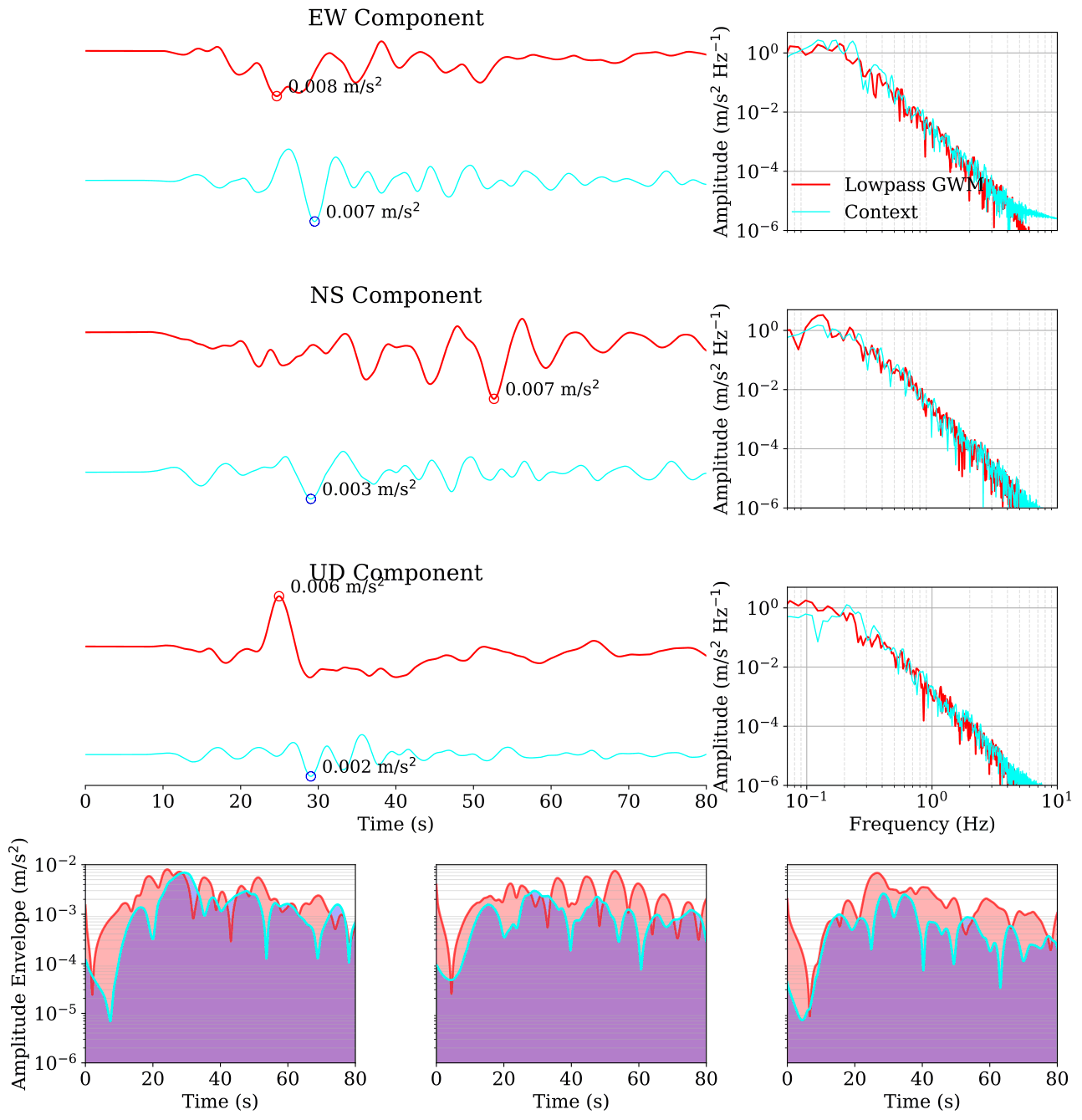}
    \caption{Comparison of low-pass-filtered traces at their corner frequency ($f_c = 0.16$ Hz) with the corresponding conditioning waveform for all three components, for conditioning parameters: $M_w=6.5$, $V_{S30}=607$ m/s; station at $12.57^\circ$E, $43.36^\circ$N, elevation 960.0 m; hypocenter depth 10 km at $13.11^\circ$E, $42.83^\circ$N; hypocentral distance 74.71 km. (upper panel) Time-domain waveforms and corresponding amplitude spectra: HighFEM-2 synthetics (red) vs. conditioning waveform (cyan). (lower panel) Amplitude envelopes for the three components (from left to right: EW, NS, and vertical), with synthetics in red and conditioning waveform in cyan. Legend terms, Context: conditioning waveform; Lowpass GWM: low-pass-filtered HighFEM-2 synthetic.}
    \label{fig:compare_low_freq_9850}
\end{figure}

\begin{figure}[!htp]
    \centering
    \includegraphics[width=\linewidth]{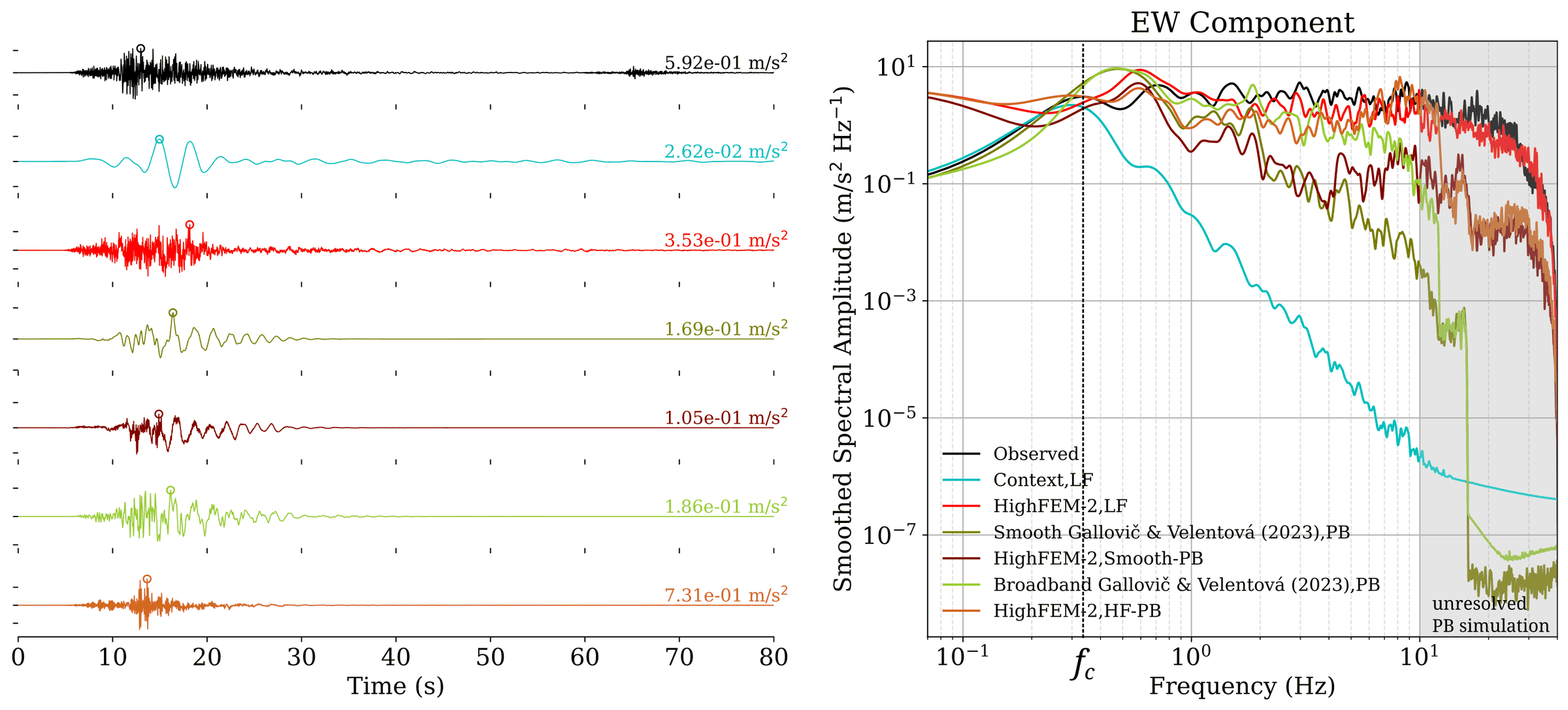}
    \caption{Waveform comparison between observations, physics-based simulations, and HighFEM-2 using different low-frequency inputs for the 2016 Amatrice earthquake ($M_w$ 6.0 used for conditioning) at station IV.TERO component East-West. Left: Time-domain comparison; all traces are aligned 5~s after the start of each waveform. Right: smoothed Fourier amplitude spectra of the corresponding waveforms. The simulations are resolved up to 10~Hz (gray area: unresolved band); the dotted line marks $f_c$. Legend terms, Context,LF: conditioning waveform from the observed record low-pass filtered at $f_c$; Smooth and Broadband Gallovi\v{c} \& Velentov\'a (2023),PB: smooth and rough physics-based simulations; HighFEM-2,LF, HighFEM-2,Smooth-PB, and HighFEM-2,HF-PB: synthetics conditioned on the filtered record, the smooth simulation, and the rough simulation.}
    \label{fig:compare_acc_ew}
\end{figure}

\begin{figure}[!htp]
    \centering
    \includegraphics[width=\linewidth]{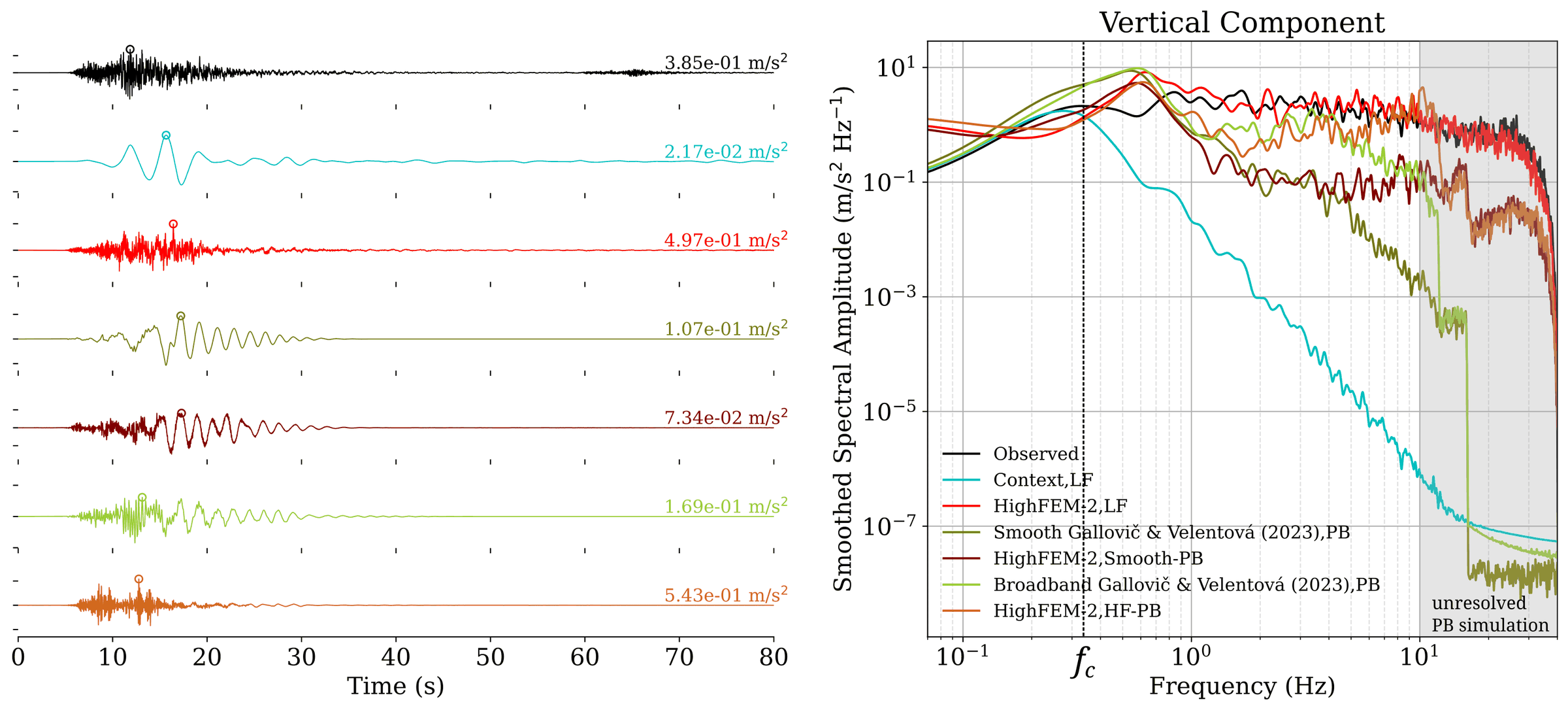}
    \caption{Same as Figure \ref{fig:compare_acc_ew} but for the vertical component.}
    \label{fig:compare_acc_ud}
\end{figure}

\begin{figure}[!htp]
    \centering
    \includegraphics[width=\linewidth]{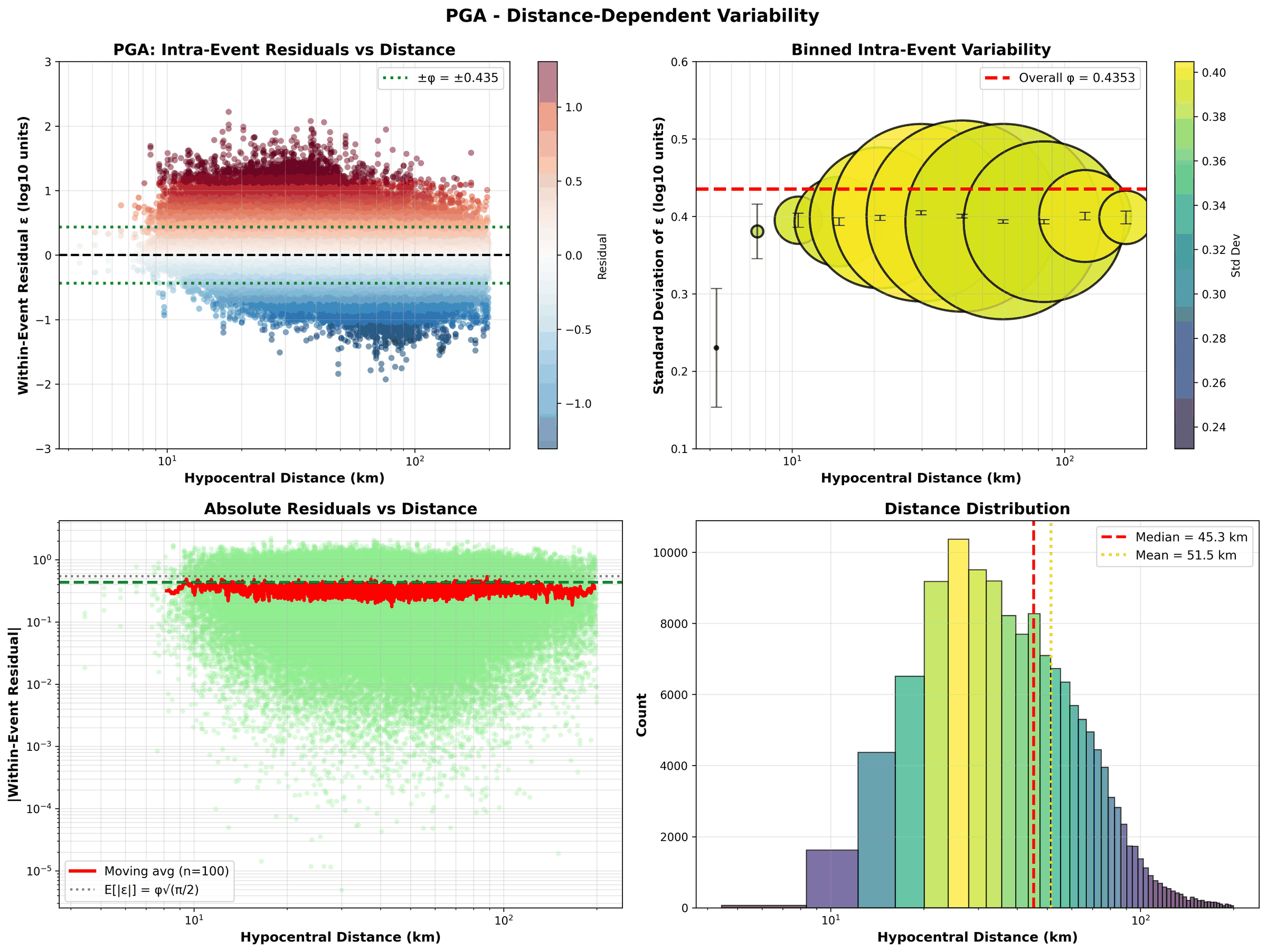}
    \caption{Intra-event residual variability of Peak Ground Acceleration (PGA) as a function of hypocentral distance using test set. HighFEM-2 is compared with the observed data.}
    \label{fig:variability_pga}
\end{figure}

\begin{figure}[!htp]
    \centering
    \includegraphics[width=\linewidth]{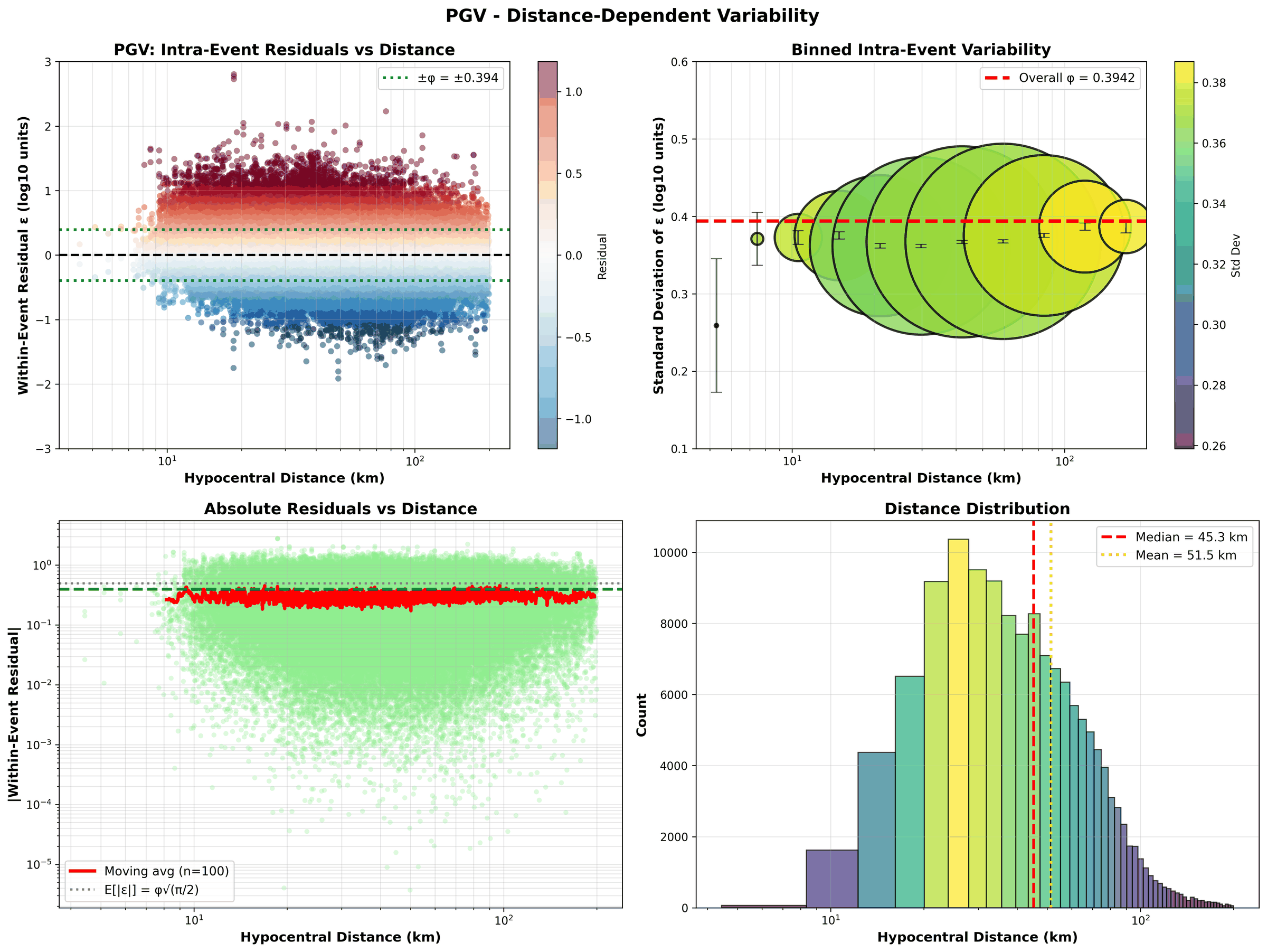}
    \caption{Intra-event residual variability of Peak Ground Velocity (PGV) as a function of hypocentral distance using test set. HighFEM-2 is compared with the observed data.}
    \label{fig:variability_pgv}
\end{figure}

\begin{figure}[!htp]
    \centering
    \includegraphics[width=\linewidth]{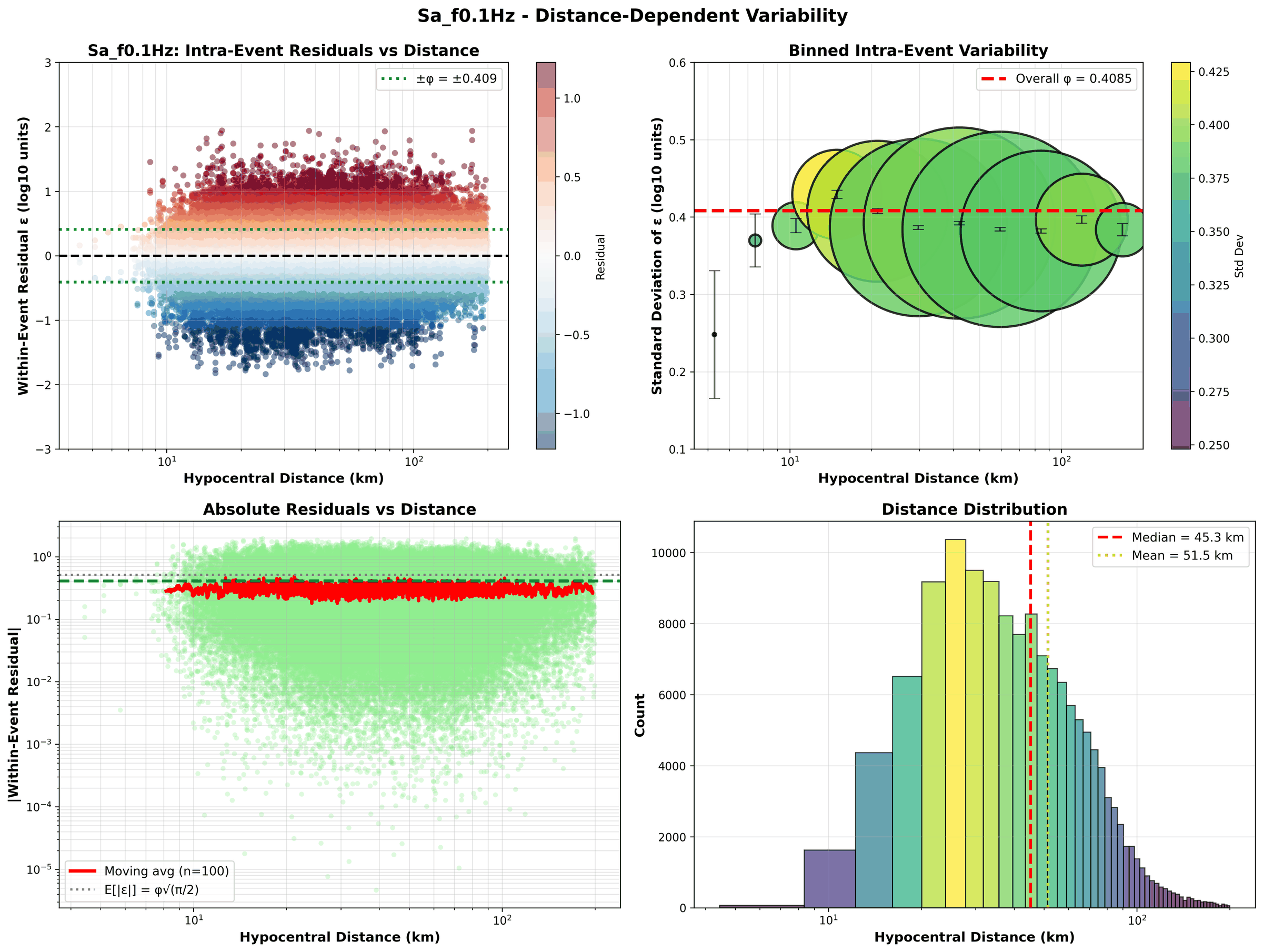}
    \caption{Intra-event residual variability of pseudo-spectral acceleration (PSA) of period 0.1 s as a function of hypocentral distance using test set. HighFEM-2 is compared with the observed data.}
    \label{fig:variability_sa01}
\end{figure}

\begin{figure}[!htp]
    \centering
    \includegraphics[width=\linewidth]{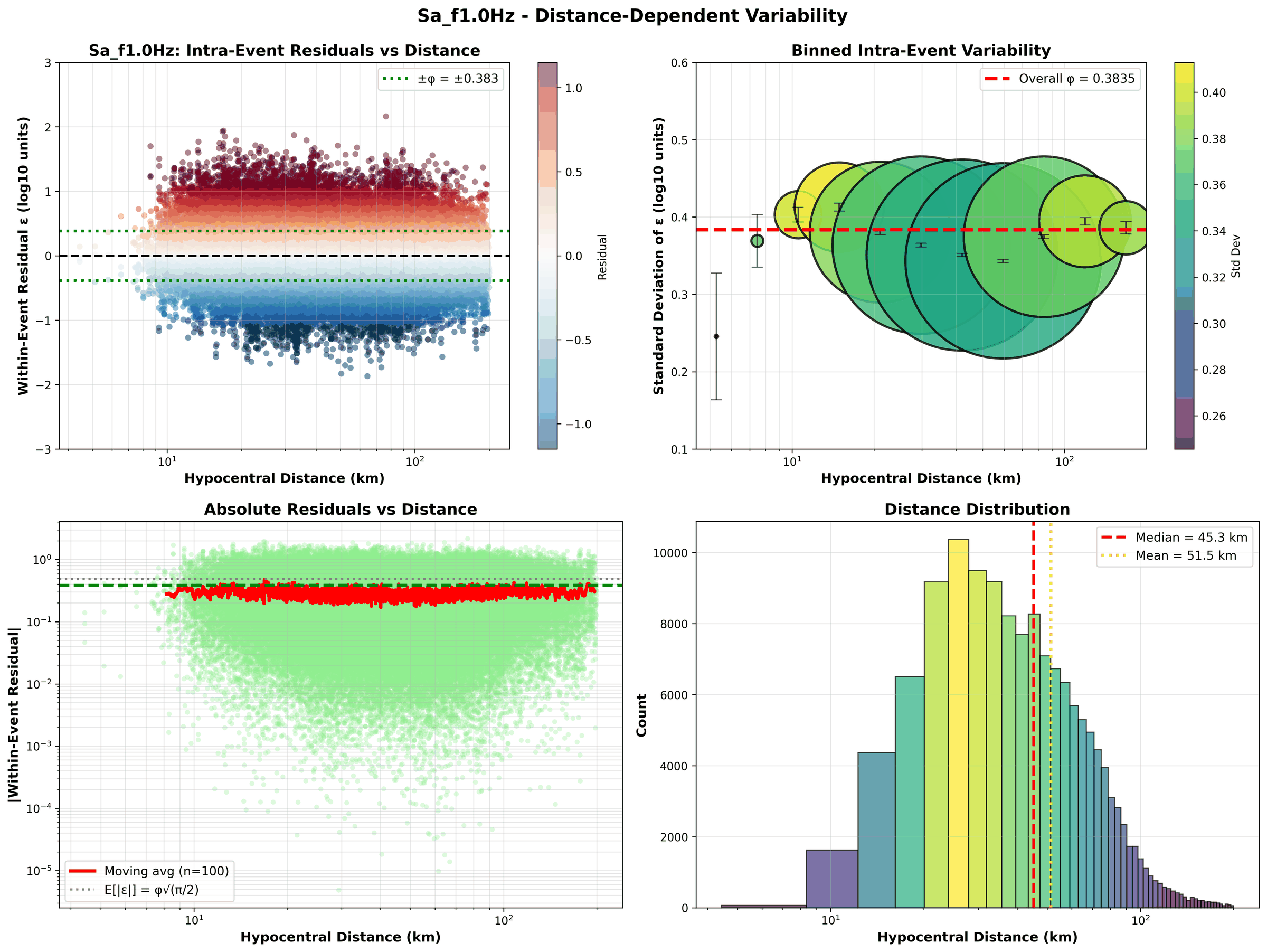}
    \caption{Intra-event residual variability of pseudo-spectral acceleration (PSA) of period 1 s as a function of hypocentral distance using test set. HighFEM-2 is compared with the observed data.}
    \label{fig:variability_sa1}
\end{figure}

\begin{figure}[!htp]
    \centering
    \includegraphics[width=\linewidth]{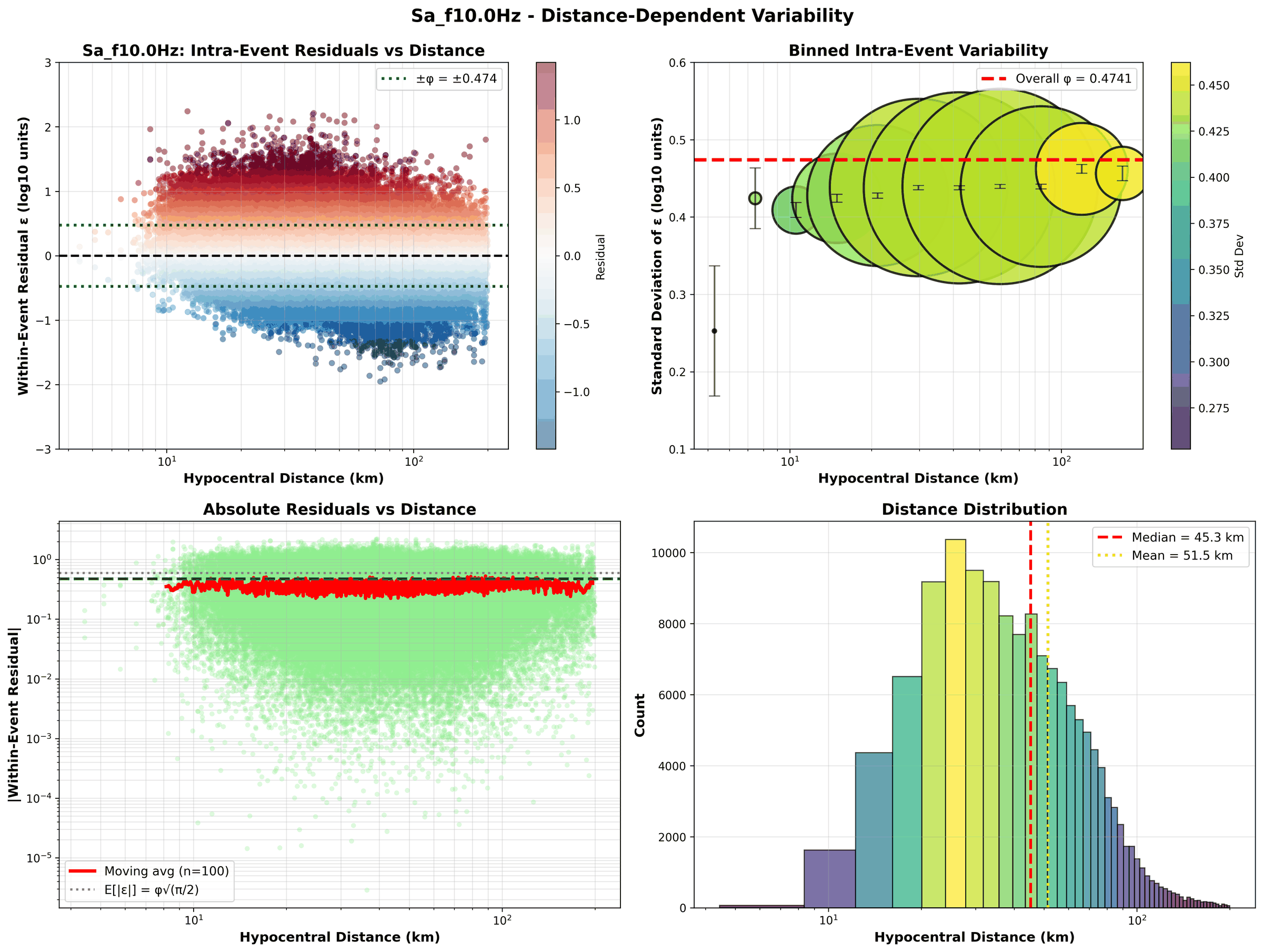}
    \caption{Intra-event residual variability of pseudo-spectral acceleration (PSA) of period 10 s as a function of hypocentral distance using test set. HighFEM-2 is compared with the observed data.}
    \label{fig:variability_sa10}
\end{figure}

The supplementary figures show:
\begin{enumerate}
  \item statistics of the selected metadata (Figure~\ref{fig:statistics_data});
  \item waveform comparisons between observed data and the four generative models (Figures~\ref{fig:compare_model_19}--\ref{fig:compare_model_10572});
  \item time-domain envelopes and Fourier amplitude spectra for all components and further distance bins (Figures~\ref{fig:envelope_supp_EW}--\ref{fig:spectra_supp_UD});
  \item integrated velocity and displacement waveforms and realization ensembles (Figures~\ref{fig:integration_waveforms_supp}--\ref{fig:wavegen_displ});
  \item further examples of low-frequency preservation (Figures~\ref{fig:compare_low_freq_12220} and~\ref{fig:compare_low_freq_9850});
  \item the East--West and vertical components of the Amatrice comparison (Figures~\ref{fig:compare_acc_ew} and~\ref{fig:compare_acc_ud});
  \item intra-event residual variability of PGA, PGV, and PSA as a function of hypocentral distance (Figures~\ref{fig:variability_pga}--\ref{fig:variability_sa10}).
\end{enumerate}

\subsection*{Additional Text}

Text S1 describes the data-filtering and preprocessing steps used in the main study (Figures~\ref{fig:filter_envelope}--\ref{fig:picks}).

\clearpage

\section*{Figures:}

\clearpage

\section*{Text S1: Steps for Data Filtering}

We filter the INSTANCE data in three steps, described below.

\section*{Step 1}
Step 1 selects records by their metadata, computes the corner frequency, and extracts fixed time windows. Because the INSTANCE dataset exceeds 400~GB, the data are processed in chunks.

\section*{Data Configuration and Initialization}
\subsection*{Dataset Structure}
The processing uses two data sources: a metadata file and waveform data. The metadata include event parameters, station information, and trace-quality metrics. The waveform data are stored in a large HDF5 file containing three-component seismic traces, organized by trace identifiers with datasets for the E (east), N (north), and Z (vertical) components.

\subsection*{Processing Parameters}
\textbf{Temporal parameters:}
\begin{itemize}
    \item Sampling rate: 100~Hz.
    \item Waveform length: 9{,}000 samples (90~s).
    \item Pre-arrival window offset: 500 samples (5~s).
\end{itemize}

\textbf{Chunking configuration:}
\begin{itemize}
    \item Chunk size: 10{,}000 records per chunk, balancing memory efficiency and I/O overhead.
    \item Maximum memory usage: 50\% of available system memory, to prevent overflow during parallel operations.
\end{itemize}

\section*{Metadata Preparation and Quality Control}
\subsection*{Selection Criteria}
We keep records that meet all of the following criteria:
\begin{itemize}
    \item Minimum magnitude: $M_w=1.4$.
    \item Maximum magnitude: $M_w=6.5$.
    \item Hypocentral distance: $\le 200$~km.
    \item Hypocentral depth: $\le 100$~km.
    \item $V_{S30}$: $\ge 116$~m/s.
    \item P-wave arrival constraints: 500–3{,}500 samples.
    \item Signal-to-noise ratio (all components): $\ge 3$.
    \item Strong-motion records only.
\end{itemize}

\section*{Corner Frequency Calculation}
The filtering approach is based on the Brune source model \cite{brune1970tectonic}, which relates earthquake magnitude to the source corner frequency. The corner frequency ($f_c$) marks the transition from a flat amplitude spectrum to $\omega^{-2}$ decay. 

\subsection*{Corner Frequency Calculation and Low-Pass Filtering}
The $f_c$ calculation uses stress-drop ($\Delta \sigma$) estimates from \cite{bindi2020reliability}, taken as $0.1$, $0.2$, $0.8$, $1.5$, and $7.9$~MPa for the magnitude ranges
$M_w<2.0$, $2.0\le M_w<3.0$, $3.0\le M_w<4.0$, $4.0\le M_w<5.0$, and $M_w\ge 5.0$, respectively. The steps are:
\begin{enumerate}
    \item Convert moment magnitude ($M_w$) to seismic moment ($M_0$):
    \begin{equation}
        M_0 = 10^{1.5 M_w + 9.1}.
    \end{equation}
    \item Compute $f_c$ from Brune’s relation:
    \begin{equation}
        f_c = k\,\beta \left(\frac{\Delta \sigma}{M_0}\right)^{1/3},
    \end{equation}
    where $k = 2.34/2\pi \approx 0.49$ \cite{brune1970tectonic} and $\beta = 3500$~m/s.
\end{enumerate}

To obtain the conditioning waveform, we low-pass filter each record at $f_c$ with a fourth-order zero-phase Butterworth filter.

\section*{Waveform Extraction and Windowing}
For each valid record, we extract a 90~s window that begins 5~s (500 samples) before the P-wave arrival:
\begin{enumerate}
    \item Read the P-wave arrival sample index from the metadata.
    \item Compute the window start and end indices.
    \item Validate boundaries: if the start index $< 0$, discard the record due to insufficient pre-event data; if the end index exceeds the total number of samples, discard the record due to incomplete recording.
\end{enumerate}

\section*{Step 2}
In the second step, we remove waveforms that contain a secondary event within the same 90~s window. To detect them, we apply an automatic quality control consisting of signal-envelope extraction, decay-curve fitting, spike detection, and quality assessment. The procedure is as follows:
\begin{enumerate}
    \item \textbf{Signal-envelope extraction}\\
    The envelope of the seismogram is extracted using the Hilbert transform, which provides the instantaneous amplitude of the signal. We compute the analytic signal from the real-valued waveform and then take its magnitude:
    \[
      \tilde{x}(t) = x(t) + \mathrm{j}\,\mathcal{H}\{x(t)\}, \qquad
      E(t) = |\tilde{x}(t)| = \sqrt{x^2(t) + \mathcal{H}\{x(t)\}^2},
    \]
    where $\tilde{x}(t)$ is the analytic signal, $x(t)$ is the original waveform, and $\mathcal{H}\{\cdot\}$ denotes the Hilbert transform.

    \item \textbf{Envelope smoothing}\\
    To reduce noise and facilitate decay-curve fitting, the envelope is smoothed with a Gaussian kernel:
    \[
      E_s(t) = \int_{-\infty}^{\infty} E(\tau)\, G(t-\tau,\sigma)\, d\tau, \qquad
      G(t,\sigma) = \frac{1}{\sigma\sqrt{2\pi}} \exp\!\left(-\frac{t^2}{2\sigma^2}\right),
    \]
    where $E_s(t)$ is the smoothed envelope, $\sigma = w/4$ is one quarter of the smoothing-window length $w$.

    \item \textbf{Exponential decay-curve fitting}\\
    The post-peak envelope is modeled with an exponential function:
    \[
      E_{\text{fit}}(t) = A\, e^{-\alpha t} + C,
    \]
    where $A$ is the initial amplitude above baseline, $\alpha$ is the decay rate (attenuation coefficient), $C$ is the baseline offset, and $t$ is time relative to the peak. Initial parameter guesses are
    $A_{\text{init}} = E(t_{\text{peak}})$, $C_{\text{init}} = P_{10}(E)$ (the 10th percentile of $E$), and
    $\alpha_{\text{init}} = \ln(2)/t_{1/2}$, where $t_{1/2}$ is the time at which $E(t) \le 0.5\,A_{\text{init}}$.
    The fitting region is
    \[
      t_{\text{start}} = \arg\max_t E(t), \qquad
      t_{\text{end}} = t_{\text{start}} + \min\!\bigl\{L_{\text{fit}},\, 0.95\,(L_{\text{total}} - t_{\text{start}})\bigr\},
    \]
    where $L_{\text{fit}}$ is the minimum decay length (e.g., 50 samples) and $L_{\text{total}}$ is the total signal length.

    \item \textbf{Goodness-of-fit assessment}\\
    Fit quality is measured with
    \[
      R^2 = 1 - \frac{SS_{\text{res}}}{SS_{\text{tot}}}, \qquad
      SS_{\text{res}} = \sum_{i=1}^n (E_i - E_{\text{fit},i})^2, \qquad
      SS_{\text{tot}} = \sum_{i=1}^n (E_i - \bar{E})^2,
    \]
    where $E_i$ are the observed envelope values, $E_{\text{fit},i}$ are the fitted values, $\bar{E}$ is the mean of $E$, and $n$ is the number of points. Fits are accepted when $R^2 \ge 0.7$.

    \item \textbf{Spike detection}\\
    Points are flagged as spikes if they exceed the fitted curve by a threshold factor $\theta$:
    \[
      \text{Spike}[i] =
      \begin{cases}
        1, & \text{if } E[i] > \theta\, E_{\text{fit}}[i],\\
        0, & \text{otherwise},
      \end{cases}
    \]
    with $\theta = 1.5$.

    \item \textbf{Spike metrics}\\
    The spike count and percentage are
    \[
      N_{\text{spikes}} = \sum_{i=1}^n \text{Spike}[i], \qquad
      P_{\text{spikes}} = \frac{N_{\text{spikes}}}{n}\times 100\%,
    \]
    where $n$ is the number of points in the fitting region.
\end{enumerate}

\begin{figure}[htp!]
    \centering
    \includegraphics[width=0.9\linewidth]{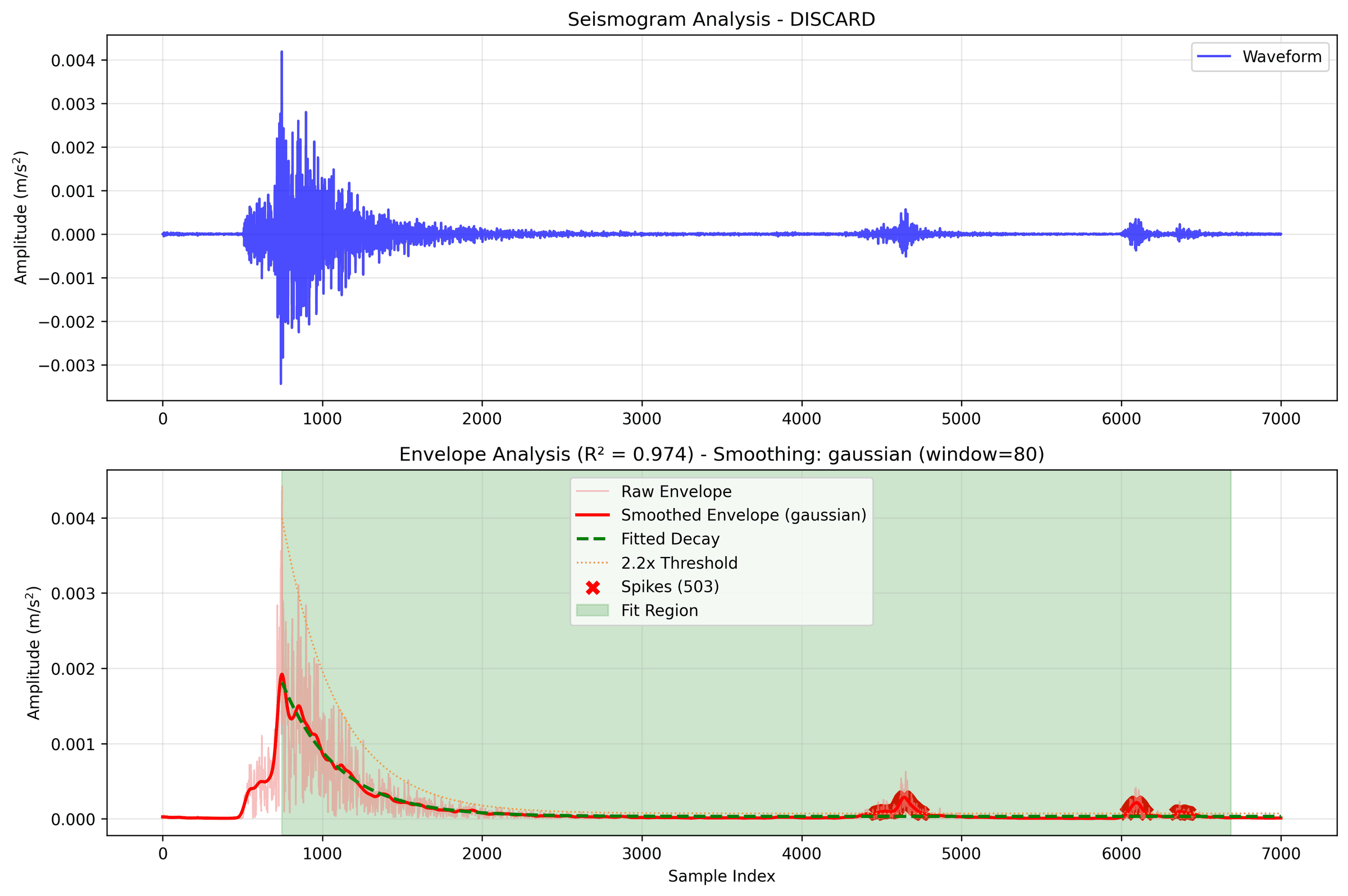}
    \caption{Step 2 analysis, example of removed traces. Top panel: waveform containing more than one event within a 90 s window. Bottom panel: envelope-based quality control. The translucent red curve shows the raw waveform envelope; the solid red curve is the Gaussian-smoothed envelope; the dashed green curve is the fitted decay; red crosses mark detected spikes; the green shaded band indicates the fitting region.}
    \label{fig:filter_envelope}
\end{figure}

Waveforms are discarded if more than one spike is detected (i.e., $N_{\text{spikes}} > 1$).
Figure~\ref{fig:filter_envelope} illustrates the procedure and an example of a discarded waveform.

\section*{Step 3}
\begin{figure}[htp!]
    \centering
    \includegraphics[width=0.9\linewidth]{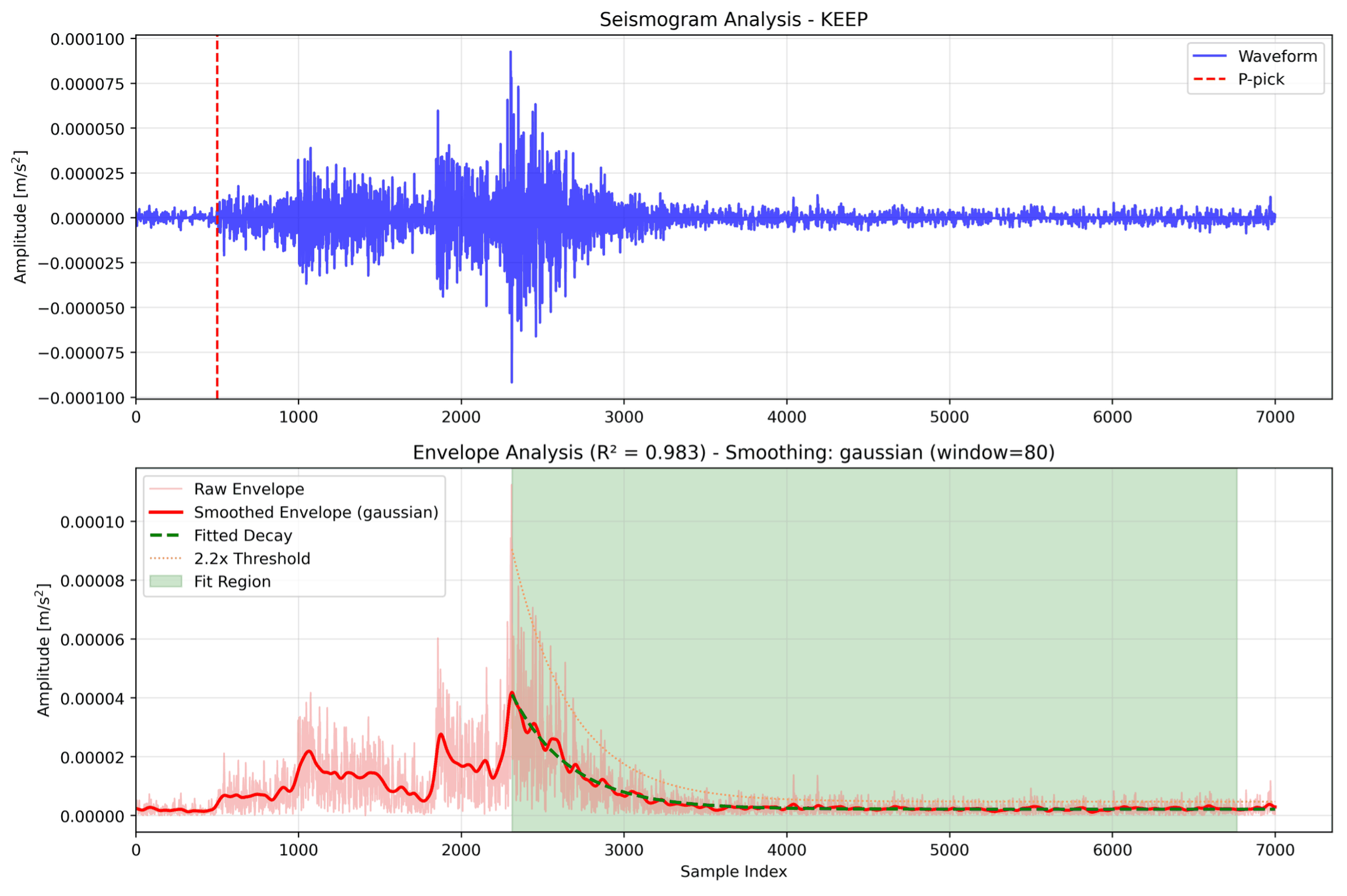}
    \caption{Example of a 90 s window containing multiple events that passed the Step 2 filter.}
    \label{fig:filter_step3}
\end{figure}

\begin{figure}
    \centering
    \includegraphics[width=0.9\linewidth]{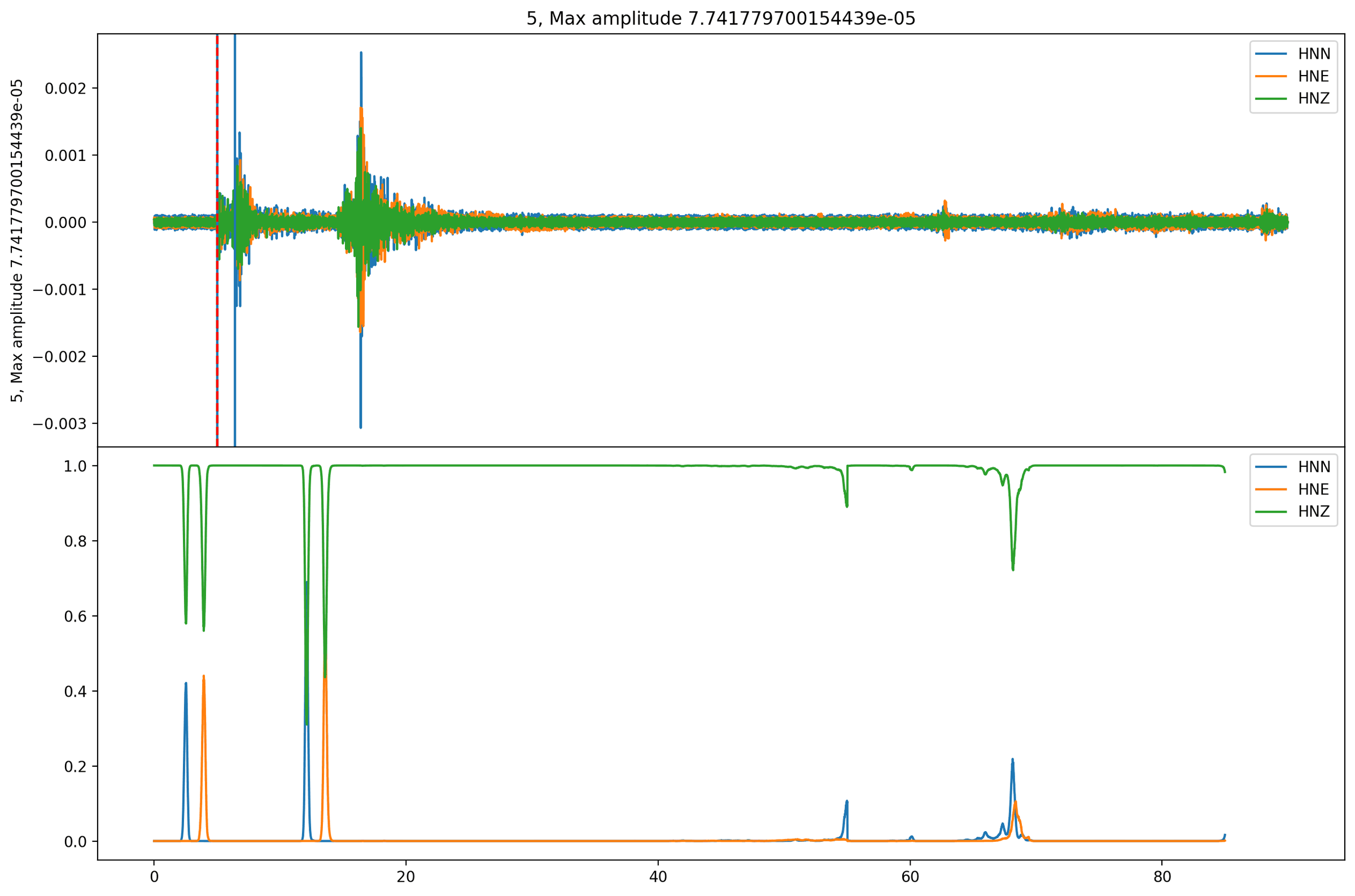}
    \caption{Example of more than one P-picks within time window of 90 s.}
    \label{fig:picks}
\end{figure}

Some problematic waveforms escape Step~2, for example when a secondary event within the window has a larger amplitude (Figure~\ref{fig:filter_step3}). To address this, we apply an additional filter based on automatic phase picking with PhaseNet \cite{zhu2019phasenet}, pre-trained on the INSTANCE dataset. We discard any trace with more than one P pick within a 5~s interval (Figure \ref{fig:picks}).

\clearpage
\bibliography{references}